\documentclass[showpacs,twocolumn,preprintnumbers,amsmath,amssymb,amsfonts,aps,prb,floatfix,superscriptaddress,longbibliography]{revtex4-1}

\usepackage{epsfig, epstopdf}
\usepackage{graphicx, color, textcomp} % Include figure files
\usepackage{dcolumn} % Align table columns on decimal point
\usepackage{bm} % bold math
\usepackage[caption=false]{subfig}

\begin{document}

\title{Theory of surface spectroscopy for noncentrosymmetric superconductors}

\author{Niclas~Wennerdal}
\affiliation{Department of Physics, Royal Holloway,
University of London, Egham, Surrey TW20 0EX, United Kingdom}

\author{Matthias~Eschrig}
\affiliation{Department of Physics, Royal Holloway,
University of London, Egham, Surrey TW20 0EX, United Kingdom}
%\affiliation{SEPnet and Hubbard Theory Consortium, Department of Physics, Royal Holloway, University of London, Egham, Surrey TW20 0EX, United Kingdom}

\date{14 December 2016}

\begin{abstract}
We study noncentrosymmetric superconductors with the tetrahedral $T_d$, tetragonal $C_{4v}$, and cubic point group $O$. The order parameter is computed self-consistently in the bulk and near a surface for several different singlet to triplet order parameter ratios. It is shown that a second phase transition below $T_c$ is possible for certain parameter values. In order to determine the surface orientation's effect on the order parameter suppression, the latter is calculated for a range of different surface orientations. For selected self-consistent order parameter profiles the surface density of states is calculated showing intricate structure of the Andreev bound states (ABS) as well as spin polarization. The topology's effect on the surface states and the tunnel conductance is thoroughly investigated, and a topological phase diagram is constructed for open and closed Fermi surfaces showing a sharp transition between the two for the cubic point group $O$.
\end{abstract}


\maketitle

\section{Introduction}

Non-centrosymmetric materials lack a center of inversion in their crystal lattice. They have attracted increasing attention in recent years due to the fact that spin-orbit interaction has a strong effect on their physical properties.\cite{SigristBook,RevModPhys.83.1057,Yip14,Ando15,Nakajima15,NCMag,thermopower}
In crystals with a center of inversion the band-diagonal elements of the spin-orbit interaction in a Bloch basis, ${\bf L}_{nn}({\bf k})$, vanish by symmetry.
This is not the case for non-centrosymmetric materials, where these diagonal elements can be non-zero and indeed large (30-300 meV).\cite{SamokhinAnnals}
Anderson, in discussing heavy fermion materials, used group classification to study the possibilities for spin-triplet superconductivity in spin-orbit coupled materials.\cite{PhysRevB.30.4000}
Experimental signatures of spin-triplet (as well as spin-singlet) pairing were found in the
non-centrosymmetric heavy-fermion superconductor CePt$_3$Si, discovered in 2004.\cite{PhysRevLett.92.027003} Since then many more non-centrosymmetric superconductors (NCSs) have been identified, including Y$_2$C$_3$,\cite{Y2C3} Li$_2$(Pd$_{1-x}$Pt$_x$)$_3$B,\cite{PhysRevLett.93.247004}  CeIrSi$_3$,\cite{JPSJ.75.043703}  UIr,\cite{0953-8984-16-4-L02}, BiPd, \cite{PhysRevB.84.064518}, and PbTaSe$_2$,\cite{PbTaSe2} amongst others.\cite{PhysRevLett.95.247004, Settai2007844,CeIrGe3,BaAl4,Rh2Ga9,Mg10Ir19B16,Re3W,KneidingerReview} These materials show signs of both spin-singlet and spin-triplet supconductivity to a varying degree. The system Li$_2$(Pd$_x$Pt$_{1-x}$)$_3$B has been studied in more detail,\cite{Badica, 1742-6596-153-1-012028} indicating that the difference between the two end compounds, $x=0$ and $x=1$, can at least in part be explained by a dominating triplet component for $x=0$, i.e. Li$_2$Pt$_3$B, whereas Li$_2$Pd$_3$B seems to have a dominating $s$-wave singlet component, indicated by the rather low value of the upper critical magnetic field extrapolated to zero temperature. Some systems, like LaNiC$_2$ and LaNiGa$_2$, are candidates for a non-unitary spin-triplet pairing state.\cite{Hillier09}

Furthermore, it has been shown that,
as spin-orbit interaction is time-reversal invariant, 
these superconductors can be topologically non-trivial.\cite{Sato09,Tanaka09,Kitaev09,Schnyder09,PhysRevB.78.195125,RevModPhys.83.1057,Ando15,Samokhin15,Scheurer15,Scheurer16}
The topology and the singlet-triplet admixture are both a consequence of the spin-orbit coupling (SOC) term in the Hamiltonian of these materials, 
which is derived from the non-relativistic limit of the Dirac equation and is proportional to the gradient of the crystal lattice potential. 
The lack of a center of inversion in the unit cell allows the gradient of the potential to be large throughout the Brillouin zone (BZ), and thus the SOC cannot be neglected.
The above-mentioned property, that diagonal elements of the SOC in a Bloch basis are in general non-vanishing in non-centrosymmetric materials,
allows to study the effects of the SOC in a minimal one-band model, \cite{SamokhinAnnals} which is not possible in centrosymmetric materials. 

In this paper we theoretically study NCSs with the emphasis on self-consistent superconducting order parameters for various surface orientations, as well as for all the topological phases of the crystal point groups $T_d$, $C_{4v}$, and $O$ with a closed Fermi surface. 
The SOC vector is expanded in terms of harmonic functions, constrained by the symmetries of the point group, to second order. The relative weight of the first and second order terms is parameterized by $g_2$. Second order terms are investigated for the point groups $C_{4v}$ and $O$: one non-zero value of $g_2$ for $C_{4v}$ and three for $O$. Besides the gapped topologically trivial phases all point groups have one non-trivial gapless phase, and $O$ has, for a closed Fermi surface, four non-trivial gapped phases, and we have chosen values of $g_2$ to correspond to these phases. 
In the literature the point group $C_{4v}$ with $g_2=0$ has been studied extensively,\cite{PhysRevLett.101.127003, EurPhysJB.10.1140, PhysRevLett.92.097001, 1367-2630-6-1-115,springerBook} as well as $O$ with values of $g_2$ equivalent to our choice of $g_2 = 0.7$.\cite{PhysRevLett.97.017006, Og2, PhysRevB.84.020501} 
All results we present in this paper are self-consistent, and for parameter combinations not discussed so far in the literature. 
Non-self-consistent results for surface spectra for various point groups and surface orientations were obtained in Ref. \onlinecite{springerBook}, and subsequently in Ref. \onlinecite{PhysRevB.84.020501}.
Topological aspects were in the focus of attention in Ref. \onlinecite{PhysRevB.85.024522}, whereas in Ref. \onlinecite{PhysRevB.91.180503} the possibility of a surface instability was suggested.

\section{Theory}
\subsection{Normal state band dispersion}
Within an effective one-band model,
the SOC term in the Hamiltonian is given by $H^\text{SO}_{\bf k} = \alpha {\bf l}_{\bf k} \cdot \boldsymbol{\sigma}$, where $\alpha$ is the SOC strength, $\boldsymbol{\sigma} = (\sigma_1,\sigma_2,\sigma_3)$ is a vector of spin Pauli matrices, and ${\bf l}_{\bf k}$ is the SOC vector which is real, 
invariant under crystal point group operations $g$, 
\begin{align}
{\bf l}_{\bf k} \equiv {\bf l}({\bf k})= g {\bf l}({g^{-1}{\bf k}}), 
\label{inv}
\end{align}
and odd in ${\bf k}$, ${\bf l}_{-{\bf k}}=-{\bf l}_{\bf k}$.
We normalize the SOC vector such that its maximum magnitude within the BZ is unity, $\max_{\bf k\in BZ}|{\bf l}_{\bf k}|=1$.

The kinetic part of the normal-state Hamiltonian can thus be written as
\begin{equation}
\label{NormalHamiltonian}
\hat{\mathcal{H}}_{\bf k} = \sum_{{\bf k}\alpha\beta}c^\dagger_{{\bf k}\alpha} \left(\xi_{\bf k} \sigma_0 + \alpha {\bf l}_{\bf k} \cdot \boldsymbol{\sigma}\right)_{\alpha\beta } c_{{\bf k}\beta}
\end{equation}
with $ \xi_{\bf k} = \epsilon_{\bf k} - \mu$, where $\epsilon_{\bf k}$ is 
the band dispersion in the absence of SOC
(we will use for simplicity a nearest-neighbor tight-binding dispersion),
$\mu$ is the chemical potential, and $c_{{\bf k}\alpha}$ ($c^\dagger_{{\bf k}\alpha}$) are fermion annihilation (creation) operators for a quasiparticle with spin $\alpha\in \left\{\uparrow,\downarrow \right\}$. 
We will study simple cubic (CUB) and body centered cubic (BCC) lattices.
The corresponding nearest-neighbor tight binding dispersions are
\begin{align}
\epsilon^\text{CUB}_{\bf k} = t_1\left[\cos\left(k_x\right) + \cos\left(k_y\right) + \cos\left(k_z\right)\right]
\label{d1}
\end{align}
and
\begin{align}
\epsilon^\text{BCC} _{\bf k} = 8t_1\cos\left(k_x/2\right) \cos\left(k_y/2\right) \cos\left(k_z/2\right),
\label{d2}
\end{align}
where $t_1$ is the hopping integral.

The point groups considered here are the cubic point group $O$, relevant for e.g. Li$_2$Pd$_x$Pt$_{3-x}$;\cite{PhysRevLett.93.247004, LiPdBLatticeStructure, PhysRevB.72.174505, Badica, 1742-6596-153-1-012028} the tetragonal point group $C_{4v}$, relevant for e.g. CePt$_3$Si;\cite{PhysRevLett.92.027003} and the tetrahedral point group $T_d$, relevant for e.g. Y$_2$C$_3$.\cite{Yuan2011577} 
We use dispersion \eqref{d1} for the cubic point group, $O$, and for sake of simplicity also for the tetragonal point group, $C_{4v}$, whereas dispersion
\eqref{d2} will be used 
for the tetrahedral point group $T_d$. 
The SOC vectors are obtained as lattice Fourier series, ${\bf l}_{\bf k} = \sum_n {\bf l}_n \sin({\bf k}\cdot {\bf R}_n)$, where ${\bf R}_n$ are Bravais lattice vectors, and where the invariance under point group operations, Eq.~\eqref{inv}, leads to restrictions on the ${\bf l}_n$.\cite{SamokhinAnnals} 

The Hamiltonian, Eq.~\eqref{NormalHamiltonian}, is diagonalized and brought to the so-called helicity basis by the canonical transformation $U_{\bf k} \left( {\bf l}_{\bf k} \cdot \boldsymbol{\sigma}\right) U^\dagger_{\bf k}=|{\bf l}_{\bf k}|\sigma_3$, where
\begin{equation}
U_{\bf k} = \begin{pmatrix} \cos\left(\frac{\theta_l}{2} \right) & e^{-i \phi_l}\sin\left(\frac{\theta_l}{2} \right) \\
-e^{i \phi_l}\sin\left(\frac{\theta_l}{2}  \right)  & \cos\left(\frac{\theta_l}{2} \right)\end{pmatrix}\; ,
\end{equation} 
with $\phi_l = \tan^{-1}(l_y/l_x)$ and $\theta_l = \tan^{-1}(\sqrt{l^2_x + l^2_y}/l_z)$ being the spherical angles of the SOC vector, ${\bf l}_{\bf k} = (l_x,l_y,l_z)^T$, yielding 
\begin{eqnarray}
\label{HelicityHamiltonian}
\hat{\mathcal{H}}_{\bf k} = \sum_{{\bf k}\lambda} \xi^\lambda_{\bf k}b^\dagger_{{\bf k}\lambda} b_{{\bf k}\lambda} &,\quad & b_{{\bf k}\lambda} = \sum_{\alpha} U_{{\bf k}\lambda\alpha} c_{{\bf k}\alpha}
\end{eqnarray}
where the helical index takes the values $\lambda = \{ +,-\}$, and the helical band dispersion is given by $\xi^\pm_{\bf k} = \xi_{\bf k} \pm \alpha |{\bf l}_{\bf k}|$. Note that $\xi^\lambda_{\bf k} = \xi^\lambda_{-{\bf k}}$ even though the SOC vector is antisymmetric. This is a consequence of Eq. \eqref{NormalHamiltonian} being time-reversal invariant. Furthermore, the quasiparticle spin is fixed with respect to its momentum on each band, being parallel ($\lambda = +$) or antiparallel ($\lambda = -$) to ${\bf l}_{\bf k}$.

%%%%%%%%%%%%%%%%%%%%%%%%%%
\subsection{Superconducting state}
\label{SectionQuasiclassical}

Superconductivity is modeled within the Nambu-Gor'kov formalism. 
Under the canonical transformation defined above the Nambu spinor $\hat C_{\bf k} = (c_{{\bf k}\uparrow}, c_{{\bf k}\downarrow}, c^\dagger_{-{\bf k}\uparrow}, c^\dagger_{-{\bf k}\downarrow})^T$ transforms into its helical equivalent 
$\hat B_{\bf k} = (b_{{\bf k}+}, b_{{\bf k}-}, b^\dagger_{-{\bf k}+}, b^\dagger_{-{\bf k}-})^T \equiv  \hat U_{\bf k} \hat C_{\bf k}$ with $\hat U_{\bf k} \equiv \text{diag}(U_{\bf k}, U^*_{-\bf k})$, and the "hat" denotes Nambu structure. 
It is straightforward to construct $4\times 4$ helical Green functions, e.g. the retarded $\hat G^\text{R}_{{\bf k}_1 {\bf k}_2}(t_1, t_2) = -i \Theta(t_1 - t_2) \langle\{\hat B_{{\bf k}_1}(t_1) , \hat B^\dagger_{{\bf k}_2}(t_2) \} \rangle_\mathcal{H}$, where $\Theta$ is the Heaviside step function, $\langle \bullet \rangle_\mathcal{H}$ denotes a grand canonical average, $\{\bullet,\bullet\}$ is an anticommutator, and $\hat B_{\bf k}(t)$ a Heisenberg operator. 
The quasiclassical propagator is obtained by integrating out fast oscillations from the full Green functions. 
In the case when the magnitude of the SOC is much smaller than the Fermi energy, $\alpha \ll E_F$,
it suffices to integrate over $\xi_{\bf k}$ and treat the SOC term perturbatively. For this case,
in Wigner coordinates the quasiclassical propagator is given by $\check g({\bf k}_F,{\bf R}, \epsilon, t) = \int \text{d}\xi_{{\bf k}} \hat\tau_3 \check G ({\bf k},{\bf R}, \epsilon, t)$, with ${\bf k}$ parameterized by $(\xi_{\bf k},{\bf k}_F)$,
$\xi_{{\bf k}} = {\bf v}_F\cdot ({\bf k} - {\bf k}_F)$, $\hat{\boldsymbol{\tau}} = (\hat\tau_1,\hat\tau_2,\hat\tau_3)$ are Pauli matrices in particle-hole space, and the "check" denotes Keldysh matrix structure.

The SOC term enters the transport equations as a source term. 
Within this approximation the Eilenberger equation \cite{Eilenberger} for the quasiclassical Green function takes the following form in the helicity basis
\begin{equation}
\label{EilenbergerEq}
i {\bf v}_F\cdot\nabla_{\bf R} \hat g^\text{R,A,M} + [z \hat \tau_3 - \hat \Delta - \hat v_\text{SO}, \hat g]^\text{R,A,M}  = \hat 0
\end{equation} 
with $z=i\epsilon_n = i\pi T(2n+1)$ for Matsubara, and $z = \epsilon \pm i 0^+$ for retarded (advanced), quantities. $[\bullet, \bullet]$ is a commutator, the SOC term is $\hat v_\text{SO} = \alpha |{\bf l}_{{\bf k}_F}|\sigma_3 \hat \tau_0$, and the gap has the form
\begin{equation}
\hat \Delta = \begin{pmatrix} 0 & \Delta \\ \tilde\Delta & 0 \end{pmatrix}
\end{equation} 
where the "tilde operation" is defined as the particle-hole conjugate, $\tilde Q({\bf k}_F, {\bf R}, z, t) \equiv Q^*(-{\bf k}_F, {\bf R}, -z^*, t)$. 
Eq. \eqref{EilenbergerEq} is supplemented by the normalization condition $(\hat g^\text{R,A,M})^2= -\pi^2 \hat 1$. 
In order to simplify notation, we will henceforth drop the subscript $F$ at the Fermi momentum; all momenta in the quasiclassical theory are Fermi momenta. The subscript will be written out only when it is necessary to avoid confusion.
We consider time-independent situations, such that the time variable $t$ will be dropped from here on.

The lack of a center of inversion allows for an admixture of spin-singlet and spin-triplet pairing.\cite{EurPhysJB.10.1140} The spin-triplet vector is set to be parallel to the SOC vector in order to maximize $T_c$.\cite{PhysRevLett.92.097001} In spin basis the order parameter is written 
\begin{align}
\Delta({\bf k}) = \mathcal{Y}_{\bf k}(\Delta_s + \Delta_t {\bf l}_{\bf k} \cdot \boldsymbol{\sigma})i\sigma_2, 
\end{align}
where $\mathcal{Y}_{\bf k}$ is a crystal basis function corresponding to irreducible representation of the dominant pairing channel, and $\Delta_s$ and $\Delta_t$ are referred to as the singlet and triplet component, respectively. 
In the helicity basis the order parameter takes the form $\Delta({\bf k}) = \mathcal{Y}_{\bf k} \cdot \text{diag}(\Delta_+({\bf k}) t_+({\bf k}), \Delta_-({\bf k}) t_-({\bf k}))$, where 
\begin{align}
\Delta_\pm ({\bf k})= \Delta_s \pm \Delta_t |{\bf l}_{\bf k}|, 
\end{align}
and the phase factors are given by 
\begin{align}
t_\pm({\bf k}) = -e^{\mp i \phi_l({\bf k})},\quad
\phi_l({\bf k}) = \tan^{-1}(l_y/l_x). 
\label{phase}
\end{align}
Note that $t_\pm(-{\bf k}) = - t_\pm({\bf k}) $.

Eq. \eqref{EilenbergerEq} can be parameterized in terms of coherence functions, $\gamma({\bf k}, {\bf R}, z)$ and $\tilde\gamma({\bf k}, {\bf R}, z)$,\cite{PhysRevB.61.9061} in such a way as to automatically fulfill the normalization condition,
\begin{eqnarray}
\hat g^\text{R,A,M} &\equiv& \begin{pmatrix} \mathfrak{g} & \mathfrak{f} \\ \tilde{\mathfrak{f}} & \tilde{\mathfrak{g}} \end{pmatrix}^\text{R,A,M} 
=\mp i\pi \left[{\cal N}^{-1} {\cal G}\right]^\text{R,A,M}
\nonumber \\
{\cal N}&=& \begin{pmatrix}(\sigma_0 - \gamma\tilde\gamma) & 0 \\ 0 & (\sigma_0 - \tilde\gamma\gamma)  \end{pmatrix} \nonumber \\
{\cal G}&=&\begin{pmatrix} (\sigma_0 + \gamma\tilde\gamma) &  2\gamma \\ -2\tilde\gamma & -(\sigma_0 + \tilde\gamma\gamma) \end{pmatrix}
\end{eqnarray}
where the top (bottom) sign corresponds to $\hat g^\text{R}$ ($\hat g^\text{A}$), and in the case of $\hat g^\text{M}$, to positive (negative) Matsubara frequencies.
With this, Eq. \eqref{EilenbergerEq} transforms into two decoupled Riccati differential equations, 
\begin{eqnarray}
\label{ricc1}
(i {\bf v}_F\cdot\nabla_{\bf R} + 2z)\gamma &=& \gamma \tilde\Delta\gamma + [\alpha |{\bf l}_{\bf k}|\sigma_3, \gamma] - \Delta \; ,\\
(i {\bf v}_F\cdot\nabla_{\bf R} - 2z)\tilde \gamma &=& \tilde \gamma \Delta\tilde \gamma + [\alpha |{\bf l}_{\bf k}|\sigma_3, \tilde \gamma] - \tilde \Delta \; .
\label{ricc2}
\end{eqnarray}
In the homogeneous case, i.e. in the bulk, the solution is 
$\gamma_h = \mathcal{Y}_{\bf k} \cdot\text{diag}(\gamma_+({\bf k}) t_+({\bf k}), \gamma_-({\bf k}) t_-({\bf k}))$ with the abbreviations $\gamma_\pm = -\Delta_\pm/(z + i\sqrt{|\mathcal{Y}_{\bf k}\Delta_\pm|^2 - z^2})$.
For this case the SOC term drops out.

The surface problem is treated by solving Eqs.~\eqref{ricc1}-\eqref{ricc2} along classical trajectories parallel to ${\bf v}_F$, using the homogeneous solutions as initial conditions at a sufficient distance from the surface. This is done by discretizing the path and treating the order parameter as a series of step functions in the middle between the desired grid points. Each step is solved analytically.\cite{PhysRevB.80.134511} Parameterizing the path as ${\bf R} = {\bf R}_0 + \rho {\bf v}_F$ and writing the order parameter $\Delta(\rho) = \Delta_0 + \Theta(\rho)(\Delta_1 - \Delta_0)$ at one of these steps, $\gamma(\rho)$ with $\rho>0$ is given by 
\begin{equation}
\gamma(\rho) = \gamma_h + e^{i\Omega_1 \rho}\delta_0 \left(e^{i\Omega_2 \rho} + C(\rho) \delta_0  \right)^{-1}
\end{equation}
with $\delta_0 = [\gamma_0 - \gamma_h]$, where $\gamma_0 \equiv \gamma(0)$ is the initial value and $\gamma_h$ is the homogeneous solution for $\rho>0$, $\Omega_1 = z - \gamma_h\tilde\Delta$ and $\Omega_2 = -z + \tilde\Delta\gamma_h$, and $C(\rho) = C_0 e^{i\Omega_1} - e^{i \Omega_2}C_0$, where $C_0$ is the solution to $C_0\Omega_1 - \Omega_2 C_0 = \tilde\Delta$. The solution for $\tilde\gamma(\rho)$ is completely analogous.

The reflection at the surface is in leading approximation (as $\alpha \ll E_F$) considered to be specular in spin space, with the momentum component parallel to the surface, $ {\bf k}_\parallel$, conserved. Writing the momentum for incoming trajectories ${\bf k} = (k_\perp, {\bf k}_\parallel)$ this gives the momentum for outgoing trajectories as $\underline{\bf k} = (-k_\perp, {\bf k}_\parallel)$. Following Ref.~\onlinecite{PhysRevB.61.9061}, incoming (outgoing) quantities are written with lowercase (uppercase) symbols and the surface boundary conditions become 
\begin{align}
U^{\dagger}_{\underline{{\bf k}}} \Gamma(\underline{{\bf k}},\varepsilon)U^*_{-\underline{\bf k}} = \Gamma^s(\underline{{\bf k}},\varepsilon) = \gamma^s({\bf k},\varepsilon) = U^{\dagger}_{\bf k}\gamma({\bf k}, \varepsilon)U^*_{-{\bf k}}
\end{align}
and 
\begin{align}
U^{T}_{{-\bf k}} \tilde{\Gamma}({\bf k},\varepsilon)U_{\bf k} = \tilde{\Gamma}^s({\bf k},\varepsilon) = \tilde{\gamma}^s(\underline{\bf k},\varepsilon) = U^T_{-\underline{\bf k}}\tilde{\gamma}(\underline{\bf k}, \varepsilon)U_{\underline{\bf k}}, 
\end{align}
where the $s$ superscript indicates that the coherence functions are expressed in the spin basis.

\subsection{Gap equation}

The pairing potential in spin space can be written as a sum of singlet, triplet, and a mixture term \cite{EurPhysJB.10.1140}
\begin{widetext}
\begin{eqnarray}
V_{s_1 s_2 s_3 s_4}({\bf k}, {\bf k}') &=& 
\frac{V}{2}  \mathcal{Y}_{\bf k} \mathcal{Y}^*_{{\bf k}'}\left \{ v_s (i\sigma_2)_{s_1 s_2}(i\sigma_2)^\dagger_{s_3 s_4} + 
%\right. \nonumber \\ && \qquad 
v_t ({\bf l}_{\bf k}\cdot \boldsymbol{\sigma}i\sigma_2)_{s_1 s_2}  ({\bf l}_{{\bf k}'}\cdot \boldsymbol{\sigma}i\sigma_2)^\dagger_{s_3 s_4} +  
\right. \nonumber \\ &&  \left.
v_m \left [ ({\bf l}_{\bf k}\cdot \boldsymbol{\sigma}i\sigma_2)_{s_1 s_2} (i\sigma)^\dagger_{s_3 s_4}  + 
(i\sigma)_{s_1 s_2} ({\bf l}_{{\bf k}'}\cdot \boldsymbol{\sigma}i\sigma_2)^\dagger_{s_3 s_4} \right] \right \} \nonumber \\
\end{eqnarray}
where $v_s$, $v_t$, and $v_m$ are free parameters that describe the relative coupling strength of each term, respectively, $V$ is the overall pairing potential strength, and $\mathcal{Y}_{\bf k}$ is the basis function of the irreducible representation with the highest $T_\text{c}$. To avoid ambiguity, we normalize the relative pairing strengths according to 
\begin{align}
v^2_s + v^2_t + v^2_m = 1 
\end{align}
and for later reference introduce spherical coordinates
\begin{eqnarray}
\label{vvv_angles}
(v_s,v_t,v_m) &=& \left(\cos(\phi_v)\sin(\theta_v), \sin(\phi_v)\sin(\theta_v), \cos(\theta_v) \right) \; . \nonumber\\
\label{phiv}
\end{eqnarray}

In helicity space the pairing potential takes the form
\begin{equation}
V({\bf k}, {\bf k}') = \frac{V}{2} \mathcal{Y}_{\bf k} \mathcal{Y}^*_{{\bf k}'}\begin{pmatrix} v_s + v_t|{\bf l}_{\bf k}| |{\bf l}_{{\bf k}'}| - v_m l_+  & v_s - v_t|{\bf l}_{\bf k}| |{\bf l}_{{\bf k}'}| - v_m l_-  \\
v_s - v_t|{\bf l}_{\bf k}| |{\bf l}_{{\bf k}'}| + v_m l_-  & v_s + v_t|{\bf l}_{\bf k}| |{\bf l}_{{\bf k}'}| + v_m l_+ \end{pmatrix}
\end{equation}
\end{widetext}
with $l_\pm = |{\bf l}_{\bf k}| \pm |{\bf l}_{{\bf k}'}|$. 
The self-consistency equation in the Matsubara formalism is expressed in terms of Fermi surface averages $\langle\bullet\rangle$, defined as
\begin{eqnarray}
\langle\bullet\rangle &=& \frac{1}{N_F}\int \frac{\text{d}^2{\bf k}_F}{(2\pi)^3|{\bf v}_F|}(\bullet) \;, 
\quad N_F = 
\int \frac{\text{d}^2{\bf k}_F}{(2\pi)^3|{\bf v}_F|} \; .\quad
\end{eqnarray}
With this, the self-consistency equation takes the form
\begin{eqnarray}
\begin{pmatrix}\Delta_+({\bf k}) \\ \Delta_-({\bf k}) \end{pmatrix} &=& T N_F\sum^{|\epsilon_n| <\epsilon_\text{c}}_{\epsilon_n} \left \langle V({\bf k}, {\bf k}') \begin{pmatrix}\mathfrak{f}_+({\bf k}', \epsilon_n)\ \\ \mathfrak{f}_-({\bf k}', \epsilon_n) \end{pmatrix} \right \rangle_{{\bf k}'}
\end{eqnarray}
where $\mathfrak{f}_\pm$ are defined by
\begin{eqnarray}
\mathfrak{f}({\bf k}, \epsilon_n) &=& \begin{pmatrix} \mathfrak{f}_+({\bf k}, \epsilon_n)t_+({\bf k}) & 0 \\ 0 & \mathfrak{f}_-({\bf k}, \epsilon_n)t_-({\bf k}) \end{pmatrix},
\end{eqnarray}
the phase factors are defined in Eq.~\eqref{phase}, and $\epsilon_c$ is the BCS technical cutoff. 
Using the relations $\Delta_s = \frac{1}{2}(\Delta_+ + \Delta_-)$ and $\Delta_t |{\bf l}_{\bf k}| = \frac{1}{2}(\Delta_+ - \Delta_-)$, the implicit form of the self-consistency equation for the singlet and triplet components of the order parameter reads 
\begin{eqnarray}
  \label{implicit_op}
\begin{pmatrix}\Delta_s \\ \Delta_t \end{pmatrix} &=& T \sum^{|\epsilon_n| < \epsilon_\text{c}}_{\epsilon_n} N_FV\left \langle  A_{\bf k} \begin{pmatrix}\mathfrak{f}_+\\ \mathfrak{f}_- \end{pmatrix} \right \rangle \; ,
  \end{eqnarray}
where
\begin{eqnarray}
A_{\bf k} &=& \frac{1}{2}\mathcal{Y}^*_{\bf  k} \begin{pmatrix}v_s - v_m|{\bf l}_{\bf k}| &  v_s + v_m|{\bf l}_{\bf k}| \\ v_t|{\bf l}_{\bf k}| - v_m & -v_t|{\bf l}_{\bf k}| - v_m \end{pmatrix} \; .
\end{eqnarray}
After elimination of the cutoff and the pairing strength $V$
in favor of the superconducting transition temperature, one obtains
\begin{widetext}
\begin{eqnarray}
\label{op_explicit}
 \begin{pmatrix}\Delta_s \\ \Delta_t \end{pmatrix} &=& \left[ \ln \left ( \frac{T^{\left \langle L_{\bf k} \right \rangle}}{T_c^{\lambda_\text{max}}}\right ) \right]^{-1} T\sum_{\epsilon_n} \left \langle  A_{\bf k} \begin{pmatrix}\mathfrak{f}_+\\ \mathfrak{f}_- \end{pmatrix} - \frac{\pi}{\left | \epsilon_n \right |} L_{\bf k}  \begin{pmatrix}\Delta_s \\ \Delta_t \end{pmatrix} \right \rangle \; ,
\end{eqnarray}
\end{widetext}
where the matrix exponent in the logarithm, $T^{\left \langle L_{\bf k} \right \rangle}$, is taken element-wise, i.e.
\begin{equation}
\left[T^{\left \langle L_{\bf k} \right \rangle}\right]_{ij} \equiv T^{\left \langle \left[ L_{\bf k}\right]_{ij} \right \rangle} \; ,
\end{equation}
and with
\begin{eqnarray}
L_{\bf k} &=& \begin{pmatrix} v_s |\mathcal{Y}_{\bf k}  |^2 & -v_m |\mathcal{Y}_{\bf k} {\bf l}_{\bf k}|^2  \\
-v_m |\mathcal{Y}_{\bf k}  |^2  & v_t |\mathcal{Y}_{\bf k} {\bf l}_{\bf k}|^2 \end{pmatrix}  \; .
\label{L_matrix}
\end{eqnarray}
Furthermore, $\lambda_\text{max} \equiv \max\{\lambda_1,\lambda_2\}$, and $\lambda_{1,2}$ are the eigenvalues of the matrix $\left\langle L_{\bf k} \right\rangle$.
We follow Ref.~\onlinecite{Grein2013} in eliminating the cut-off dependence in close vicinity to the surface as well.
For details on the numerical procedure to achieve self-consistency see appendix \ref{SectionTemperatureDependence}.

\subsection{Bulk superconducting phase}

At $T = T_c$ the self-consistency equation reduces to
\begin{equation}
\label{nucleate}
\ln\left(\frac{2e^{\gamma}\epsilon_c}{\pi T_c}\right)\langle L_{\bf k}\rangle \left(\begin{matrix} \Delta_s \\ \Delta_t \end{matrix} \right) = \frac{1}{N_FV}\left(\begin{matrix} \Delta_s \\ \Delta_t \end{matrix} \right) \; 
\end{equation}
where $\gamma = 0.5772...$ is the Euler-Mascheroni constant.
\begin{figure}[b]
\centering
{\includegraphics[width=0.7\columnwidth]{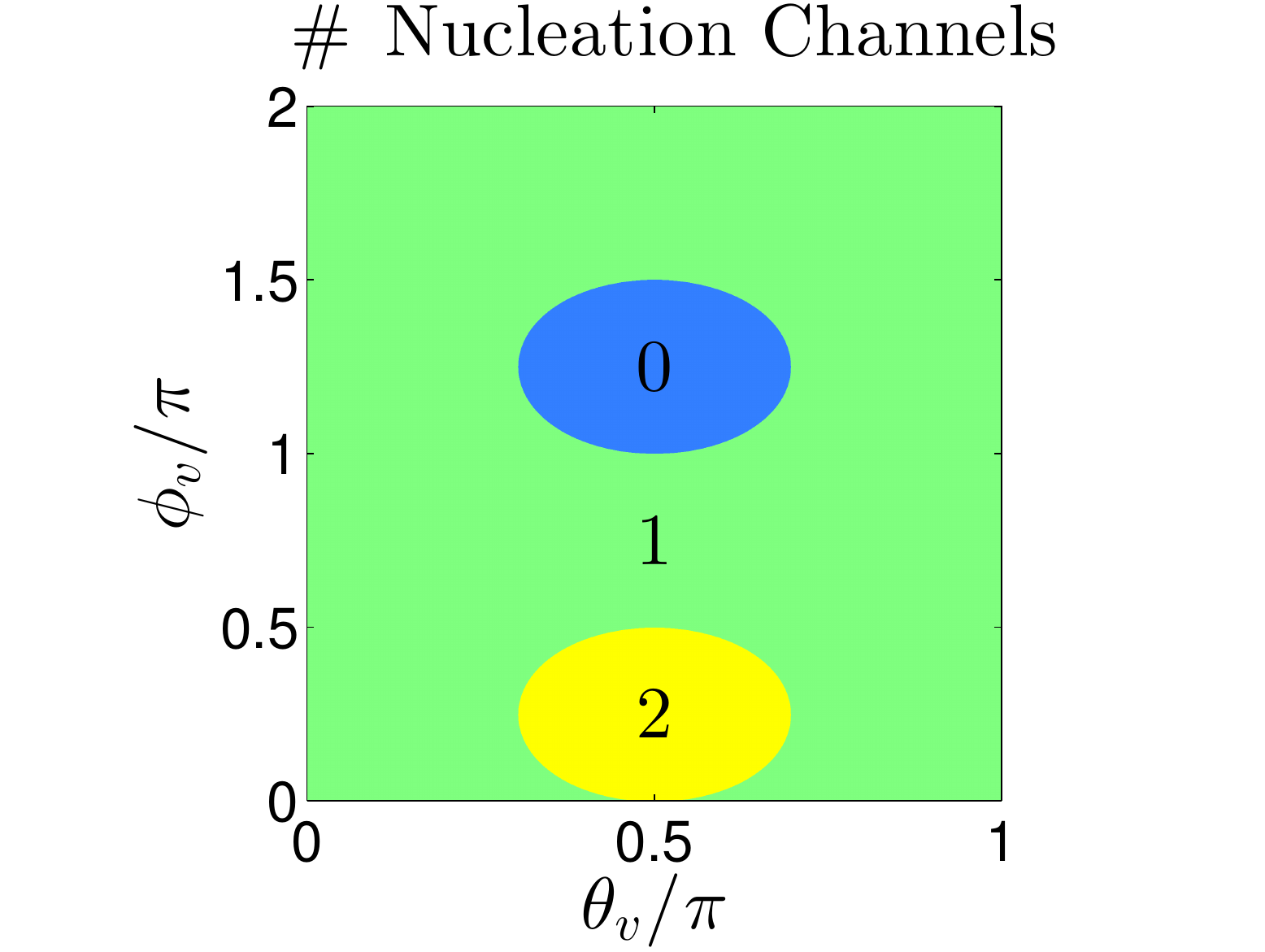}}%
\caption{\label{NucleationChannels} Dependence of the number of nucleation channels, i.e. positive eigenvalues to the matrix defined in Eq. \eqref{L_matrix}, on the angles $\phi_v = \tan^{-1}(v_t/v_s)$ and $\theta_v = \tan^{-1}(\sqrt{v^2_s + v^2_t}/v_m)$. The ovals are given by $2\cot(\theta_v) \leq \sin(2\phi_v)$. The number of channels is independent of the SOC as long as it is finite.}
\end{figure}
The number of positive eigenvalues of $\langle L_{\bf k}\rangle$ determines the number of nucleation channels, with $T_c$ determined by the largest eigenvalue $\lambda_{\rm max}$. 
Using Eq.~\eqref{phiv}, the eigenvalues can be mapped onto the unit sphere.
How the number of nucleation channels depends on the spherical angles $\phi_v = \tan^{-1}(v_t/v_s)$ and $\theta_v = \tan^{-1}(\sqrt{v^2_s + v^2_t}/v_m)$ can be seen in Fig. \ref{NucleationChannels}. When both eigenvalues are positive there are two possible nucleation channels, the \emph{dominant} and the \emph{subdominant} one. The dominant channel is responsible for the transition to superconductivity due to its larger critical temperature. The dominant channel also determines the singlet to triplet order parameter ratio, $\Delta_s/\Delta_t$, and their relative sign. The subdominant channel nucleates at a lower temperature $T^\text{sub.}_c \leq T_c$. With a finite mixing, $v_m \not = 0$, an admixture of singlet and triplet components is obtained. For certain choices of the parameters $(v_s,v_t,v_m)$ it is possible to achieve a cross-over from dominating singlet component at $T=T_c$ to a dominating triplet component at $T=0$. An example in the single-channel region is $(v_s,v_t,v_m) = (1, 0, a/(\langle|{\bf l}_{{\bf k}}|^2\rangle-a^2))$ (ignoring normalization) with the parameter $a$ being slightly larger than the maximum value of the SOC vector on the Fermi surface, e.g. $a = 1.01\max|{\bf l}_{{\bf k}}|$. This means that the topology of the system can be sensitive to its sub-critical temperature.

\begin{table} \centering
\begin{tabular}{c|c|ccc ccc ccc}
\hline\hline
& $g_2$ & &  & & & $r^\text{bulk}_\Delta$ & & & & \\ \hline
$O$ & $0.0$ & 0 & 0.26 & 0.38 & 0.50 & 0.62 & 0.74 & 0.86 & 0.98 & 1.1 \\
$O$ & $0.7$ & 0 & 0.20 & 0.33 & 0.46 & 0.59 & 0.72 & 0.84 & 0.97 & 1.1 \\
$O$ & $1.03$ & 0 & 0.09 & 0.23 & 0.38 & 0.52 & 0.67 & 0.81 & 0.96 & 1.1 \\
$O$ & $2.5$ & 0 & 0.17 & 0.30 & 0.44 & 0.57 & 0.70 & 0.83 & 0.97 & 1.1 \\ 
$C_{4v}$ & $0.0$ & 0 & 0.14 & 0.28 & 0.41 & 0.55 & 0.69 & 0.83 & 0.96 & 1.1 \\
$C_{4v}$ & $4.0$ & 0 & 0.14 & 0.28 & 0.41 & 0.55 & 0.69 & 0.83 & 0.96 & 1.1 \\
$T_d$ & N/A & 0 & 0.14 & 0.28 & 0.41 & 0.55 & 0.69 & 0.83 & 0.96 & 1.1 \\ \hline
\end{tabular}
\caption{\label{RatiosPointGroups}The scaled bulk singlet to triplet ratios, $r^\text{bulk}_\Delta \equiv \Delta_s/(\Delta_t \max|{\bf l}({\bf k}_F)|)$, chosen for the different point groups, $O$, $C_{4v}$, and $T_d$, and their respective $g_2$ values used in this work. }
\end{table}
In certain parameter ranges for $(v_s,v_t,v_m)$
it is possible to construct a configuration with two active channels in which the subdominant channel has a lower free energy at $T=0$, thus inducing a second phase transition below $T_c$. The simplest way to get a second phase transition is to choose $(v_s,v_t,v_m)$ in such a way as to get a dominant channel with a large triplet component, as well as a rather large subdominant critical temperature. A example for such a choice is $(v_s,v_t,v_m) = (0.999\langle |{\bf l}_{{\bf k}}|^2\rangle, 1, 0)$ (ignoring normalization) giving a dominant pure triplet channel, and a subdominant pure singlet channel. The subdominant critical temperature is $T^\text{sub.}_c = 0.996T_c$ for the point groups and SOC vectors in table \ref{RatiosPointGroups} (assuming $\mathcal{Y}_{\bf k}=1$). The condensation energy at zero temperature, assuming the same density of states on both Fermi surface sheets (which is the approximation employed here as the splitting is small), is given by
\begin{equation}
\delta\Omega = -\frac{N_F}{2}\left(|\Delta_s|^2 + 2 | \Delta_s \Delta_t|  \langle |{\bf l}_{{\bf k}_F}|\rangle + |\Delta_t|^2 \langle |{\bf l}_{{\bf k}_F}|^2\rangle\right) \; ,
\label{dW}
\end{equation}
and yields a lower value for the subdominant channel, for all three point groups and SOC vectors considered in this work, with this choice of $(v_s,v_t,v_m)$. 
\subsection{Angle-resolved density of states}
The angle-resolved surface density of states (DOS) is given by 
$N({\bf k},\epsilon) = -(2\pi)^{-1}N_F\text{Im}\text{Tr}_\lambda [\mathfrak{g}^\text{R}({\bf k},\epsilon) ]$, 
or explicitly in terms of coherence functions 
\begin{align}
\frac{N({\bf k},\epsilon)}{N_F} = \text{Re}\text{Tr}_\lambda \left\{[\sigma_0 - \gamma({\bf k},\epsilon) \tilde \Gamma({\bf k},\epsilon)]^{-1} - \frac{1}{2}\sigma_0 \right\}. 
\end{align}
The spin-resolved DOS along the quantization axis $j \in \{x,y,z\}$ is given by $N^{(j)}_\pm({\bf k},\epsilon)  = N({\bf k},\epsilon)  \pm N^{(j)}({\bf k},\epsilon) $, where
\begin{equation}
\label{Nxyz}
\frac{N^{(j)}({\bf k},\epsilon)}{N_F} = \text{Re}\text{Tr}_s \left\{\sigma_j[\sigma_0 - \gamma^s({\bf k},\epsilon) \tilde \Gamma^s({\bf k},\epsilon)]^{-1} - \frac{1}{2}\sigma_j \right\}. 
\end{equation}
Note that all quantities in Eq.~\eqref{Nxyz} are expressed in the spin basis. Using a non-self-consistent order parameter, with the bulk solution all the way to the surface, it is straightforward to show that there are two classes of trajectories giving rise to Andreev bound states (ABS) at zero energy (see appendix \ref{SectionZBCPs} for details). Introducing the notation $\Upsilon_{\bf k} \equiv \text{sign}[\mathcal{Y}_{\bf k}\Delta_-({\bf k})]$ the first class of trajectories is simply given by $\Upsilon_{\underline{\bf k}} = - \Upsilon_{\bf k} \not = 0$. With the spherical angles $(\phi_l, \theta_l)$ and $(\phi_{\underline{l}}, \theta_{\underline{l}})$ corresponding to ${\bf l}_{\bf k}$ and ${\bf l}_{\underline{\bf k}}$ respectively, the second class is given by solutions to 
\begin{equation}
\label{Feq}
F(\phi_l, \theta_l, \phi_{\underline{l}}, \theta_{\underline{l}})=-1 \;,
\end{equation}
with the definition
$F(\phi_l, \theta_l, \phi_{\underline{l}}, \theta_{\underline{l}})= \cos(\theta_l) \cos(\theta_{\underline{l}}) + \cos(\phi_l - \phi_{\underline{l}}) \sin(\theta_l) \sin(\theta_{\underline{l}})$, 
provided that 
$(\Upsilon_{\bf k}, \Upsilon_{\underline{\bf k}}) = (0,-1)$, $(\Upsilon_{\bf k}, \Upsilon_{\underline{\bf k}}) = (-1,0)$, or $(\Upsilon_{\bf k}, \Upsilon_{\underline{\bf k}}) = (-1,-1)$. 
This second class of bound states arises due to the phase factors $t_\pm ({\bf k})$ defined in Eq.~\eqref{phase}, which can yield an extra phase shift of $\pi $.
These results remain true for self-consistent order parameters as long as the gap does not completely close at some distance from the surface. 
\subsection{Point contact spectra }
The point contact conductance between a normal metal and an NCS is computed using the following assumptions: the size of the point contact is much smaller than the coherence length but much larger than the Fermi wavelength, the Fermi surfaces on both sides of the interface are considered to be equal, and the proximity effect is ignored. The normal metal having index 1, and the NCS index 2, the scattering matrix of the interface in the spin/helicity basis is given by
\begin{equation}
{\bf S } = \begin{pmatrix} S_{11} & S_{12} \\ S_{21} & S_{22} \end{pmatrix}= \begin{pmatrix} r\sigma_0 & tU^{\dagger}_{\bf k} \\ t^*U_{-\underline{\bf k}} & -rU_{\underline{\bf k}}U^{\dagger}_{\bf k} \end{pmatrix}
\end{equation}
with the transmission amplitude 
\begin{equation}
t(\alpha_{\bf k}) = \frac{t_0 \cos(\alpha_{\bf k}) }{\sqrt{1 - t^2_0 \sin^2(\alpha_{\bf k}) }}
\end{equation}
where $t_0$ is the tunneling parameter and $\alpha_{\bf k}$ is the angle between the surface normal and the Fermi velocity of the outgoing trajectories in the normal metal. The reflection amplitude is given by $r = \sqrt{1- t^2}$. The zero-temperature tunnel conductance is given by \cite{springerBook}
\begin{eqnarray}
\label{condEq}
G(eV) &=& \left\langle {\bf n} \cdot {\bf  v}_{F_1} \left[\left|\left| B(\epsilon) \right|\right|^2  - \left|\left|S_{12}A_2(\epsilon)\right|\right|^2\right]\right\rangle_\text{out} \nonumber \\
&& + \left\langle {\bf n} \cdot {\bf  v}_{F_1} \left|\left| B(-\epsilon)\gamma_2(-\epsilon)\tilde S_{21}\right|\right|^2 \right\rangle_\text{out}
\end{eqnarray}
where the expression is evaluated at $\epsilon = eV$, ${\bf v}_{{F}_1}$ is the Fermi velocity in the normal metal, $\langle\bullet\rangle_\text{out}$ indicates that the average is only for outgoing trajectories in the normal metal, $B(\epsilon) = S_{12}\left(\sigma_0 + A_2(\epsilon)S_{22}\right)$,
\begin{equation}
\label{A2}
A_2(\epsilon) = \left (\sigma_0 - \gamma_2(\epsilon)\tilde S_{22} \tilde \gamma_2(\epsilon)S_{22}   \right )^{-1} \gamma_2(\epsilon)\tilde S_{22} \tilde\gamma_2(\epsilon)
\end{equation}
and $||\bullet||^2\equiv \frac{1}{2}\text{Tr}\left [ (\bullet) (\bullet)^\dagger\right]$. The normal state conductance, $G_N$, is simply obtained by setting the coherence functions to zero.

%%%%%%%%%%%%%%%%%%%
\subsection{Topology}

We characterize the topology of a system by computing three topological invariants. The starting point is the Bogolioubov-de Gennes (BdG) Hamiltonian
\begin{equation}
H({\bf k})  = \begin{pmatrix} h({\bf k}) & \Delta({\bf k}) \\ \Delta^\dagger({\bf k}) & -h^T(-{\bf k}) \end{pmatrix}
\end{equation}
obeying time-reversal symmetry, $\mathcal{T}$, particle-hole symmetry, $\mathcal{C}$ , as well as the combined 'chiral' symmetry $\mathcal{S} = i\mathcal{TC}$. The BdG Hamiltonian is thus of the symmetry class DIII.\cite{Arxiv1502.03746} It anticommutes with $\mathcal{S}$ and in the basis where $\mathcal{S}$ is block diagonal $H$ becomes block off-diagonal, $\bar H = V H V^\dagger$. The flat-band block off-diagonal Hamiltonian $Q({\bf k})$ is constructed by projecting all bands above (below) the gap to $+1$ ($-1$)
\begin{equation}
Q({\bf k})  = \begin{pmatrix} 0 & q({\bf k})  \\ q^\dagger({\bf k})  & 0 \end{pmatrix}
\end{equation}
where $q({\bf k})$ is a $2\times 2$ matrix in the one-band model (we set for simplicity $\mathcal{Y}_{\bf k}=1$)
\begin{eqnarray}
q({\bf k}) &=& \frac{1}{2} \left [ A|{\bf l}_{\bf k}| \lambda_1 + B_{\bf k}\lambda_2 \right]\sigma_0  + \nonumber \\
&&\frac{1}{2}  \left [ A|{\bf l}_{\bf k}| \lambda_2 + B_{\bf k}\lambda_1 \right] \frac{{\bf l}_{\bf k} }{|{\bf l}_{\bf k}| } \cdot \boldsymbol{\sigma}
\end{eqnarray}
with $A = \alpha + i\Delta_t$, $B_{\bf k} = \xi_{\bf k} + i\Delta_s$, $\lambda_1 = \lambda^{-1}_+ - \lambda^{-1}_- $, $\lambda_2 = \lambda^{-1}_+ + \lambda^{-1}_- $, where $\lambda_\pm = \left | A|{\bf l}_{\bf k}| \pm B_{\bf k}\right |$. Note that $Q({\bf k})$, and thus $q({\bf k})$, is ill-defined for nodal order parameters.

Fully gapped systems are classified by calculating the 3D winding number which is defined as
\begin{equation}
\nu = \int_\text{BZ}  \frac{\text{d}^3 {\bf k}}{24\pi^2} \varepsilon^{abc} \text{Tr} \left [ (q^{-1}\partial_a q)  (q^{-1}\partial_b q)  (q^{-1}\partial_c q)  \right ]
\end{equation}
where Einstein summation is implied, $\varepsilon^{abc}$ is the Levi-Civita pseudo-tensor, $a,b,c \in \{ k_x, k_y, k_z \}$, and the integral is over the entire first BZ. From the definition of $q$ it is clear that $\nu$ is only well-defined if the order parameter on the negative helical Fermi surface does not have nodes, i.e. $\Delta_-({\bf k}^-_F) \not = 0$. There are two ways this can be true; either $\text{sign}[\Delta_-({\bf k}^-_F)] = +1 \; \forall {\bf k}^-_F \Longrightarrow \Delta_s/\Delta_t > \max|{\bf l}({\bf k}^-_F)|$, or $\text{sign}[\Delta_-({\bf k}^-_F)] = -1 \; \forall {\bf k}^-_F \Longrightarrow  \Delta_s/\Delta_t < \min|{\bf l}({\bf k}^-_F)|$.
We calculate $\nu $ numerically using the procedure in appendix \ref{appendix3DWindingNumber}.

Nodal systems are classified by calculating the 1D winding number which is defined as
\begin{equation}
N_\mathcal{L} = \oint_\mathcal{L} \frac{\text{d}l}{2\pi i} \text{Tr} \left [ q^{-1}\nabla_l q \right ]
\end{equation}
where $l$ parameterizes the loop $\mathcal{L}$ in the BZ, and $\nabla_l$ is the directional gradient along this loop. The loop $\mathcal{L}$ cannot pass through nodes of the order parameter, but is other than that arbitrary. The 1D Hamiltonian for this loop is in general not time-reversal invariant and is thus of symmetry class AIII.\cite{Arxiv1502.03746}  In order to characterize a nodal phase the loop needs to be constructed in such a way as to always encircle a line node of $\Delta_-({\bf k}^-_F)$ for any Fermi surface geometry.

With increasing singlet to triplet ratio the first nodes appear at the points where $\Delta_s/\Delta_t = \min|{\bf l}({\bf k}^-_F)|$. Increasing $\Delta_s/\Delta_t$ further the nodal rings continue to be positioned around these points until they connect with one another. At this stage the nodal rings become positioned around the points where they eventually disappear $\Delta_s/\Delta_t = \max|{\bf l}({\bf k}^-_F)|$. Thus a general loop should pass through the points where the nodal rings appear and disappear. This is accomplished by the loop
\begin{equation}
\label{LoopPath}
\mathcal{L} \: :\: \Gamma \rightarrow \min|{\bf l}({\bf k}^-_F)| \rightarrow \partial\text{BZ} \rightarrow \max|{\bf l}({\bf k}^-_F)| \rightarrow \Gamma
\end{equation}
where $\partial\text{BZ}$ is the BZ boundary, and the arrows do not necessarily imply straight lines.

In order to study the topology's effect on the surface states the 1D winding number is also computed for straight noncontractible loops, i.e. loops traversing one or several of the three circles making up the BZ torus ${\bf T}^3=S^1\times S^1\times S^1$, that are perpendicular to the surface. Writing the momentum ${\bf k} = ({\bf k}_\parallel, k_\perp)$ and the surface normal ${\bf n} = (l,m,n)$ the 1D winding number is written
\begin{equation}
N_{(lmn)}({\bf k}_\parallel) = \int \frac{\text{d}k_\perp}{2\pi i} \text{Tr} \left [ q^{-1} \nabla_\perp q \right ] \; .
\end{equation}

Restricting ourselves to time-reversal invariant noncontractible loops another topological invariant can be defined. Namely the $\mathbb{Z}_2$ invariant 
\begin{equation}
W_{(lmn)}({\bf K}_\parallel) = \prod_{\bf K}\frac{\text{Pf}[i\sigma_2 q^T({\bf K})]}{\sqrt{\det[i\sigma_2 q^T({\bf K})]}}
\end{equation}
where ${\bf K}$ are time-reversal invariant momenta on the loop, and $\text{Pf}[\bullet]$ denotes the Pfaffian of an antisymmetric matrix $\bullet$. The 1D Hamiltonian for this loop is of the symmetry class DIII.\cite{Arxiv1502.03746}

The singlet (triplet) component is said to be dominant if the inequality $\Delta_s/\Delta_t > \max|{\bf l}({\bf k}^-_F)|$ is true (false). With a dominant singlet component the material is fully gapped. Increasing $\Delta_s$ and/or decreasing $\Delta_t$ the material becomes nodal and eventually fully gapped again if $\min|{\bf l}({\bf k}^-_F)|> 0$. As is shown below the dominance of either component is temperature dependent.

%%%%%%%%%%%%%%%%%%%
\subsection{Surface band structure}

The surface band structure is computed by first Fourier transforming the BdG Hamiltonian in the relative momentum coordinate $k_\perp $ in the direction of the surface normal ${\bf n}$, 
\begin{align}
H({\bf k}_\parallel, k_\perp, {\bf R}) \rightarrow H({\bf k}_\parallel, \rho, {\bf R}). 
\end{align}
The helical dispersion, $\xi^{\lambda }_{\bf k}$, contains 
for the tight-binding approximation we use
trigonometric functions whose Fourier transform give rise to a series of delta functions
\begin{align}
H({\bf k}_\parallel, \rho, {\bf R}) = \sum_j H_j({\bf k}_\parallel, {\bf R} + \frac{1}{2}\rho{\bf n})\; \delta(j - \rho/\rho_0)
\label{eq49}
\end{align}
where $H_{-j}({\bf k}_\parallel, {\bf R} - \frac{1}{2}\rho{\bf n}) = H^\dagger_j({\bf k}_\parallel, {\bf R} + \frac{1}{2}\rho{\bf n})$, 
$j$ is a layer index, and $\rho_0$ is the length one needs to move along the direction of the surface normal in order to return to a translation-equivalent point in the lattice unit cell.
The sum has a finite number of terms, i.e. there exist a number $j_c$ such that $H_j = 0\; : \; |j| > j_c$. 
The terms $H_j$ with $j\ne 0$ can be interpreted in terms of hopping across the layers.
Discretizing the center-of-mass coordinate ${\bf R}$ in steps of $\rho_0$, the Schr\"odinger equation for $L$ layers can be written 
\begin{align}
\sum^{j(l)}_{j=-j(l)} H_j\left({\bf k}_\parallel, {\bf n}\rho_0(l + \frac{1}{2}j)\right)\psi_j({\bf k}_\parallel) = E_l({\bf k}_\parallel) \psi_l({\bf k}_\parallel), 
\label{Heff}
\end{align}
where $l = 0,1,\dots,L-1$ and $j(l) = \min\{j_c, l \}$ which takes care of the boundary conditions, i.e. no hopping across the boundary.
Eq.~\eqref{Heff} can be written more compactly as a matrix equation
$H_\text{eff}({\bf k}_\parallel)\psi({\bf k}_\parallel) = E({\bf k}_\parallel)\psi({\bf k}_\parallel)$, and the band structure is 
given by the eigenvalues of $H_\text{eff}$,
Non-trivial topology gives rise to zero-energy ABS. We are therefore mainly interested in the band structure close to zero energy. This allows us to avoid diagonalizing $H_\text{eff}$, and instead only compute the smallest magnitude eigenvalues using the Lanczos method.
Note that the order parameter 
is suppressed at both surfaces.

%%%%%%%%%%%%%%%%%

\section{Numerical Results}
In this work the SOC strength entering the quasiclassical calculations is considered to be much smaller than the Fermi energy, $\alpha \ll E_F$. 
In this case the Fermi surface is only weakly split. Ignoring this splitting, and the Fermi velocity renormalisation, the quasiparticles with opposite helicity are assigned to a single, common Fermi surface, and move coherently along classical trajectories. In addition, for the quasiclassical part of the numerical calculations, the Fermi surface is approximated to be spherical, with $|{\bf k}_F|$ being equal to the average of the Fermi surface defined by $\xi\left ( {\bf k}_F \right )=0$ with $(t_1,\mu) = (-40\alpha, -50\alpha)$. 
Here, $t_1$ determines the bandwidth, which must be large compared to the Fermi-surface splitting in order for the approximation of equal Fermi surfaces for both helicities to be valid, and $\mu <0 $ must smaller than $t_1 - \alpha $ in order for the Fermi surface to be closed. The chosen values are consistent with the approximation of an approximately spherical Fermi surface.
The SOC term enters the transport equations as a source term. 
In the following, we restrict our discussion to the maximally symmetric basis function corresponding to the irreducible representation $A_{1}$, i.e. $\mathcal{Y}_{\bf k} = 1$. 

%%%%%%%%%%%%%%%%%%%%%%
\subsection{The Cubic Point Group $O$}

\begin{figure}[b] \centering
%{\includegraphics[width=0.9\columnwidth]{Fig2_gray.pdf}}
{\includegraphics[width=0.9\columnwidth]{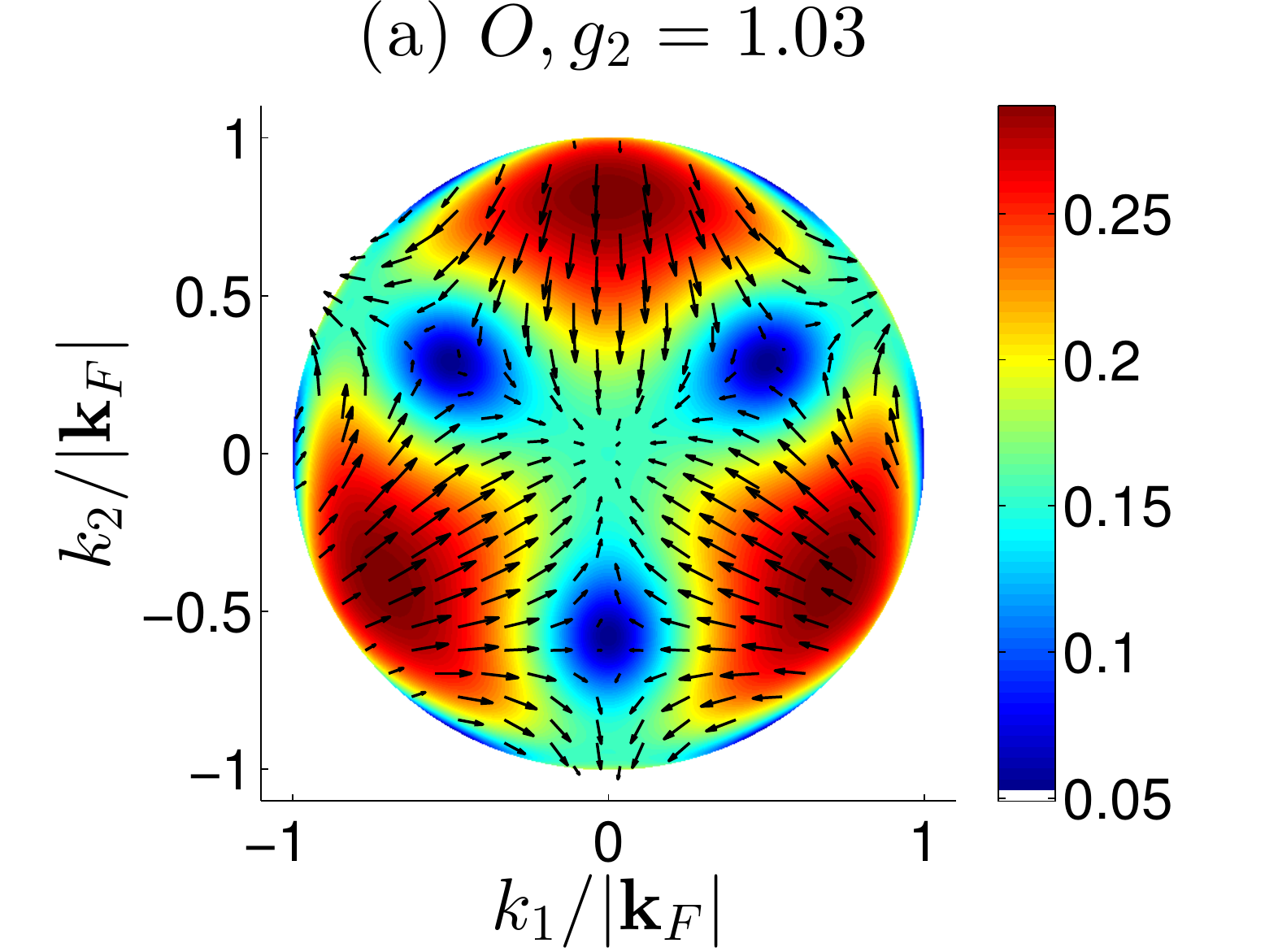}}
\caption{\label{SOCplotO}The magnitude (color) and direction (arrows) of the SOC vector corresponding to the point group $O$ defined in Eqs. \eqref{SOC_O}, with the $g_2 = 1.03$. The SOC is shown upon the spherical Fermi surface defined by the average Fermi momentum given by $\xi({\bf k}_F) = 0$, where $\xi$ is the corresponding tight-binding dispersion in the absence of SOC with $(t_1,\mu) = (-40\alpha, -50\alpha)$. The Fermi surface is seen from the ${\bf k} = (1,1,1)$ direction. }
\end{figure}

To next-nearest neighbors in the sum over Bravais lattice sites \cite{SamokhinAnnals} the SOC vector corresponding to the cubic point group $O$, takes the form
\begin{equation}
\label{SOC_O}
{\bf l}_{\bf k} = \begin{pmatrix} \sin(k_x)\left[1 - g_2\left(\cos(k_y) + \cos(k_z) \right) \right]  \\  \sin(k_y)\left[1 - g_2\left(\cos(k_z) + \cos(k_x) \right) \right]  \\  \sin(k_z)\left[1 - g_2\left(\cos(k_x) + \cos(k_y) \right) \right]  \end{pmatrix}
\end{equation}
where $g_2$ is a free parameter which determines the relative weight between the first and second order contributions. Its magnitude and direction is illustrated in Fig.~\ref{SOCplotO}.

\begin{figure*}[t] \centering
\includegraphics[width=1.3\columnwidth]{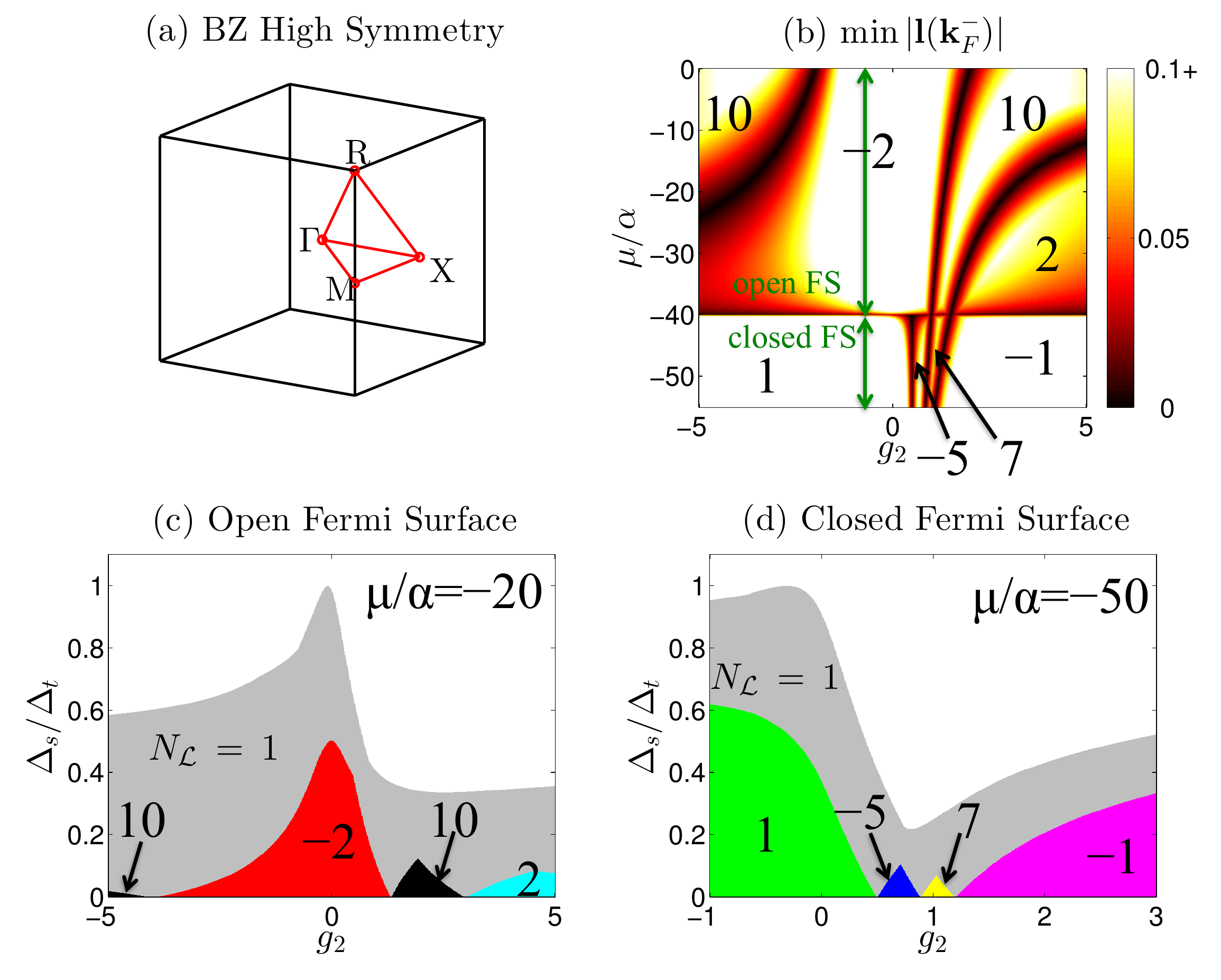}
\caption{\label{CubicStuff}(a) The high symmetry points and axes in the BZ for a simple cubic crystal. (b) The minima of the SOC vector on the negative helical Fermi surface, with $t_1 = -40\alpha$. Note the transition between a closed and open Fermi surface at $\mu=t_1$. The topological phase diagram for an open and closed FS is shown in (c) and (d) respectively. White areas indicate a gapped phase with trivial topology, $(N_\mathcal{L}, \nu)= (0,0)$; grey a nodal phase with $N_\mathcal{L}=1$ with a loop defined by Eq. \eqref{LoopPath}; colored areas gapped non-trivial phases with $\nu$ taking the values $(\text{black, red, cyan})=(+10,-2,+2)$ in (c), and $(\text{green, blue, yellow, magenta})=(+1,-5,+7,-1)$ in (d). }
\end{figure*}

An important property of the SOC vector corresponding to the cubic point group is its lack of line nodes in the BZ, it only vanishes at specific points. With $g_2=0$ these points are simply $\Gamma$, $\text{X}$, $\text{M}$, and $\text{R}$ [for the notation see Fig. \ref{CubicStuff}(a)]. A finite value of $g_2$ brings about two more points. With $g_2> 0$ they are positioned somewhere on the paths $\Gamma \rightarrow \text{R}$, and $\Gamma \rightarrow \text{M}$, and with $g_2 < 0$ on $\Gamma \rightarrow \text{R}$, and $\text{X} \rightarrow \text{R}$, in Fig. \ref{CubicStuff} (a). The exact positions, ${\bf k}^*$, of these points depend on the value of $g_2$, and are given by
\begin{eqnarray}
\Gamma \rightarrow R &:& {\bf k}^* = \cos^{-1}\left( \frac{1}{2g_2}\right)(1,1,1)^T \; ,\\
\Gamma \rightarrow M &:& {\bf k}^* = \cos^{-1}\left( \frac{1}{g_2} -1\right)(1,1,0)^T \; ,\\
X \rightarrow R &:& {\bf k}^* = (b,\pi,b)^T \; , \; b = \cos^{-1}\left( \frac{1}{g_2} + 1 \right) .
\end{eqnarray}
The lack of line nodes means that it is easy to construct a Fermi surface for which the minimum value of the SOC on the negative helical FS, $\min|{\bf l}({\bf k}^-_F)|$, is not zero. The dependence of $\min|{\bf l}({\bf k}^-_F)|$ on the chemical potential and the SOC parameter $g_2$ is shown in Fig. \ref{CubicStuff} (b). The SOC minimum is zero along certain lines in this parameter space. The line at $\mu=t_1$ marks the transition between open and closed FS, i.e. the FS is tangent to the X-point in the BZ. These lines in Fig. \ref{CubicStuff} (b) also mark the boundaries of fully gapped regions with different values of the 3D winding number $\nu$. This is demonstrated in figs. \ref{CubicStuff} (c) and (d) in which the topological phase diagram is shown for an open, $\mu=-20\alpha$, and closed, $\mu = -50\alpha$, Fermi surface respectively. White indicates that the system is fully gapped and topologically trivial, $\nu = 0$, whereas the colored regions (excluding grey) indicate that the system is fully gapped and topologically non-trivial, $\nu \not = 0$. Grey indicates a topologically non-trivial nodal phase, $N_\mathcal{L} = 1$, with loops defined by Eq. \eqref{LoopPath}.

The self-consistent order parameter is calculated for four different values of $g_2$, namely $g_2 \in \{0, 0.7, 1.03, 2.5  \}$, one for each distinct gapped topologically non-trivial phase with a closed Fermi surface, i.e. the colored regions in Fig. \ref{CubicStuff} (d). This is done for nine values of the scaled bulk singlet to triplet ratio, $r^\text{bulk}_\Delta \in [0, 1.1]$, with one active channel. These values are shown in table \ref{RatiosPointGroups}.

\begin{figure*} 
%\centering
\begin{tabular}{llll}
\subfloat{{\includegraphics[width=0.5\columnwidth,trim=30 -10 18 0, clip]{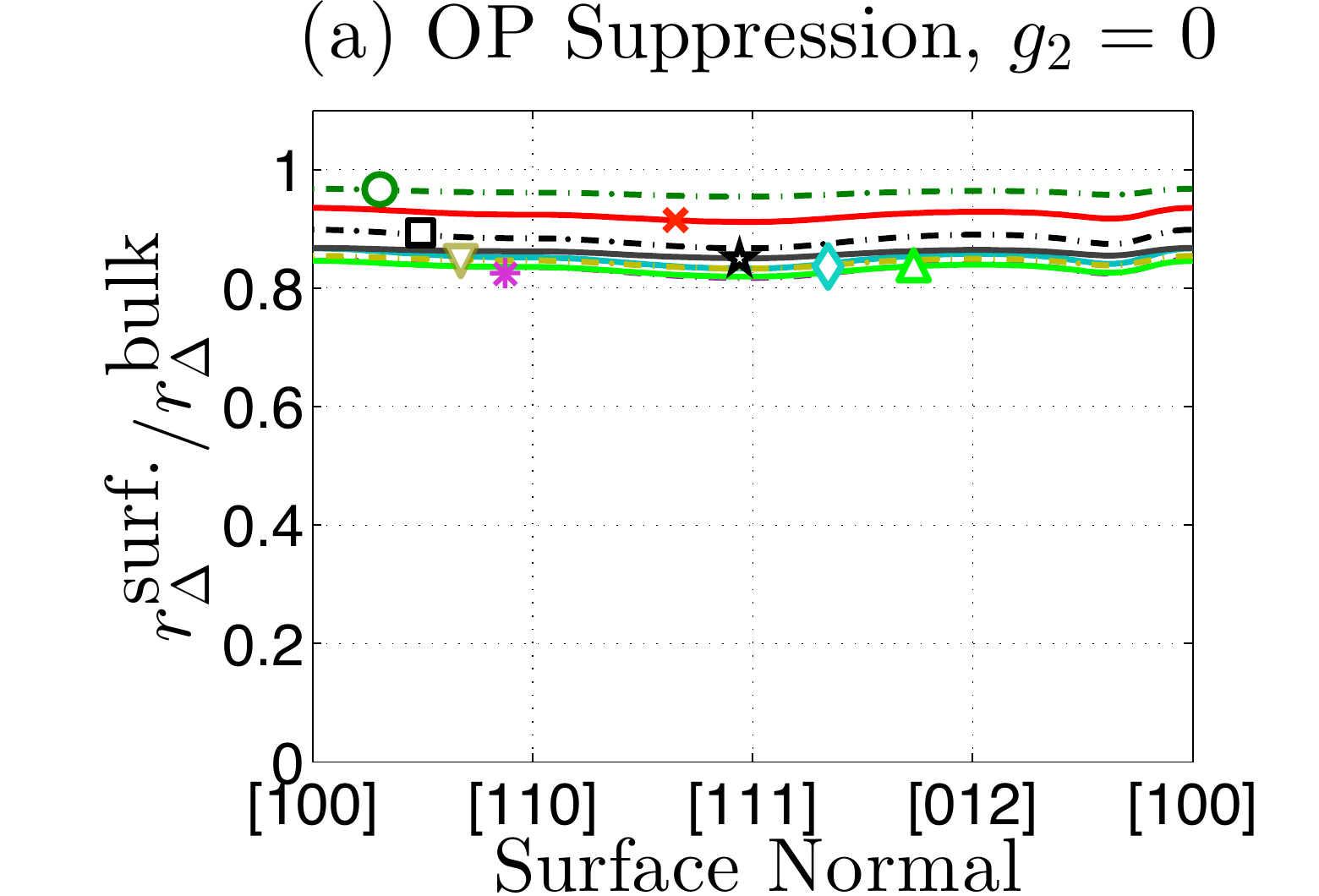}}} &
\subfloat{{\includegraphics[width=0.5\columnwidth,trim=30 -10 18 0, clip]{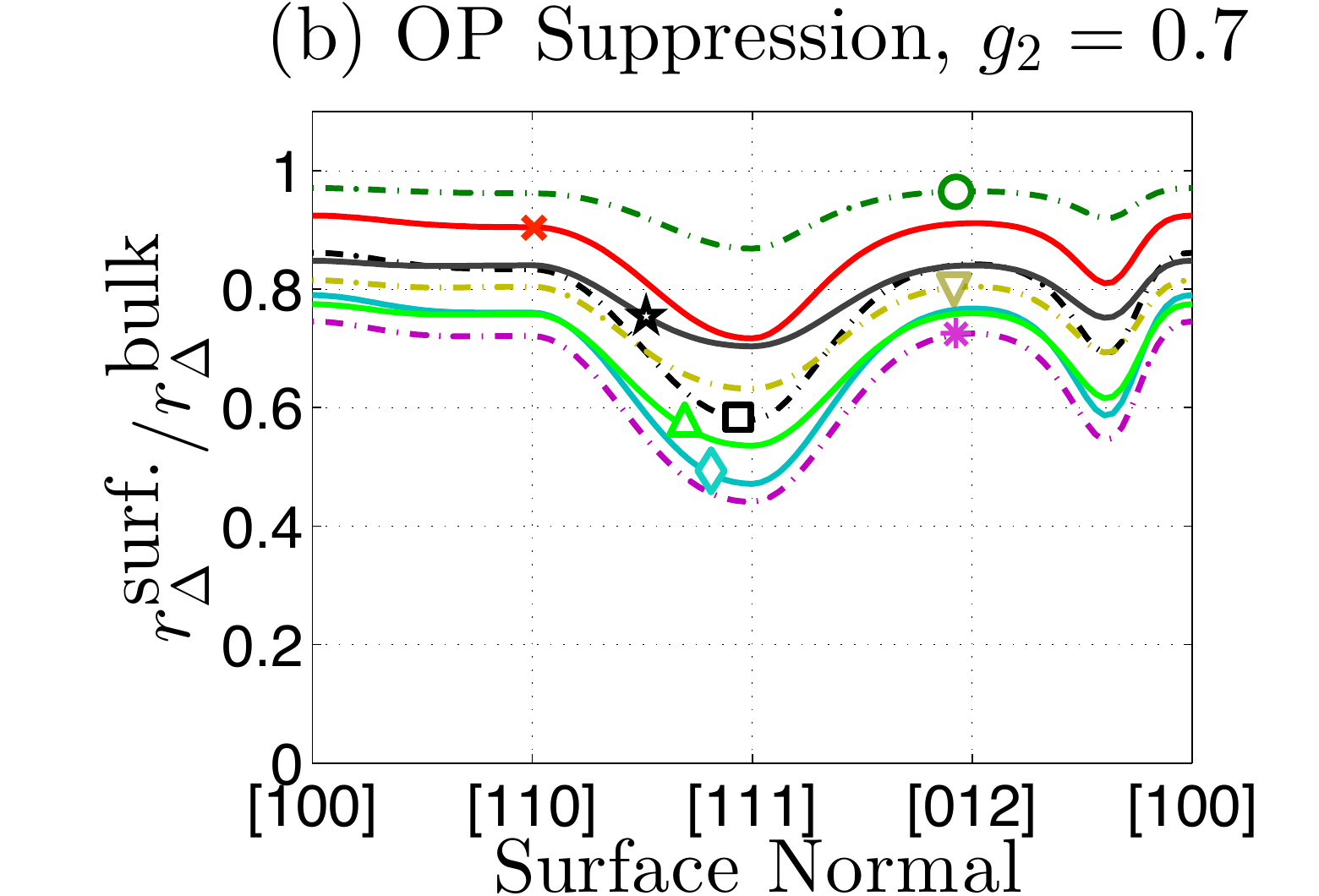}}} &
\subfloat{{\includegraphics[width=0.5\columnwidth,trim=30 -10 18 0, clip]{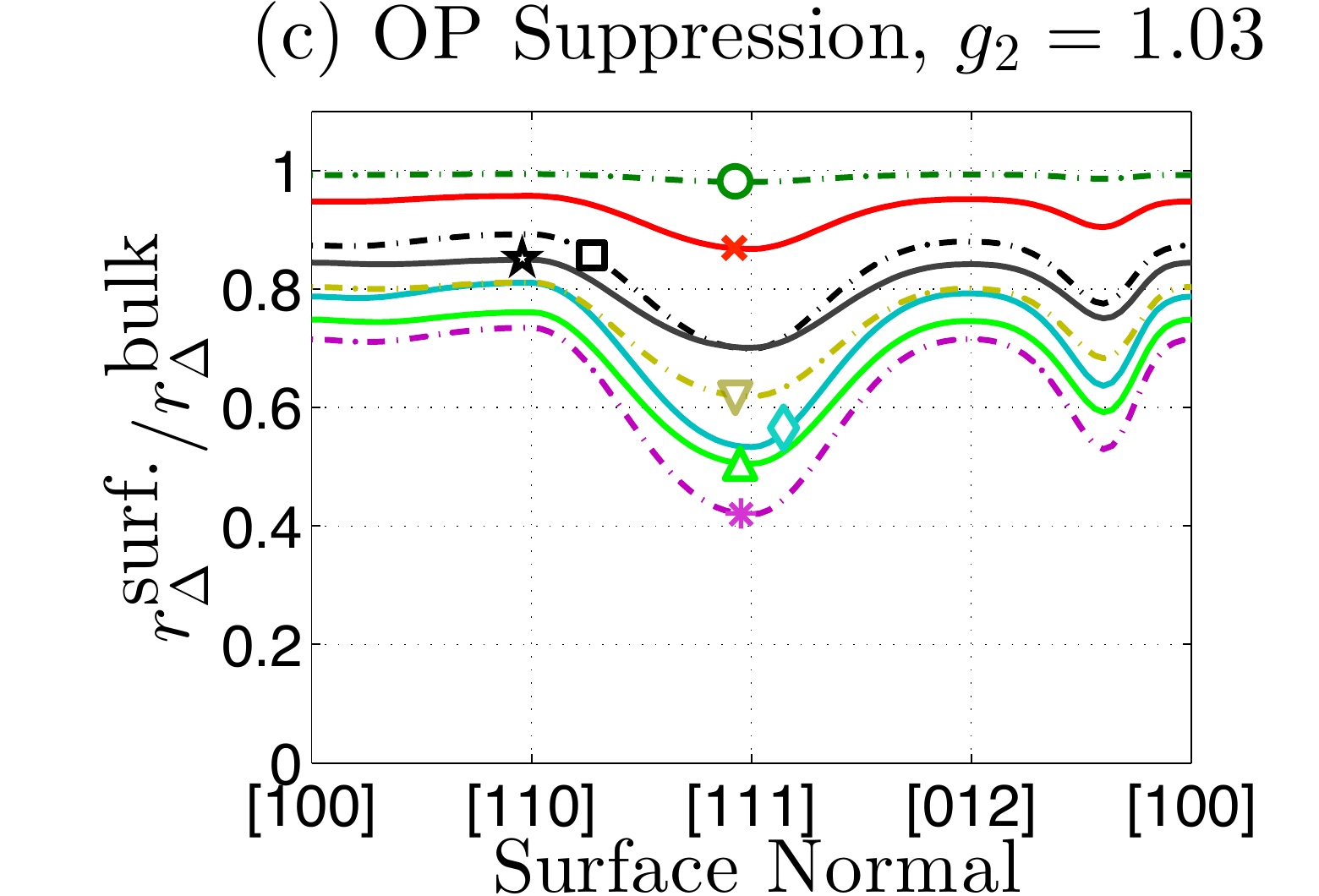}}} &
\subfloat{{\includegraphics[width=0.5\columnwidth,trim=30 -10 18 0, clip]{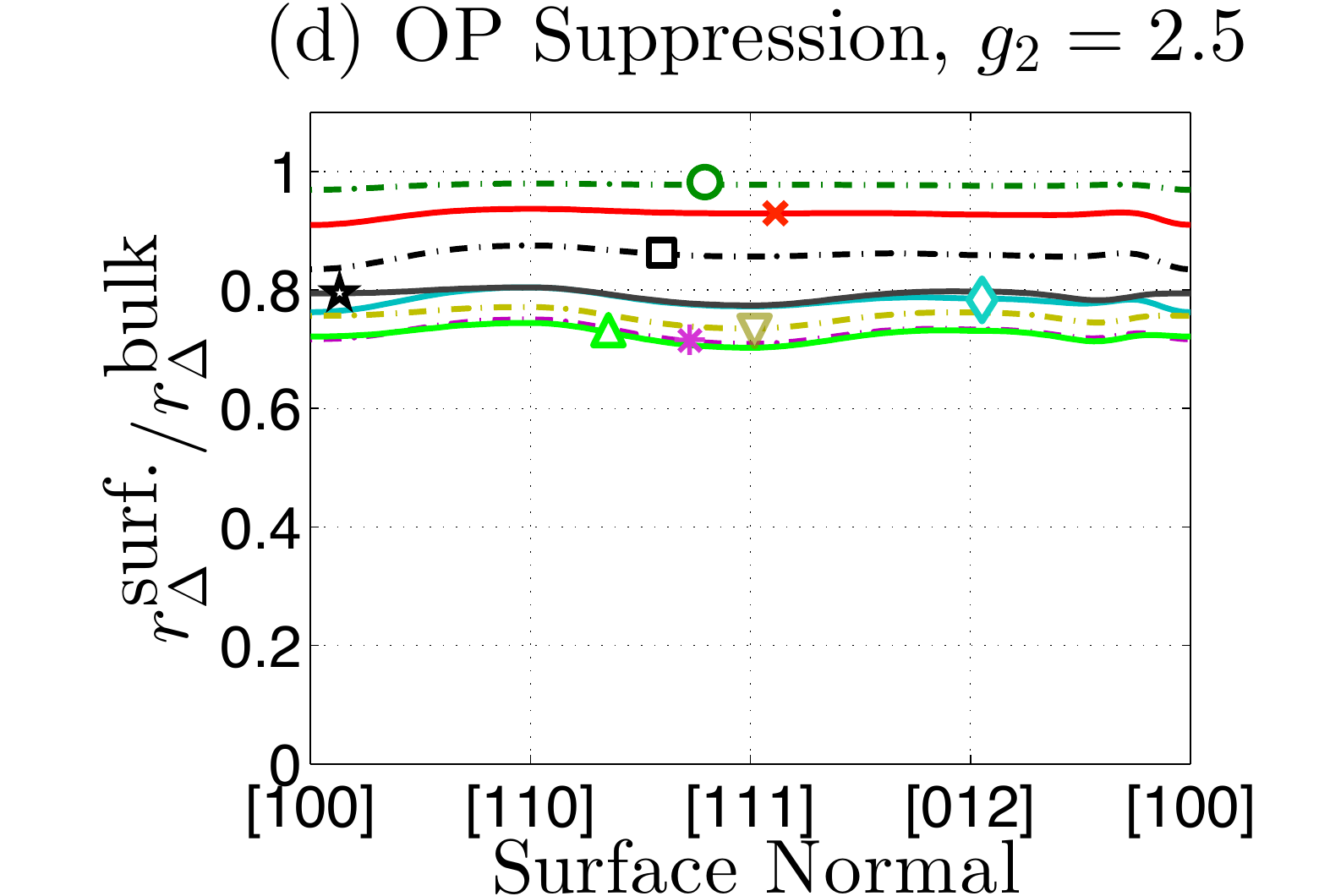}}}
\\
\subfloat{{\includegraphics[width=0.50\columnwidth,trim=5 0 18 0, clip]{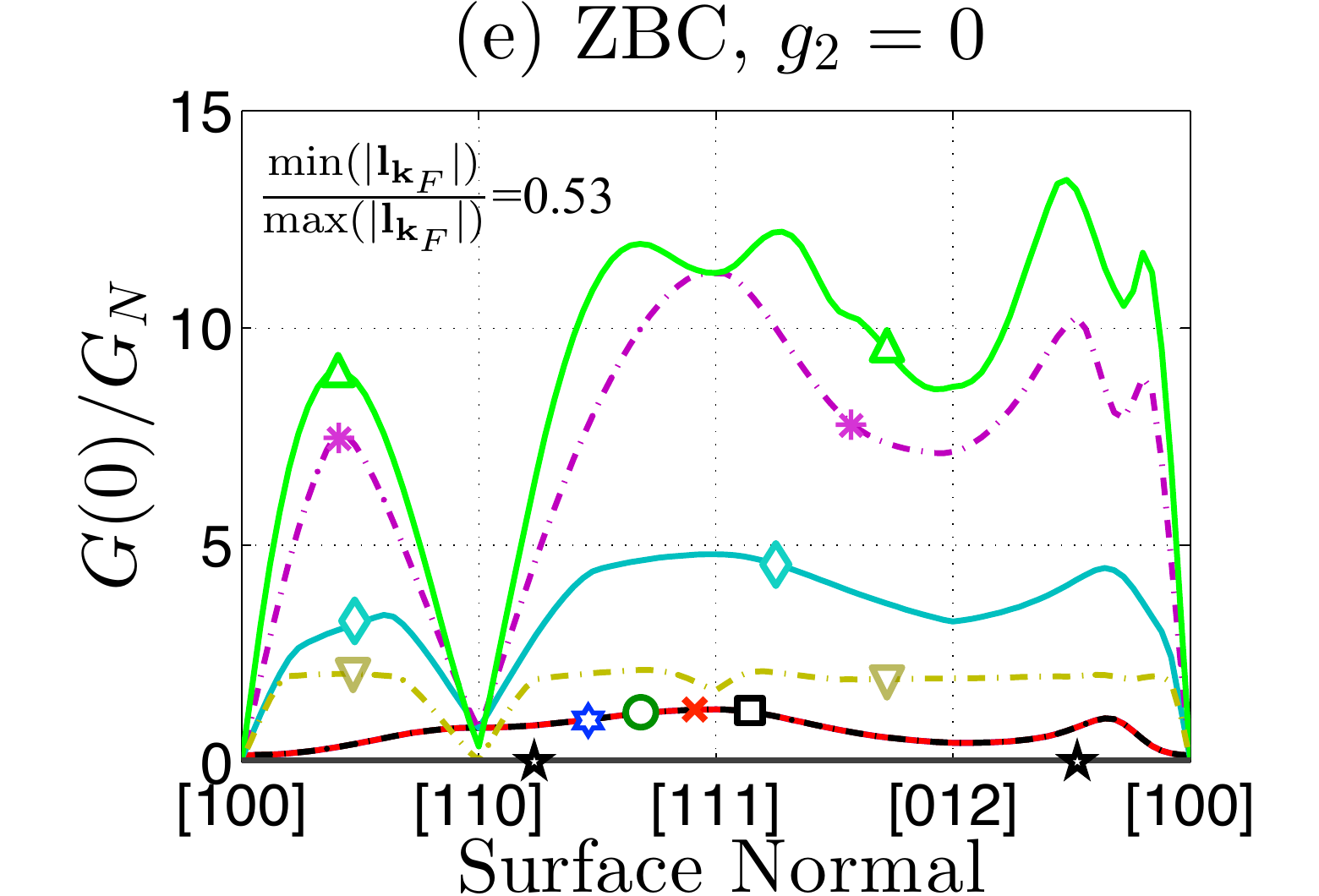}}} &
\subfloat{{\includegraphics[width=0.50\columnwidth,trim=5 0 18 0, clip]{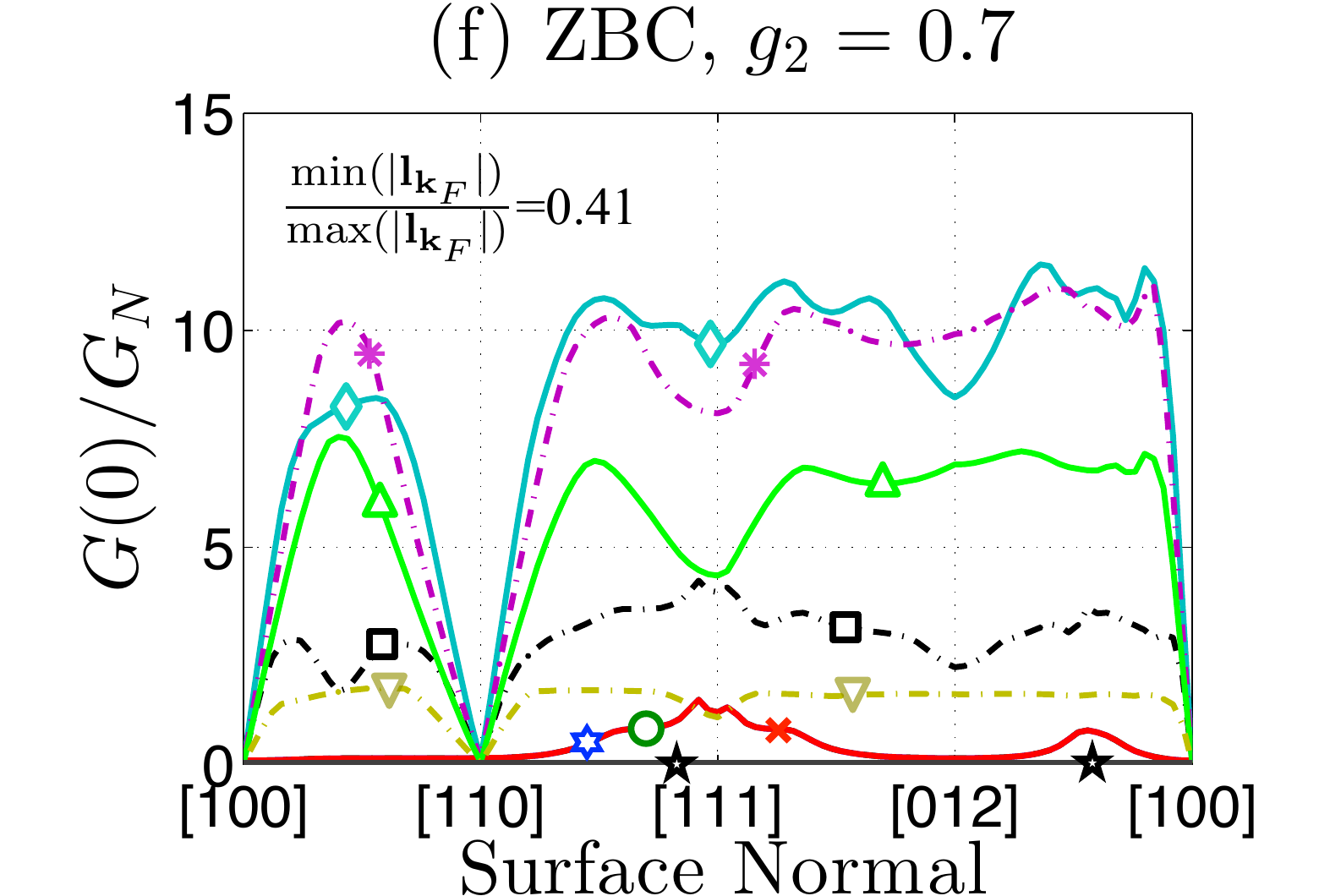}}} &
\subfloat{{\includegraphics[width=0.50\columnwidth,trim=5 0 18 0, clip]{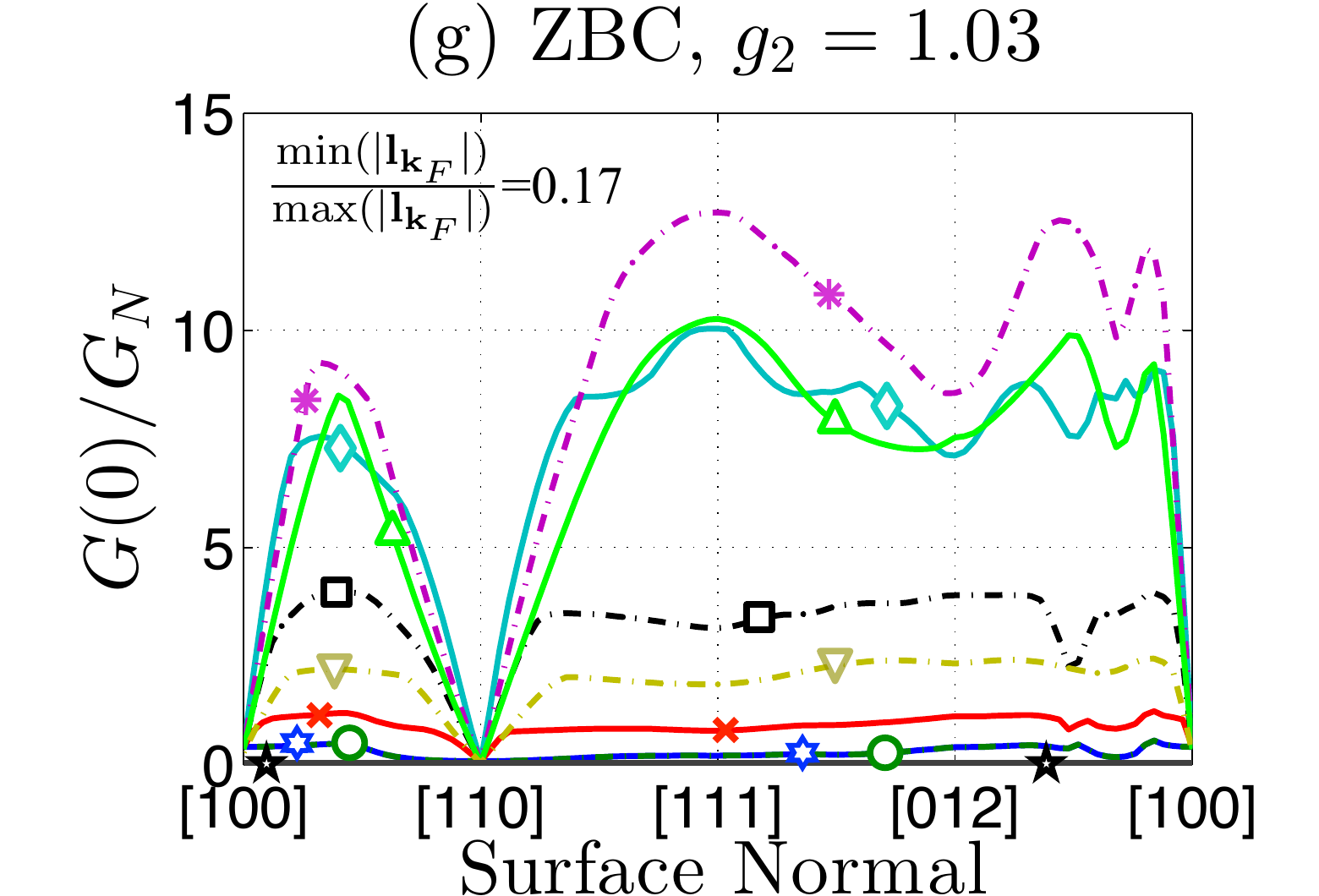}}} &
\subfloat{{\includegraphics[width=0.52\columnwidth,trim=5 0 0 0, clip]{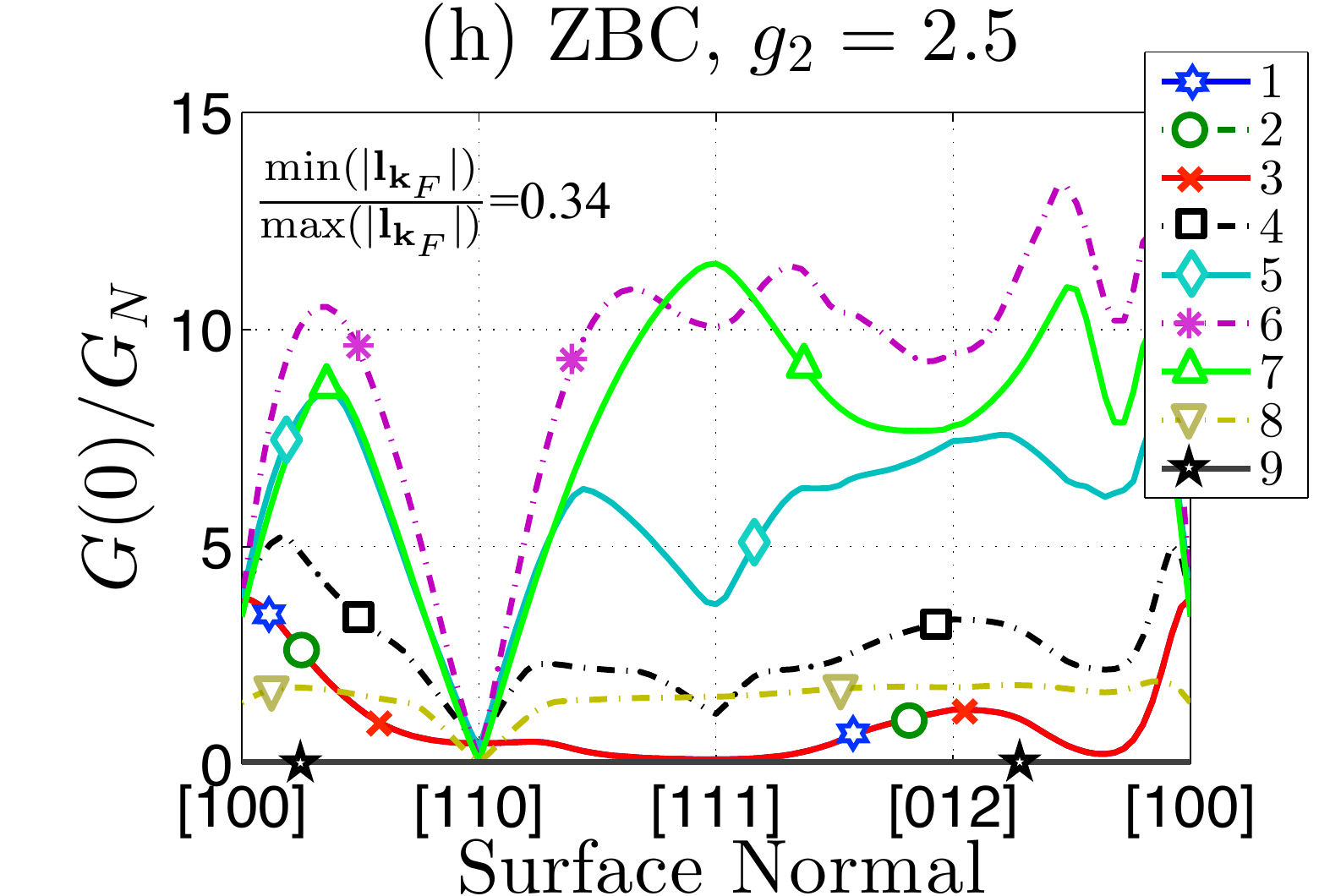}}}
\end{tabular}
\caption{\label{CubicSuppressionAndZBCP}Plots (a) - (d) show the quantity $r^\text{surf.}_\Delta/r^\text{bulk}_\Delta = \left[ \Delta_s/\Delta_t \right]^\text{surf.} \cdot \left[\Delta_t/\Delta_s \right]^\text{bulk}$ as a measure of the order parameter surface suppression. This is done for a range of different surface normals along the path ${\bf n} = (1,0,0) \rightarrow (1,1,0) \rightarrow (1,1,1) \rightarrow (0,1,2) \rightarrow (1,0,0)$. In plots (e) - (h) the zero-bias conductance, computed with $t_0 = 10^{-\frac{1}{2}}$, is shown for the same surface normals. The numbers in the legend hold for all plots and correspond to the columns in table \ref{RatiosPointGroups} showing the scaled singlet to triplet ratios.}
\end{figure*}

In order to investigate how the order parameter suppression depends on the surface orientation the order parameter is computed for a range of different surface normals, tracing out the path ${\bf n} = (1,0,0) \rightarrow (1,1,0) \rightarrow (1,1,1) \rightarrow (0,1,2) \rightarrow (1,0,0)$. As a measure of the suppression the ratio $r^\text{bulk}_\Delta/r^\text{surf.}_\Delta$, where $r^\text{surf.}_\Delta \equiv \Delta^\text{surf.}_s/(\Delta^\text{surf.}_t \max|{\bf l}_{{\bf k}_F}|)$ is the scaled surface singlet to triplet ratio, is plotted in Fig. \ref{CubicSuppressionAndZBCP} (a) - (d). The suppression is seen to be the largest for the surface normal ${\bf n} = (1,1,1)$

Along the same path of surface orientations the zero-bias conductance, computed with Eq. \eqref{condEq}, is plotted in Fig. \ref{CubicSuppressionAndZBCP} (e) - (h). For singlet to triplet ratios in the interval $\min|{\bf l}_{{\bf k}_F}| < \Delta_s/\Delta_t < \max|{\bf l}_{{\bf k}_F}|$ very large zero-bias conductance is seen for all surface orientations except the two high symmetry axes, ${\bf n} = (1,0,0)$ and ${\bf n} = (1,1,0)$. This is due to there being no trajectories for which $\Delta_-$ changes sign upon reflection for these surface orientations. For all other surface orientations this is not the case, including the high symmetry axis ${\bf n} = (1,1,1)$. Note that all lines for which $\Delta_s/\Delta_t < \min|{\bf l}_{{\bf k}_F}|$ are degenerate, and the zero-bias conductance is zero for $\Delta_s/\Delta_t > \max|{\bf l}_{{\bf k}_F}|$. Furthermore, the surface suppression due to self-consistency does not affect the zero-bias conductance. This reflects the fact that the gap does not go to zero at some distance inwards from the surface for the obtained gap profiles.

\begin{figure*}[t] \centering
%\subfloat{{\includegraphics[width=0.65\columnwidth]{Fig5a_gray.pdf} }}%
\subfloat{{\includegraphics[width=0.65\columnwidth]{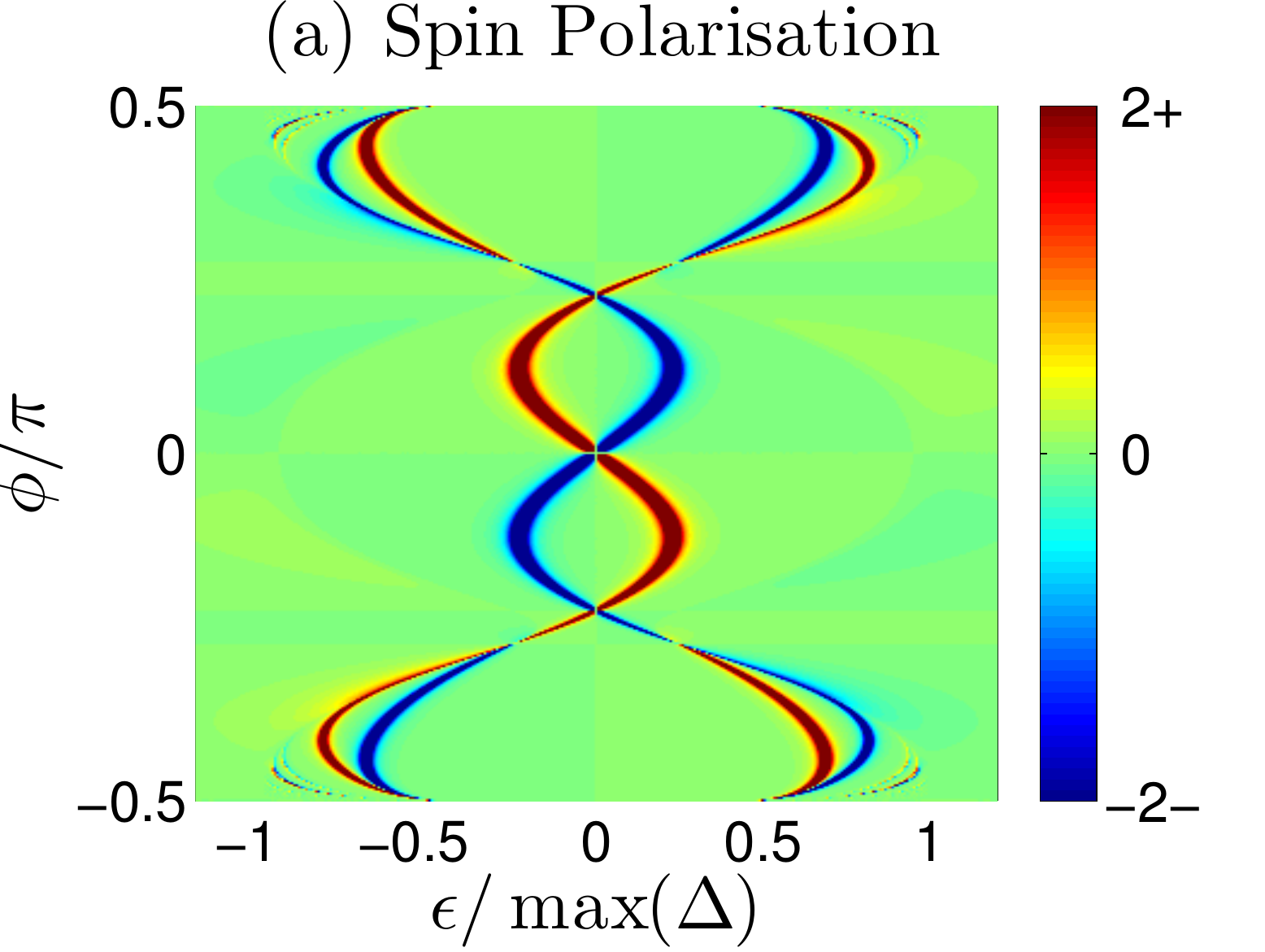} }}%
\subfloat{{\includegraphics[width=0.65\columnwidth]{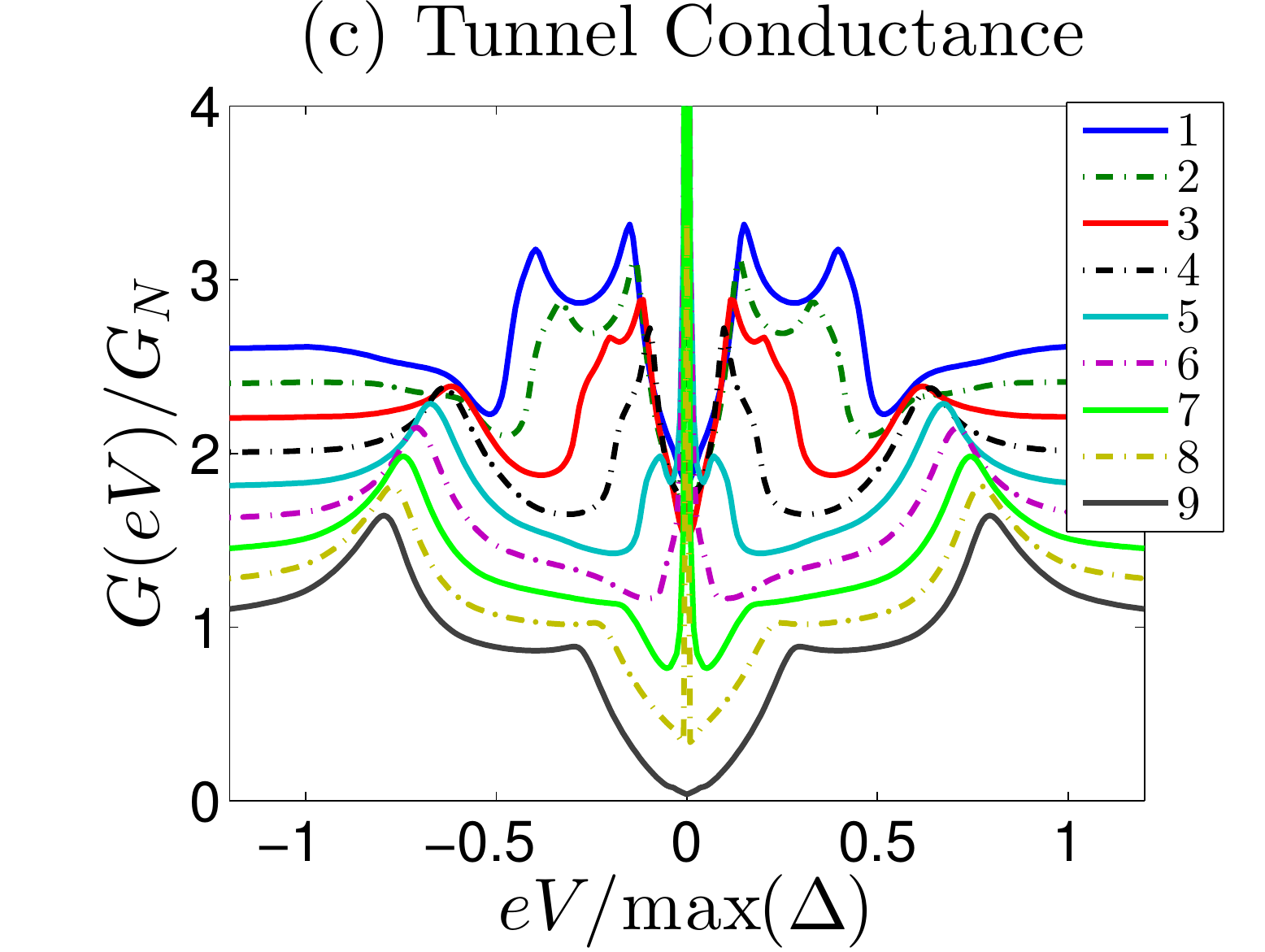} }}%
\subfloat{{\includegraphics[width=0.65\columnwidth]{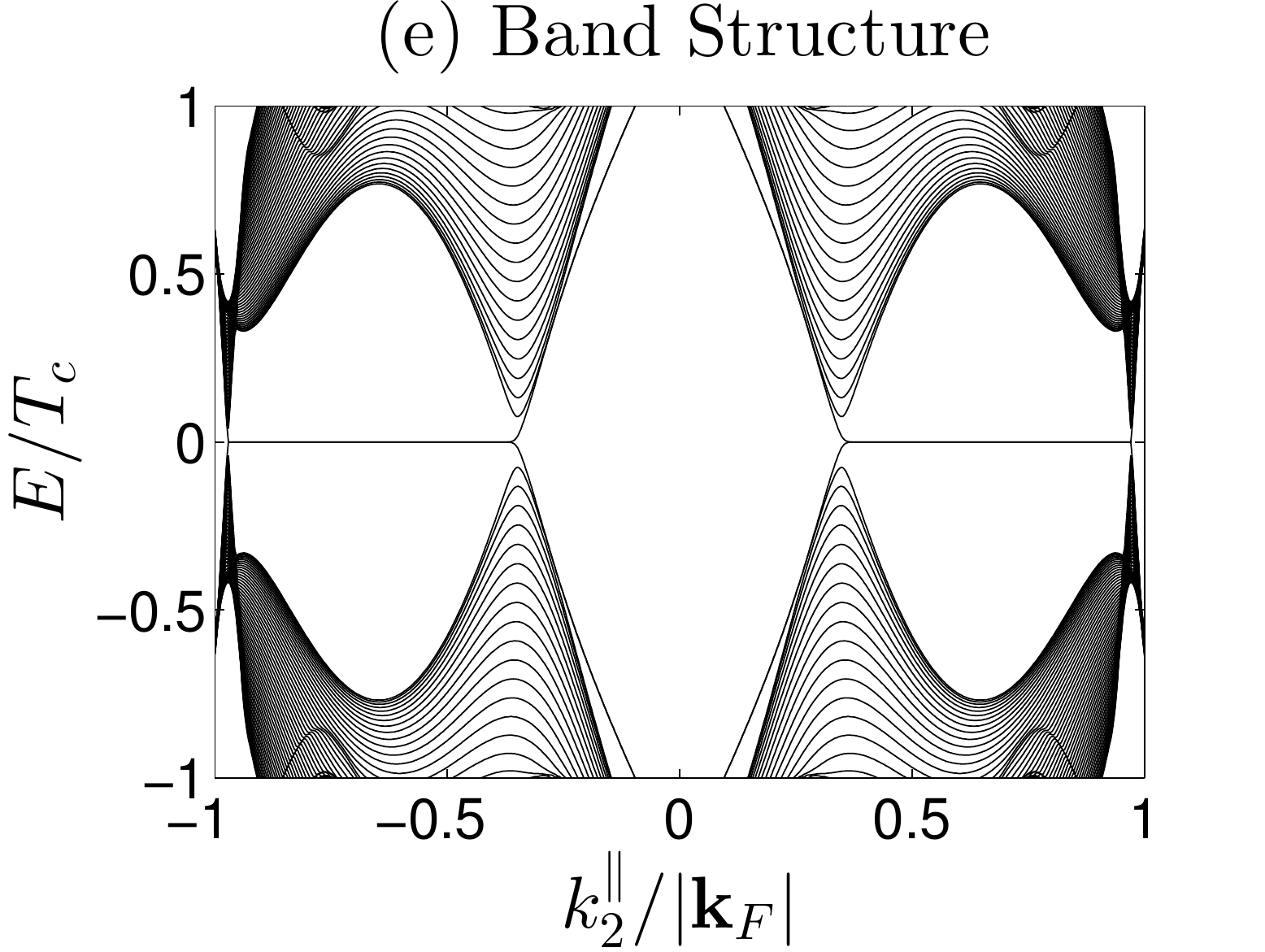} }}%
\\
\subfloat{{\includegraphics[width=0.65\columnwidth]{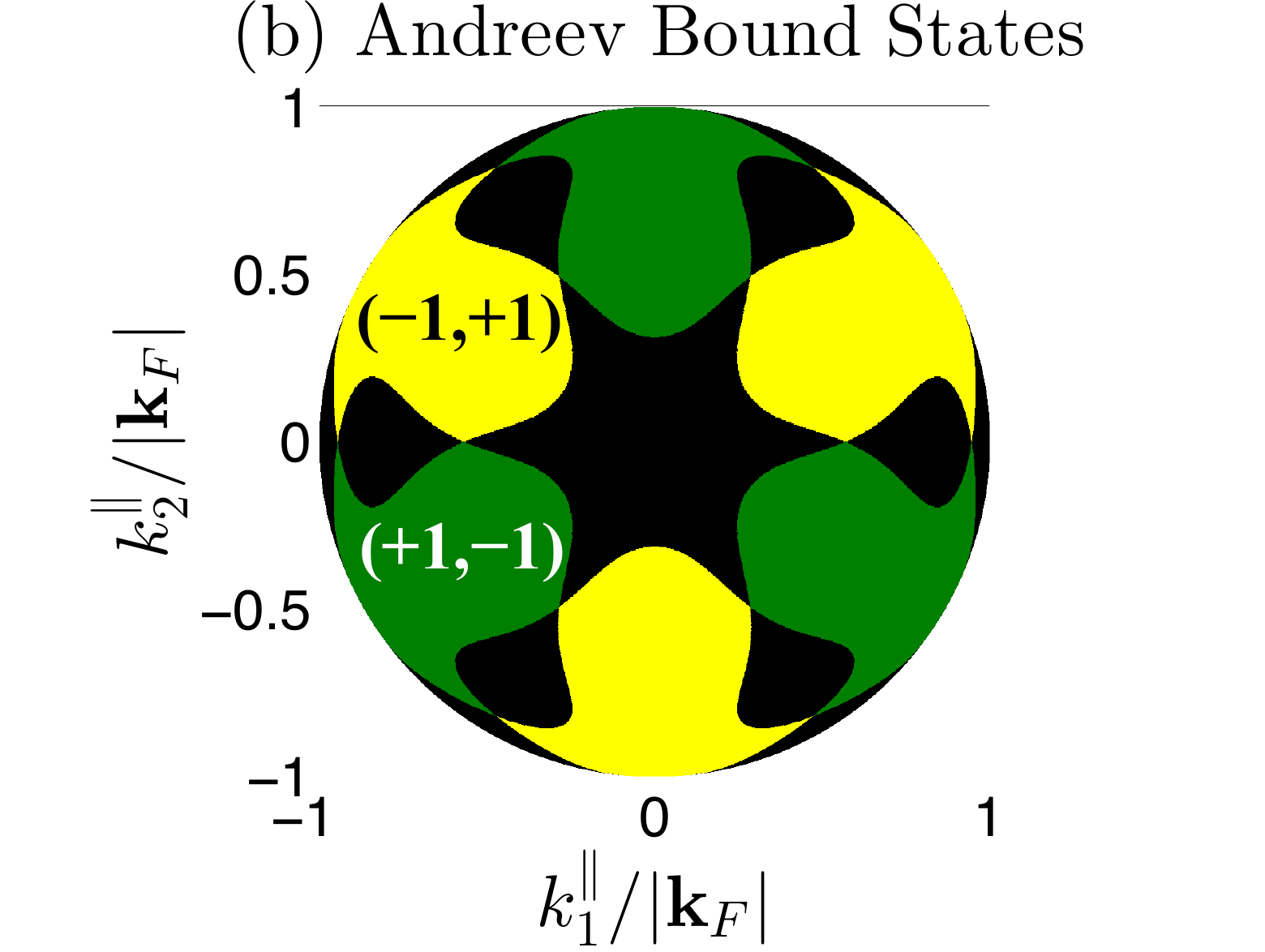} }}%
\subfloat{{\includegraphics[width=0.65\columnwidth]{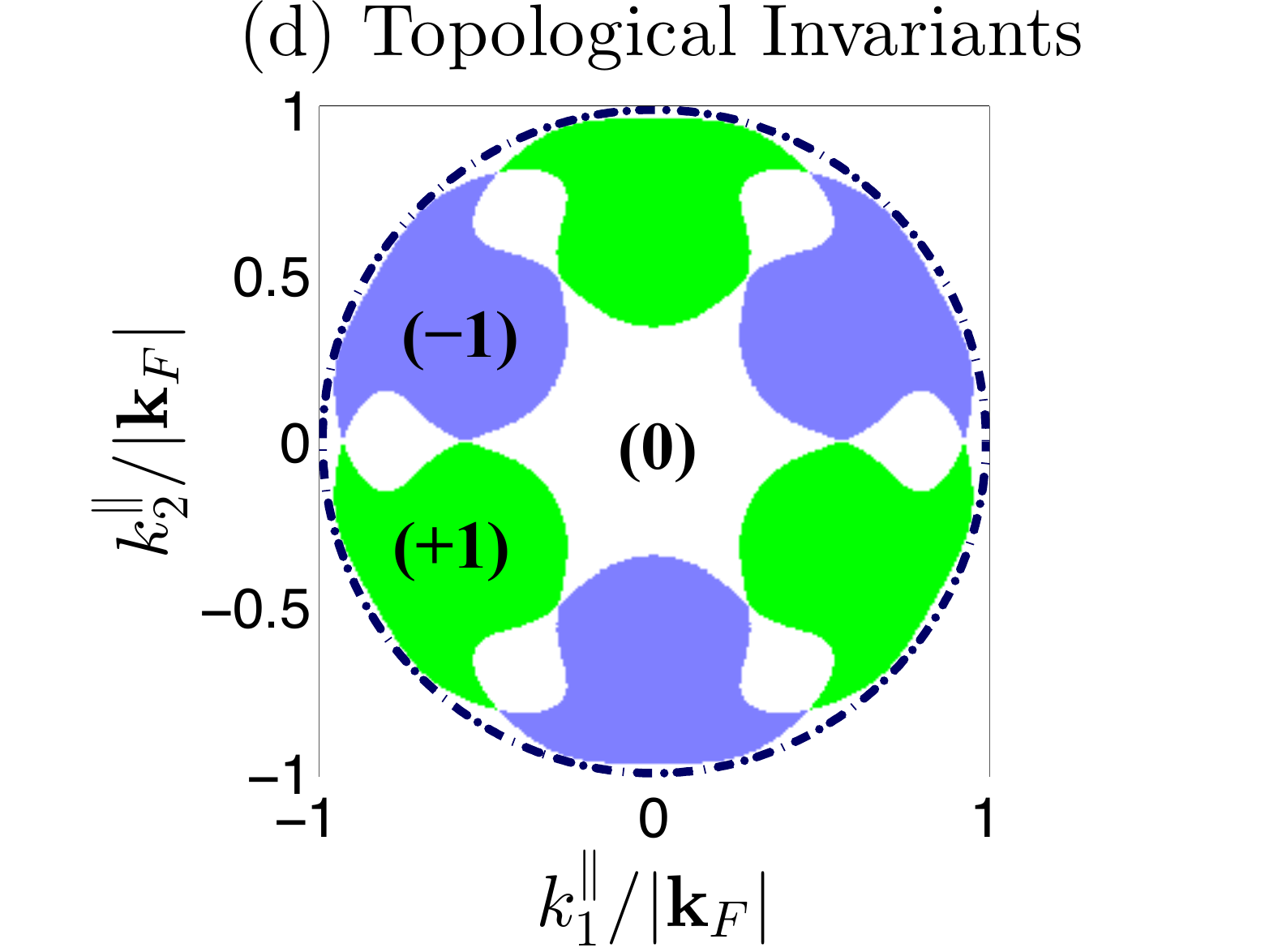} }}%
\subfloat{{\includegraphics[width=0.65\columnwidth]{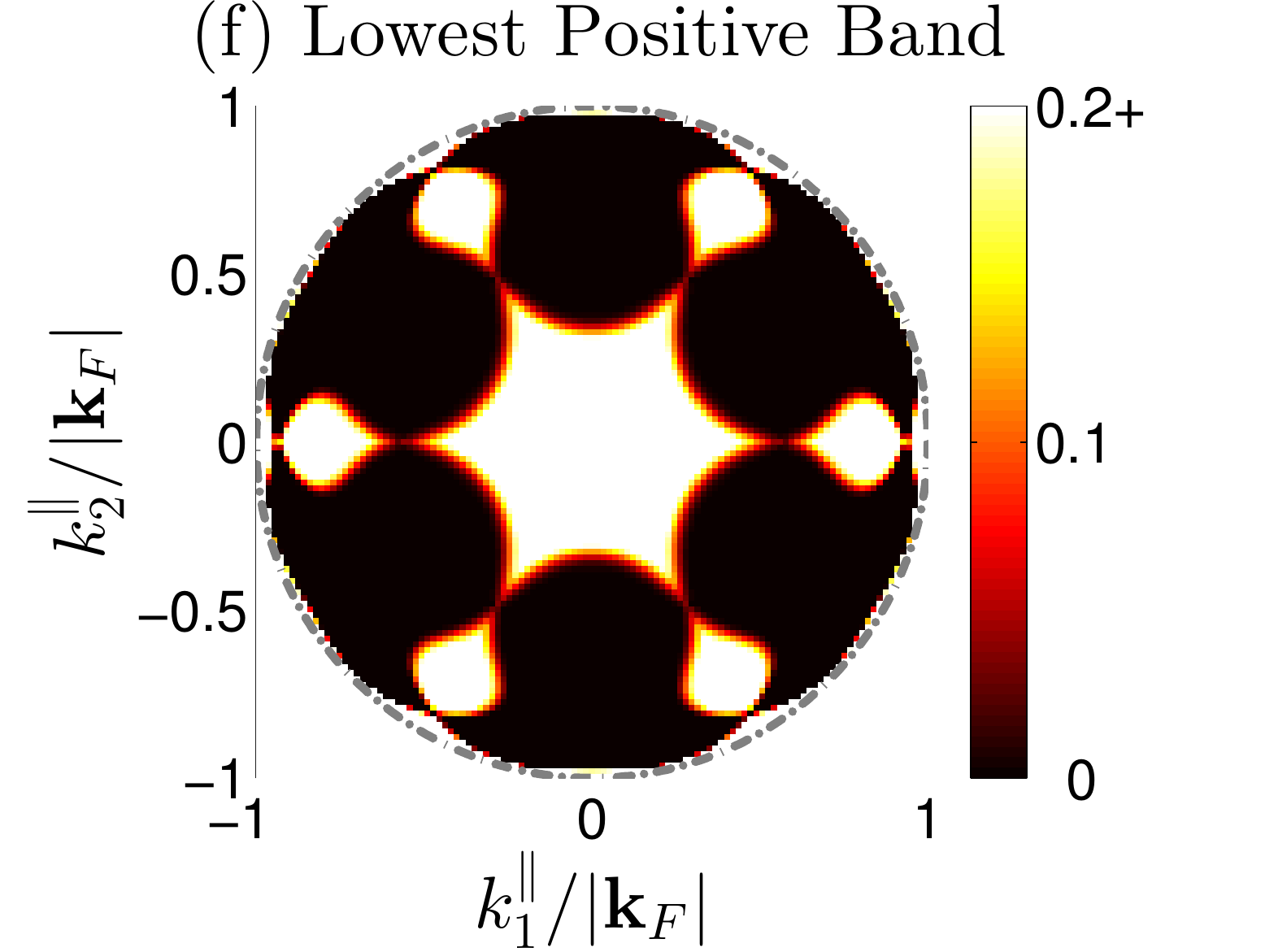} }}%
    \caption{\label{MoreCubicStuff} All plots are for the cubic point group $O$ with $g_2 = 1.03$. (a) $N^{(z)}({\bf k}, \epsilon)$, defined in Eq.~\eqref{Nxyz}, which is a measure of the spin polarization along the $z$-axis. It is shown for a self-consistent pure triplet order parameter and
for momentum directions in the $xy$ plane (i.e. $\theta = \pi/2$), at the surface with the surface normal ${\bf n} = (1,0,0)$. 
(b) Momentum-resolved ABS at zero energy computed assuming a constant order parameter with $r^\text{bulk}_\Delta = 0.67$. The disk is the projection of the Fermi surface onto the slab surface with ${\bf n} = (1,1,1)$. Green regions corresponds to ABS for which $(\Upsilon_{\bf k},  \Upsilon_{\underline{\bf k}}) = (+1,-1)$, and yellow regions to $(\Upsilon_{\bf k},  \Upsilon_{\underline{\bf k}}) = (-1,+1)$. Momenta of trajectories not yielding ABS are colored black. 
(c) Point contact conductance spectra along ${\bf n}=(1,1,1)$ for self-consistent order parameters (the numbers refer to columns for $r^\text{bulk}_\Delta$ in table \ref{RatiosPointGroups}), and with $t_0 = 10^{-\frac{1}{2}}$. 
(d) The topological invariant $N_{(111)}$, with $r^\text{bulk}_\Delta = 0.67$, where light green/blue corresponds to $N_{(111)} = \pm 1$, and white to trivial topology. 
(e) The surface band structure with $k^\parallel_1 = 0$, and $r^\text{bulk}_\Delta = 0.67$. 
(f) The lowest positive eigenvalues of $H_{\rm eff}$ for self-consistent order parameter with $r^\text{bulk}_\Delta = 0.67$. Black regions correspond to zero energy. 
Dashed circles in (d) and (f) show for comparison the projection of the spherical Fermi surface used in the quasiclassical calculations.}
\end{figure*}
The Andreev bound states (ABS) of NCSs have intricate structures and are spin polarized.\cite{PhysRevLett.101.127003} 
This is a consequence of the SOC being antisymmetric, $l_{\bf k} = - l_{-{\bf k}}$. States corresponding to different Andreev bound state branches have opposite spin polarization, and this spin polarization changes sign for reversed trajectories. As a result, the Andreev states carry spin current along the interface.\cite{PhysRevLett.101.127003}  The existence of a surface spin current is a direct consequence of the spin-orbit coupling in the system.

As an example, the momentum angle-resolved and spin-resolved local density of states, $N^{(z)}(\phi, \epsilon)$, computed with Eq. \eqref{Nxyz}, is plotted in Fig. \ref{MoreCubicStuff}(a) for momenta in the $xy$-plane (parameterized by the azimuthal angle $\phi$, the polar angle is $\theta=\pi/2$), at the surface with surface normal ${\bf n} = (1,0,0)$, for $g_2=1.03$ and a self-consistent pure triplet order parameter. An energy broadening $\epsilon \to \epsilon+i\delta $ with $\delta=10^{-2}$ was used, and the self-consistent order parameter was computed at $T=0.2T_c$.
Red (blue) indicate relative polarization for spin up (down) quasiparticles. The spin polarization axis is along the $z$-axis and $N^{(x)} =  N^{(y)} = 0$. This is true for all values of $g_2$ with a pure triplet order parameter. However, the ABS structure is very different for the four $g_2$ values. Furthermore the spin polarization axis is found to be dependent on the singlet to triplet ratio, in addition to surface orientation.

The momentum-resolved zero-energy ABS for ${\bf n} = (1,1,1)$
are shown in Fig. \ref{MoreCubicStuff} (b), 
computed with the bulk value of the order parameter all the way to the surface,
assuming $r^\text{bulk}_\Delta = 0.67$. The tunneling parameter was set to $t_0 = 10^{-\frac{1}{2}}$ 
(or $t_0^2=0.1$, making sure to be in the tunneling regime), and the broadening of the energies, $\epsilon \leftarrow \epsilon + i\delta$, with $\delta=10^{-3}$.
The disk is the projection of the spherical Fermi surface onto the slab surface. Black indicates that there are no ABS for those momenta, green indicates ABS for which $(\Upsilon_{\bf k},  \Upsilon_{\underline{\bf k}}) = (+1,-1)$, and yellow $(\Upsilon_{\bf k},  \Upsilon_{\underline{\bf k}}) = (-1,+1)$. For this choice of surface orientation and singlet to triplet ratio these two types of trajectories are the only ones yielding ABS. This is not the case for lower singlet to triplet ratios, other $g_2$ values, and/or other surface orientations. Then there can exist solutions to Eq. \eqref{Feq}. Indeed, for $\Delta_s/\Delta_t < \min|{\bf l}_{{\bf k}_F}|$ they are the only solutions yielding ABS. For $\Delta_s/\Delta_t > \max|{\bf l}_{{\bf k}_F}|$ no zero-energy ABS are seen.

Point contact conductance spectra for $g_2 = 1.03$, $t_0 = 10^{-\frac{1}{2}}$ and ${\bf n} = (1,1,1)$ are shown in Fig. \ref{MoreCubicStuff} (c).
A small energy broadening $\epsilon \to \epsilon + i\delta $ with $\delta=10^{-2}$ was used for the plot, except close to zero energy, where $\delta=10^{-5}$ was used (and 2.5 times as many momentum directions in the momentum average) in order to show the sharp zero-bias conductance peak. 
The transmission parameter is set to $t_0=10^{-\frac{1}{2}}$. Furthermore, $\max(\Delta) \equiv \Delta^\text{bulk}_s + \Delta^\text{bulk}_t\max(|{\bf l}_{{\bf k}_{\text{F}}}|)$, and the plots are shifted $0.2$ upwards from each other for the sake of visibility. The order parameters used are computed self-consistently at $T=0.2T_\text{c}$, and with only one active channel.
The point contact conductance spectra differ widely between surface orientations and the values of $g_2$, in addition to the less pronounced difference between singlet to triplet ratios. The most striking difference is the appearance of zero-bias conductance peaks (ZBCPs) which are present for all singlet to triplet ratios in the interval $\min|{\bf l}_{{\bf k}_F}| < \Delta_s/\Delta_t < \max|{\bf l}_{{\bf k}_F}|$ provided there are trajectories with $\text{sign}[\Delta_-({\bf k})] = - \text{sign}[\Delta_-(\underline{\bf k})]$. 

In Fig. \ref{MoreCubicStuff} (d) the topological invariants $N_{(111)}$ and $W_{(111)}$ are plotted for $r^\text{bulk}_\Delta = 0.67$. However, $W_{(111)} = 1$, i.e. trivial, for this choice of parameters, and trivial topology is colored white. Light green/blue corresponds to $N_{(111)} = \pm 1$. The dashed circle indicates the projection of the spherical Fermi surface used in the quasiclassical calculations, i.e. Fig. \ref{MoreCubicStuff} (a) - (c). Even though the Fermi surface is not spherical, it is clear that the zero-energy ABS are directly related to the topology. 
As is shown for the tetragonal point group $C_{4v}$ below, the ABS given by solutions to Eq. \eqref{Feq}, for the relevant values of $(\Upsilon_{\bf k}, \Upsilon_{\underline{\bf k}})$, is directly related to the $\mathbb{Z}_2$ invariant being non-trivial (i.e. $W_{(111)} = -1$).

Zero-energy states are present in the band structure whenever the aforementioned topological invariants have non-trivial values. In Fig. \ref{MoreCubicStuff} (e) the surface band structure is shown for $r^\text{bulk}_\Delta = 0.67$  along the $k^\parallel_2$-axis with $k^\parallel_1 = 0$, and $L = 1.3\cdot 10^{4}$ layers. $N_{(111)} \not = 0$ gives rise to singly degenerate zero-energy flat bands, one on each surface, with the corresponding wavefunctions decaying exponentially into the bulk. The surface momenta of the zero-energy flat bands are given by $N_{(111)}({\bf k}_\parallel) \not = 0$, which can be seen in Fig. \ref{MoreCubicStuff} (f) where the lowest positive eigenvalue of $H_{\rm eff}$ [see Eq.~\eqref{Heff}] is plotted for self-consistent order parameter.
Note that the zero-energy flat-bands are given by the projection of non-trivial values of the 1D winding number.

%%%%%%%%%%
\subsection{The Tetragonal Point Group $C_{4v}$}

\begin{figure}[t] \centering
%{\includegraphics[width=0.9\columnwidth]{Fig6_gray.pdf}}
{\includegraphics[width=0.9\columnwidth]{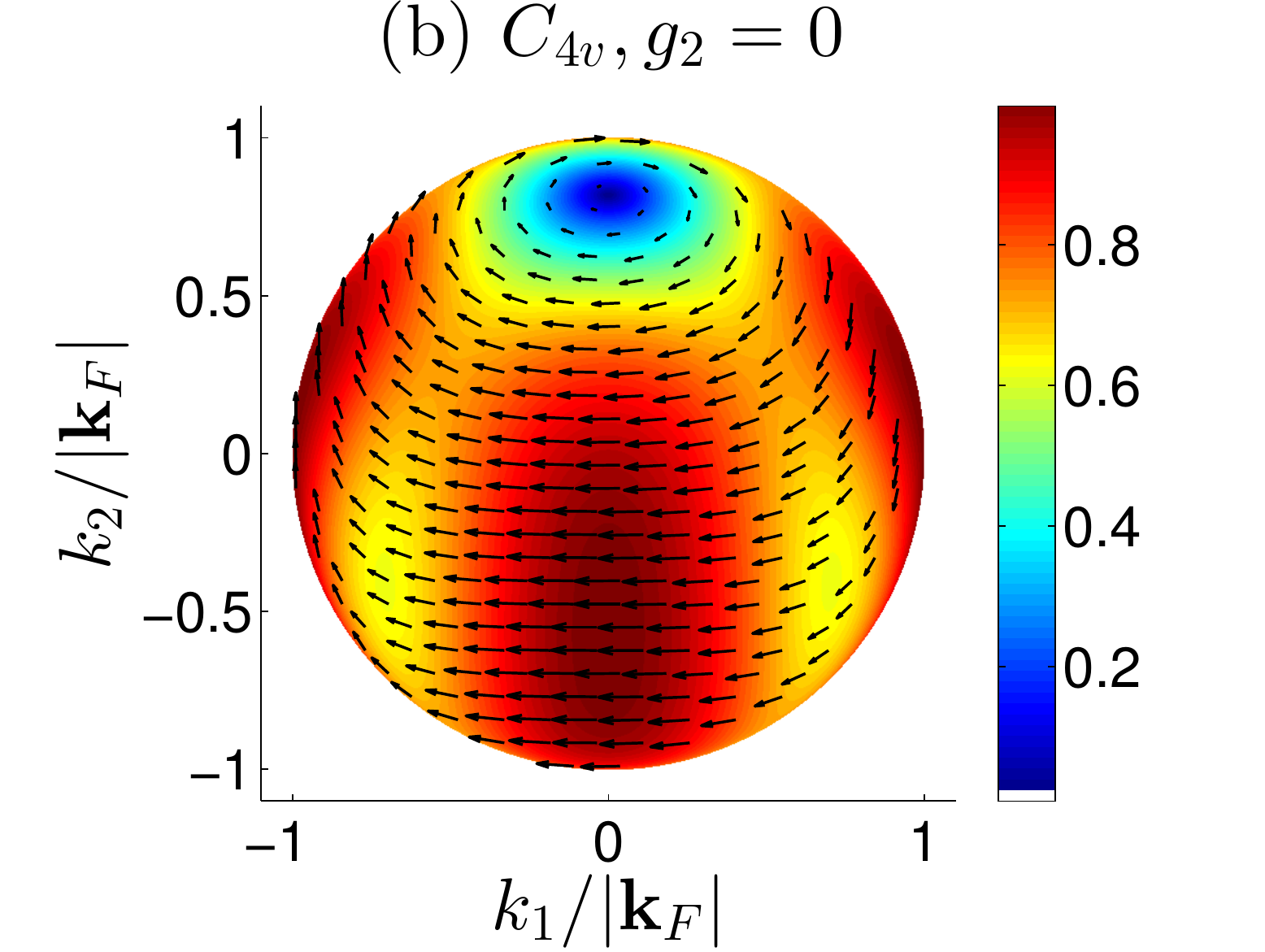}}
\caption{\label{SOCplotC4v}The SOC vector, defined by Eq. \eqref{SOC_C4v}, with $g_2 = 0$. See the caption of Fig. \ref{SOCplotO}.}
\end{figure}

To next-nearest neighbors in the sum over Bravais lattice sites \cite{SamokhinAnnals} the SOC vector corresponding to the tetragonal point group $C_{4v}$ takes the form
\begin{equation}
\label{SOC_C4v}
{\bf l}_{\bf k} = \begin{pmatrix} \sin(k_y) \\ -\sin(k_x) \\ g_2 \sin(k_x) \sin(k_y) \sin(k_z) \left[\cos(k_y) - \cos(k_x) \right]\end{pmatrix}
\end{equation}
where $g_2$ determines the relative weight between first and second order contributions, just like for the cubic point group $O$.
Its magnitude and direction on the Fermi surface is illustrated in Fig.~\ref{SOCplotC4v}.

\begin{figure}[b] \centering
    \subfloat{{\includegraphics[width=0.5\columnwidth]{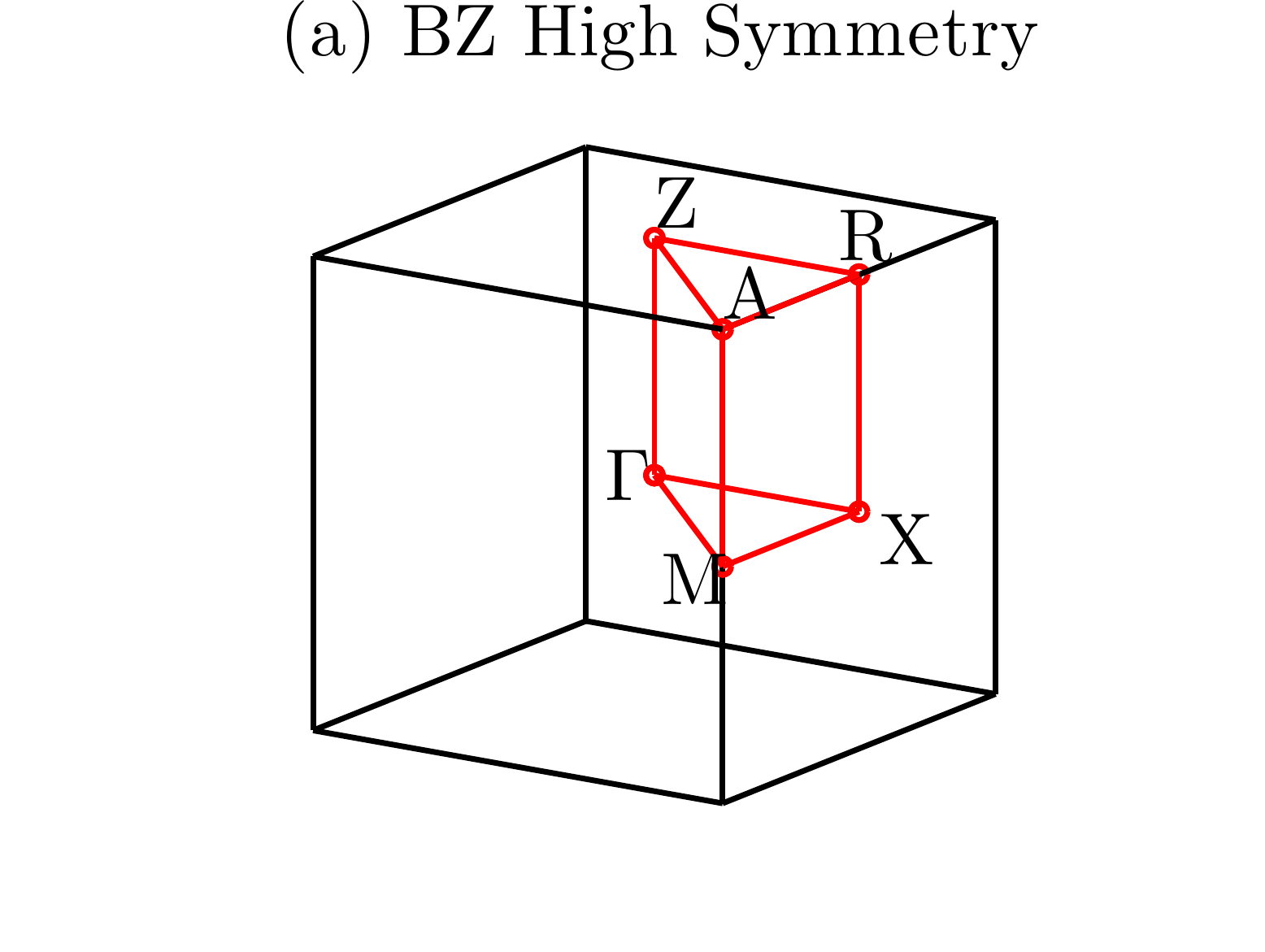} }}%
    %\qquad
    \subfloat{{\includegraphics[width=0.5\columnwidth]{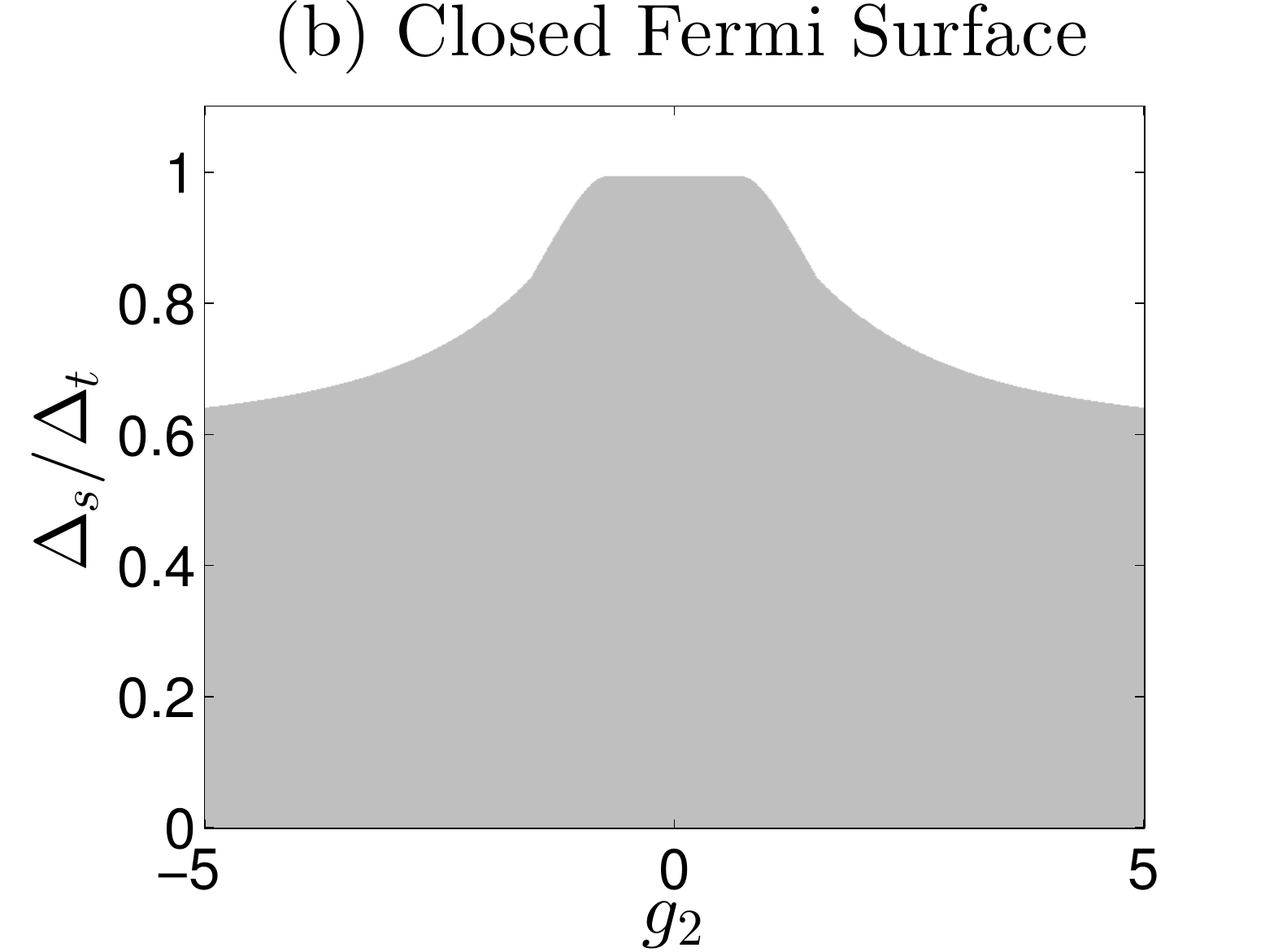} }}%
    \caption{(a) High symmetry points and axes in the BZ of a tetragonal crystal. (b) Topological phase diagram for a closed Fermi surface with $\mu = -50\alpha$ and $t_1 = -40\alpha$. 
White areas: gapped phase with trivial topology, $(N_\mathcal{L}, \nu)= (0,0)$; grey: nodal phase with $N_\mathcal{L}=1$ [loop defined by Eq. \eqref{LoopPath}].
}%
    \label{TetragonalStuff}%
\end{figure}

But unlike $O$, this point group has line nodes of the SOC in the BZ. For all values of $g_2$ the SOC is identically zero along the three paths parallel to the $z$-axis, $\Gamma \rightarrow \text{Z}$, $\text{X} \rightarrow \text{R}$, and $\text{M} \rightarrow \text{A}$ in Fig. \ref{TetragonalStuff} (a). Given the simple cubic first order tight-binding dispersion, 
for the range of $\mu $ we study
the line node $\Gamma \rightarrow \text{Z}$ intersects all closed, and the line node 
$\text{X} \rightarrow \text{R}$ 
intersects all open Fermi surfaces. 
Thus $\min|{\bf l}({\bf k}^-_F)| = 0$ for both cases. The transition between the two is therefore seamless, and there are no fully gapped phases with $\text{sign}[\Delta_-({\bf k}^-_F)] = -1$. The only two distinct phases is a topologically trivial, $\nu=0$, and a nodal non-trivial phase, $N_\mathcal{L}=1$, shown in white and grey respectively in Fig. \ref{TetragonalStuff} for a closed Fermi surface, $\mu = -50\alpha$.

\begin{figure}[t] 
\begin{tabular}{ll}
\subfloat{{\includegraphics[width=0.5\columnwidth,trim=30 -10 18 0, clip]{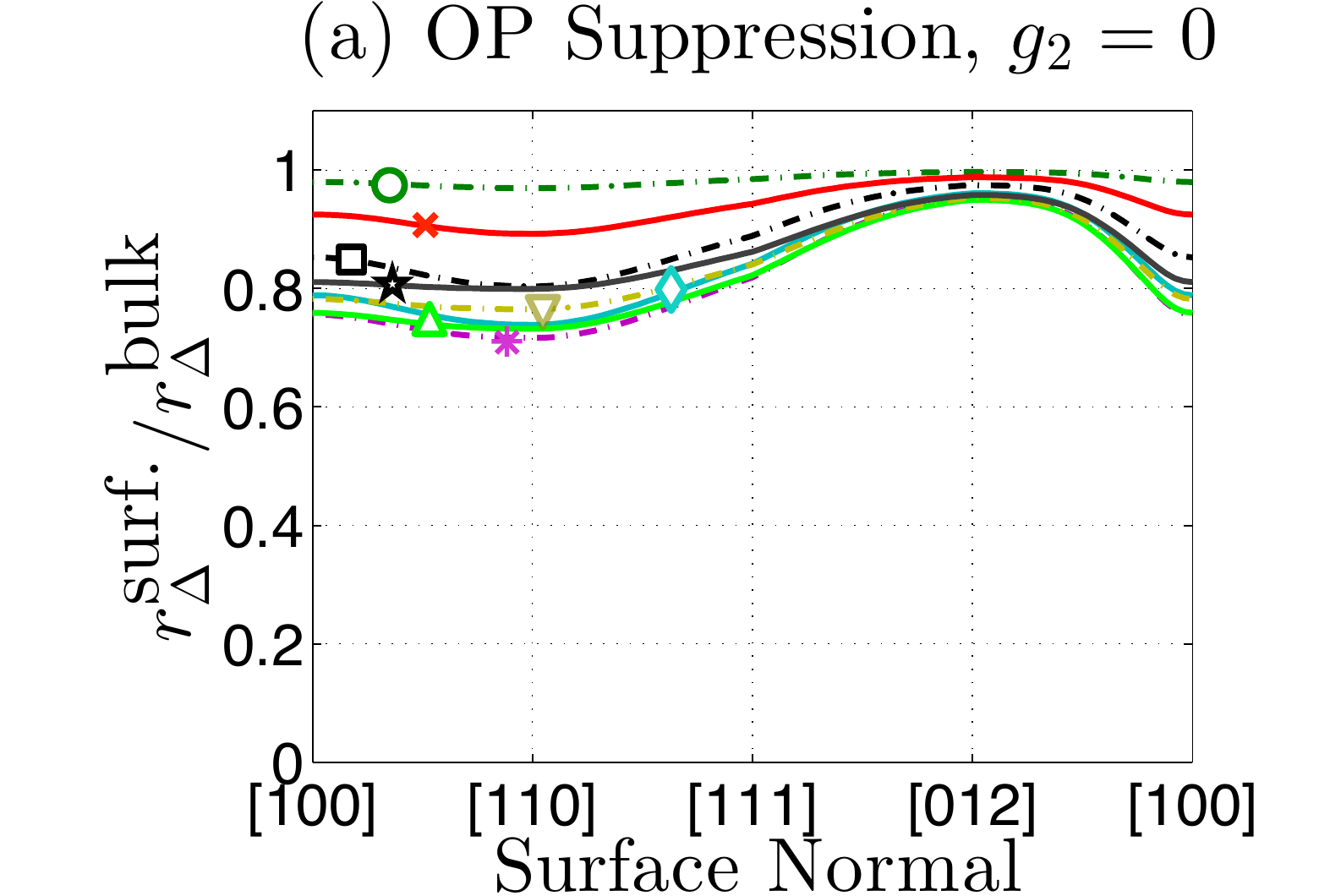} }} &
\subfloat{{\includegraphics[width=0.5\columnwidth,trim=30 -10 18 0, clip]{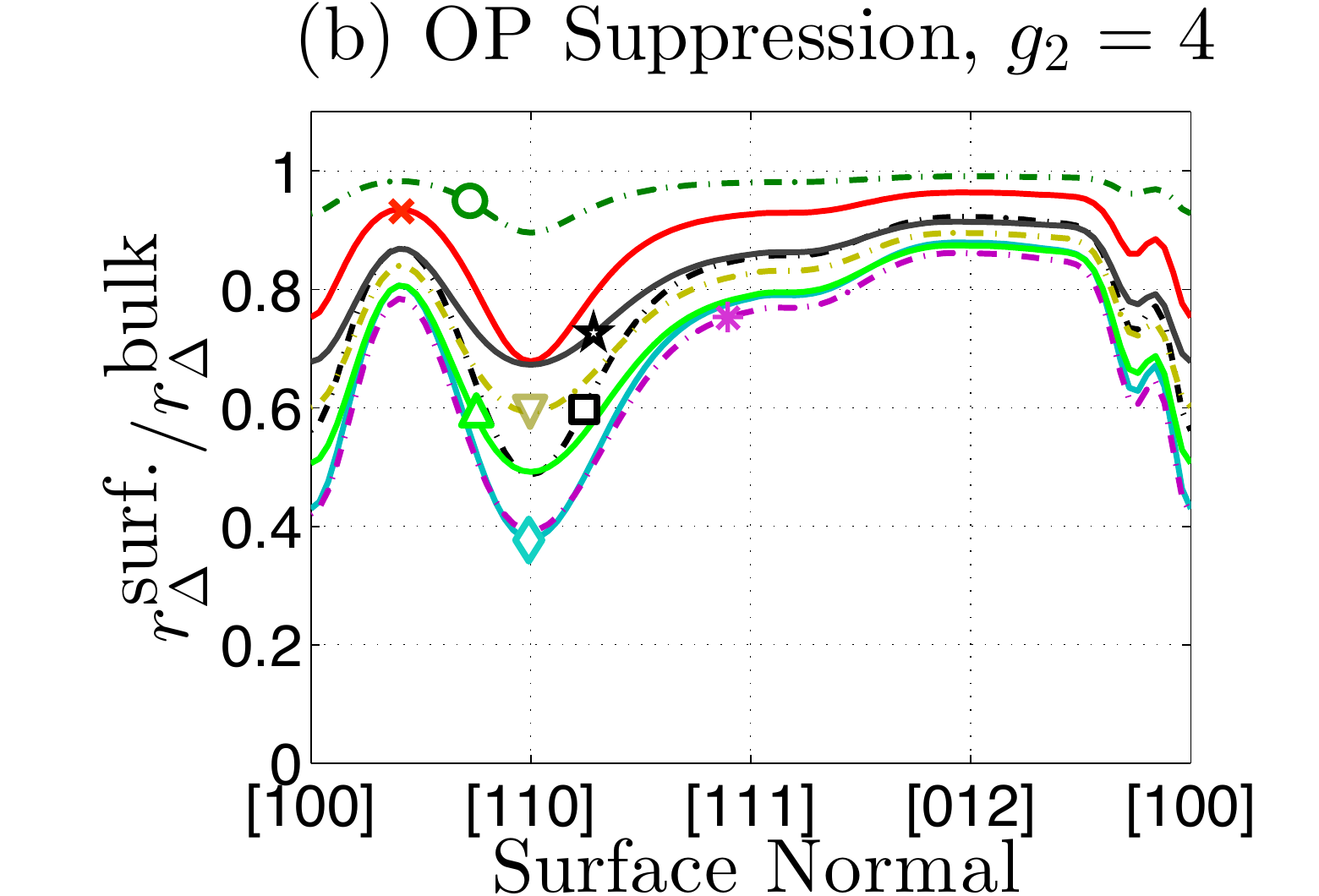} }}
     \\
\subfloat{{\includegraphics[width=0.5\columnwidth,trim=5 0 18 0, clip]{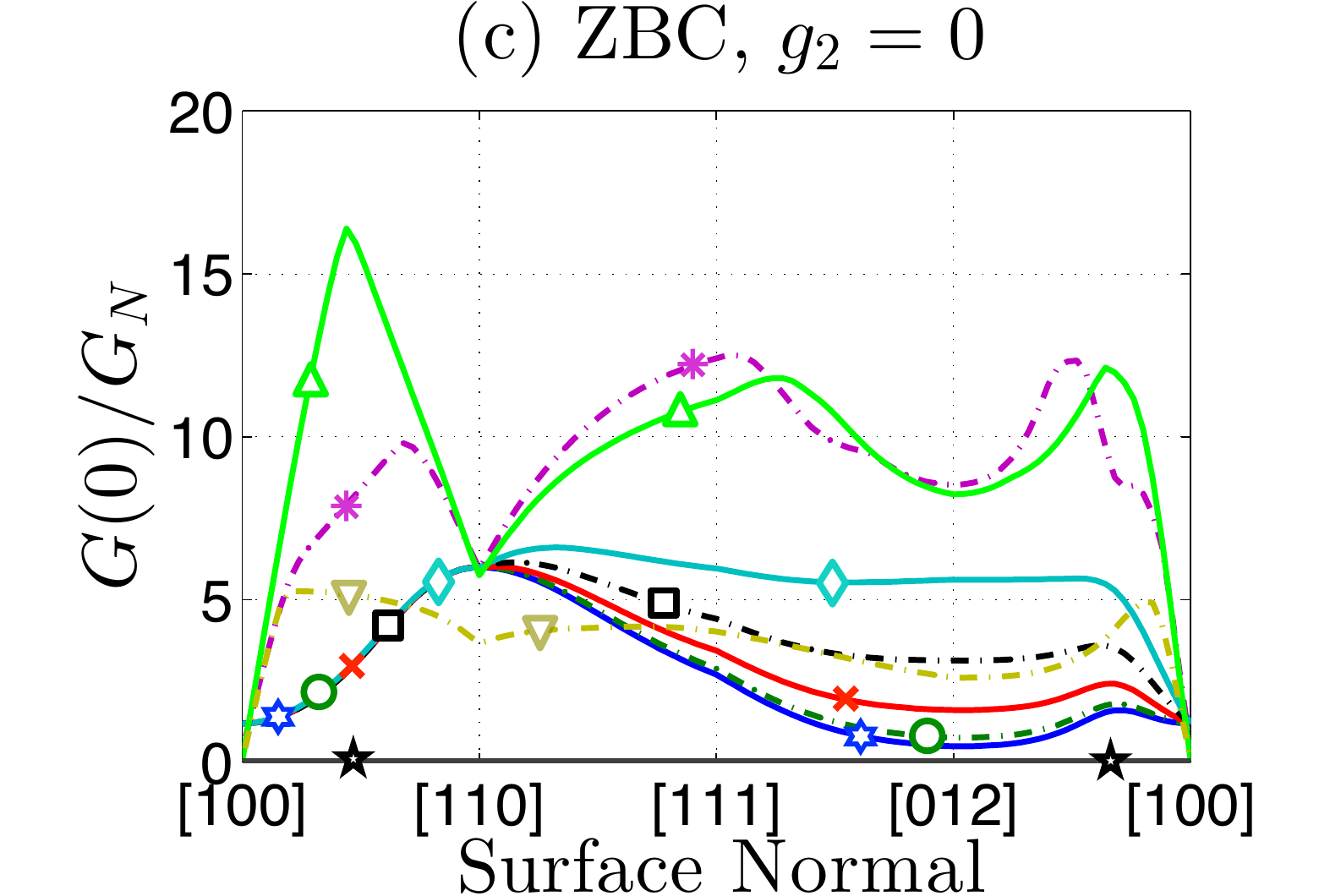} }} &
\subfloat{{\includegraphics[width=0.52\columnwidth,trim=5 0 0 0, clip]{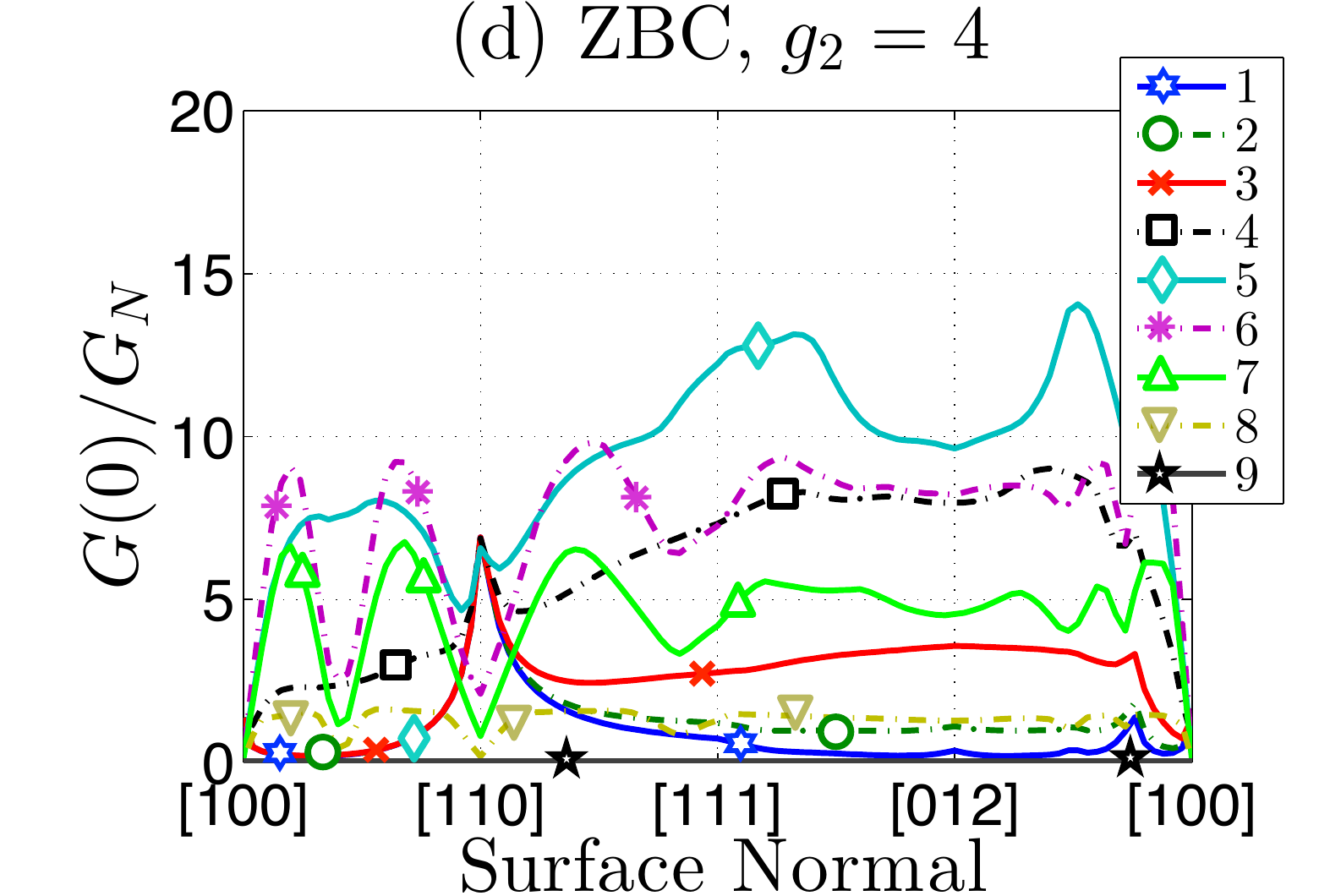} }}
\end{tabular}
\caption{\label{TetragonalSuppressionAndZBCP}Plots (a) - (b) show the quantity $r^\text{surf.}_\Delta/r^\text{bulk}_\Delta = \left[ \Delta_s/\Delta_t \right]^\text{surf.} \cdot \left[\Delta_t/\Delta_s \right]^\text{bulk}$ as a measure of the order parameter surface suppression. This is done for a range of different surface normals along the path ${\bf n} = (1,0,0) \rightarrow (1,1,0) \rightarrow (1,1,1) \rightarrow (0,1,2) \rightarrow (1,0,0)$. In plots (c) - (d) the zero-bias conductance, computed with $t_0 = 10^{-\frac{1}{2}}$, is shown for the same surface normals. The numbers in the legend holds for all plots and correspond to the columns in table \ref{RatiosPointGroups} showing the scaled singlet to triplet ratios.}
\end{figure}
Despite there only being a single topologically non-trivial phase the order parameter is calculated self-consistently for the two values $g_2 \in \{0, 4 \}$ in order to study the effect of second order contributions to the SOC vector. This is done for nine values of the scaled bulk singlet to triplet ratio, $r^\text{bulk}_\Delta \in [0, 1.1]$, with one active channel. The exact values are shown in table \ref{RatiosPointGroups}. The order parameter is calculated with the same surface normals as for the cubic point group. How the order parameter suppression depends on the surface orientation can be seen in Fig. \ref{TetragonalSuppressionAndZBCP} (a) - (b). Here the greatest suppression is not for the surface normal ${\bf n} = (1,1,1)$, but rather ${\bf n} = (1,1,0)$, and ${\bf n} = (0,1,2)$ shows very little suppression.

\begin{figure*} \centering
%\subfloat{{\includegraphics[width=0.65\columnwidth]{Fig9a_gray.pdf} }}%
\subfloat{{\includegraphics[width=0.65\columnwidth]{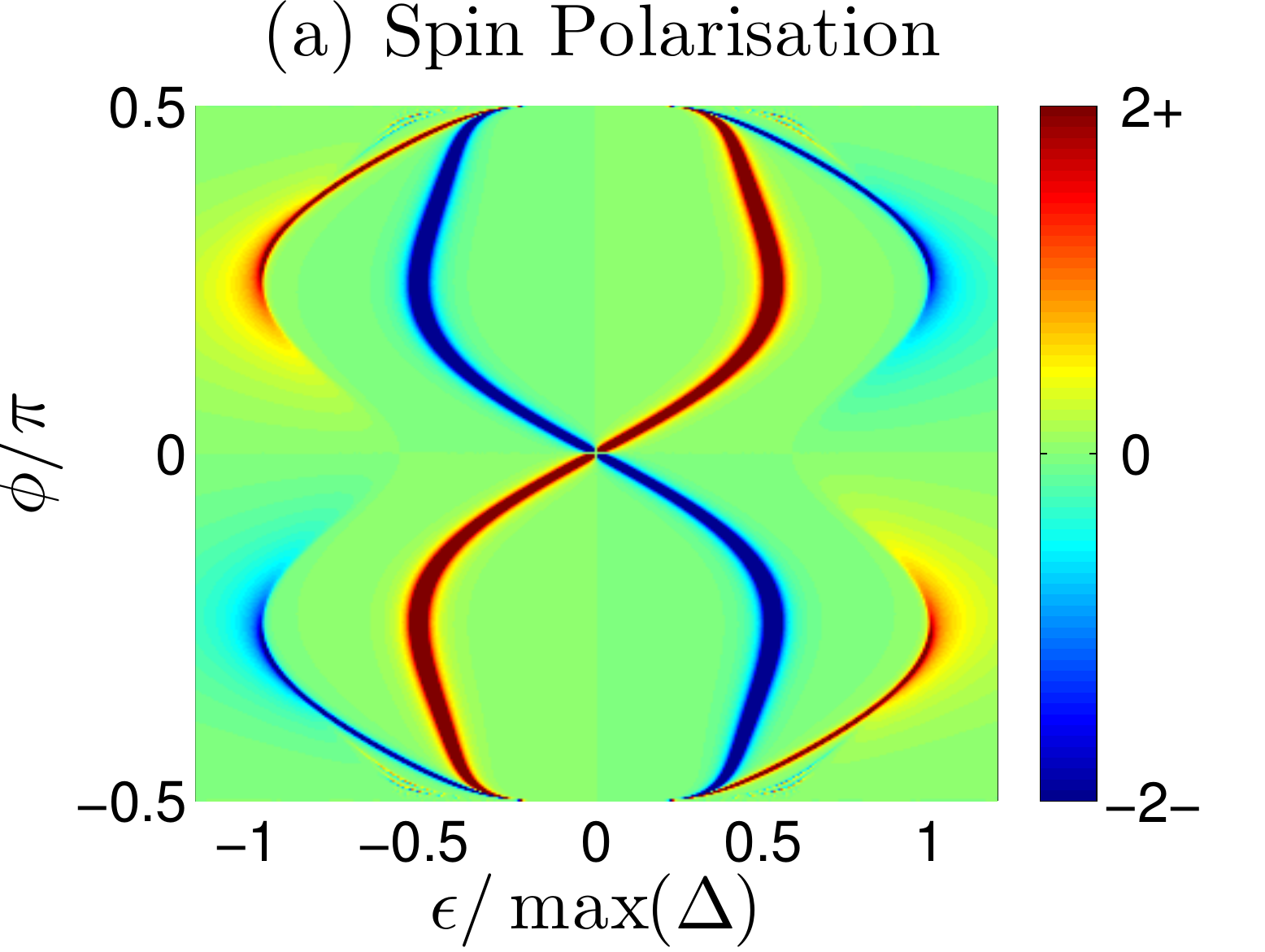} }}%
\subfloat{{\includegraphics[width=0.65\columnwidth]{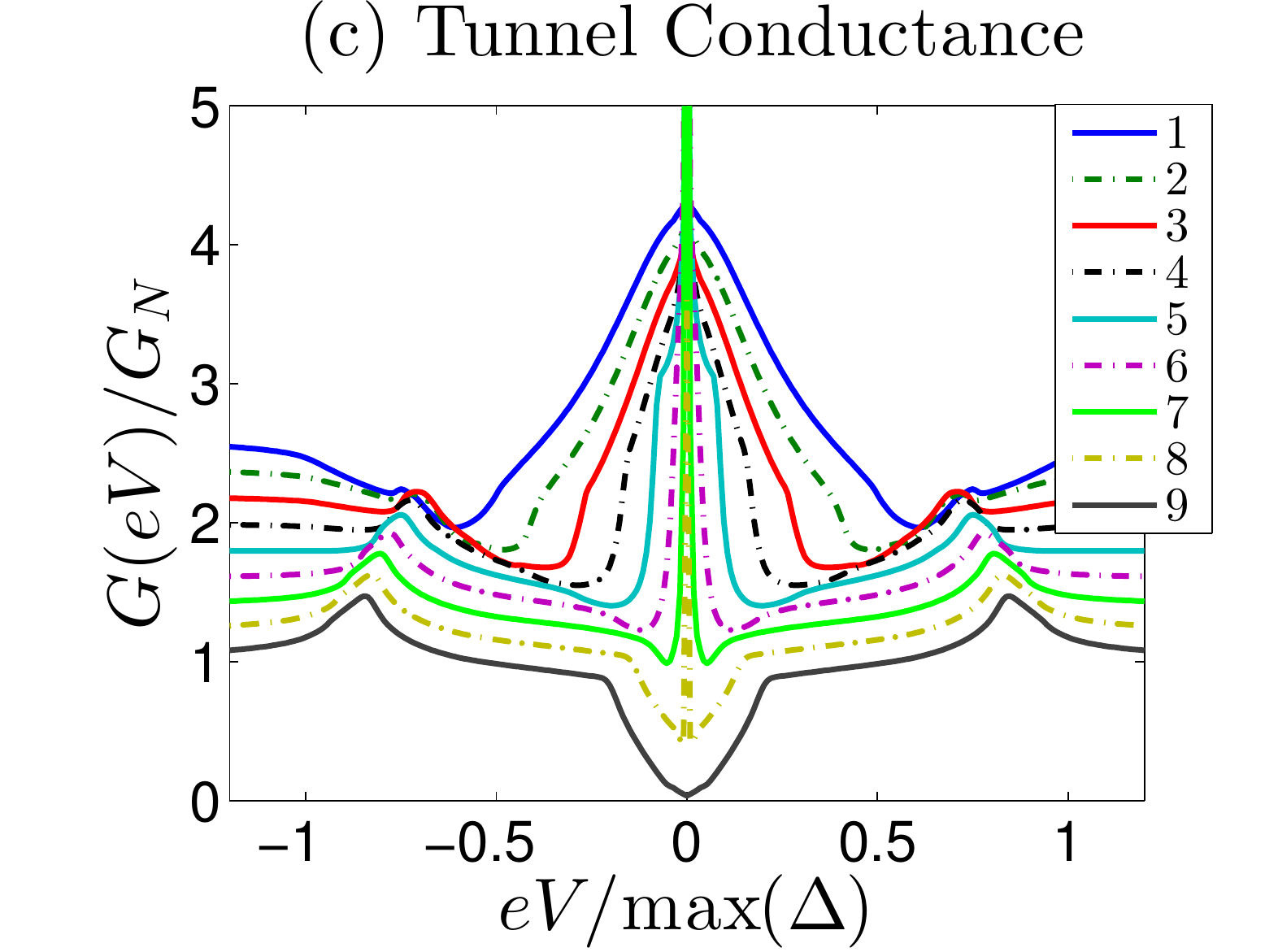} }}%
\subfloat{{\includegraphics[width=0.65\columnwidth]{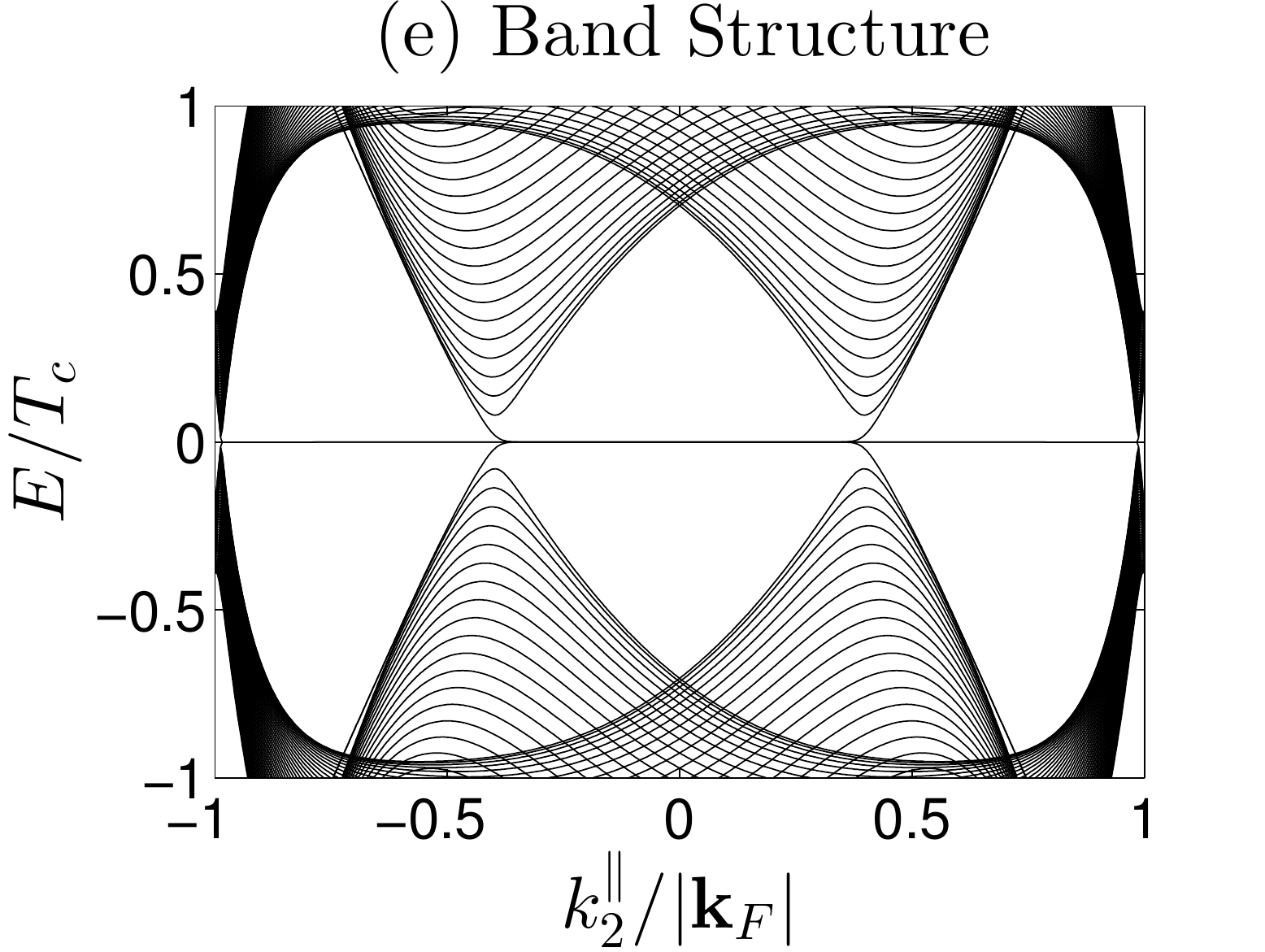} }}%
\\
\subfloat{{\includegraphics[width=0.65\columnwidth]{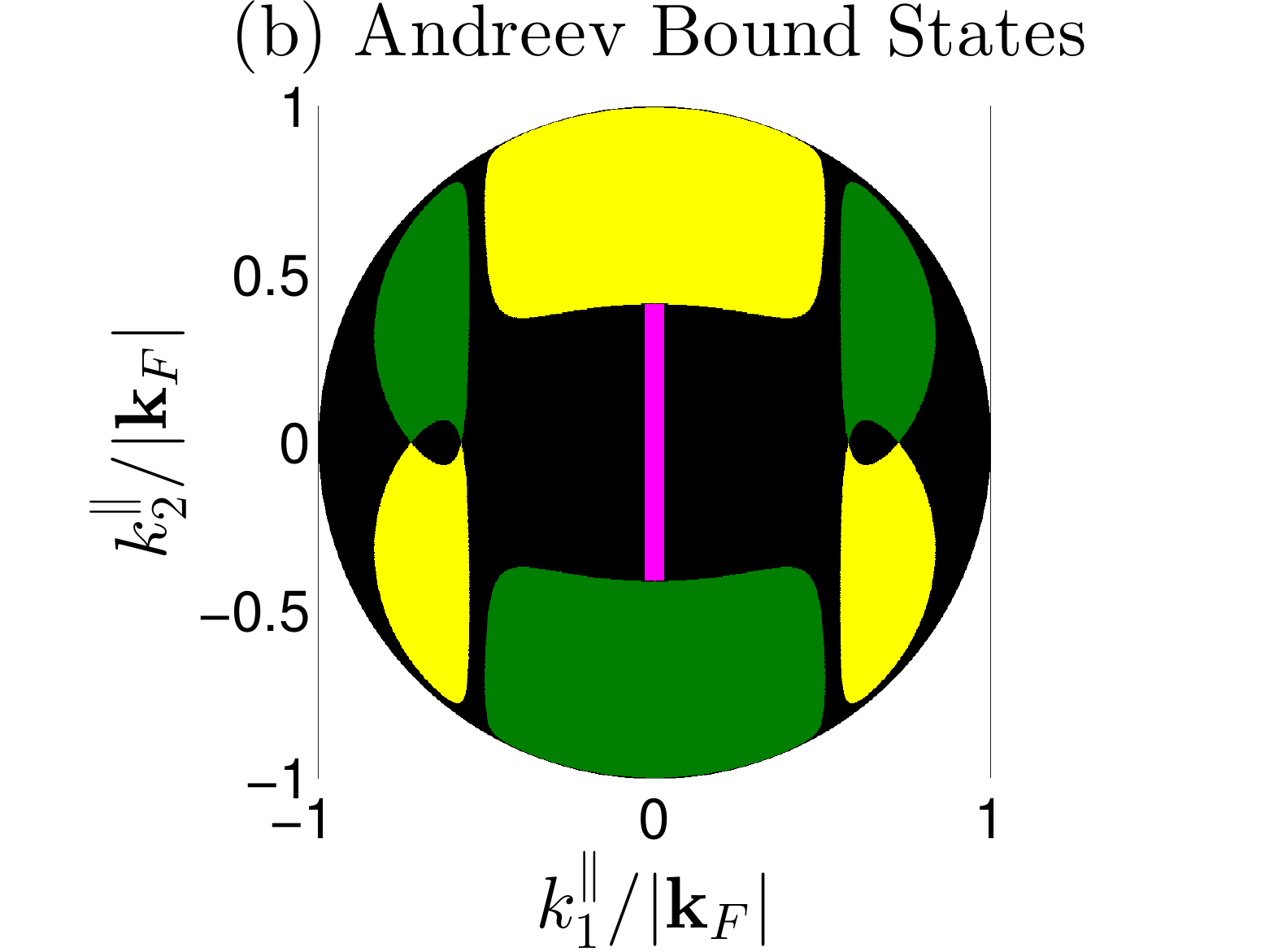} }}%
\subfloat{{\includegraphics[width=0.65\columnwidth]{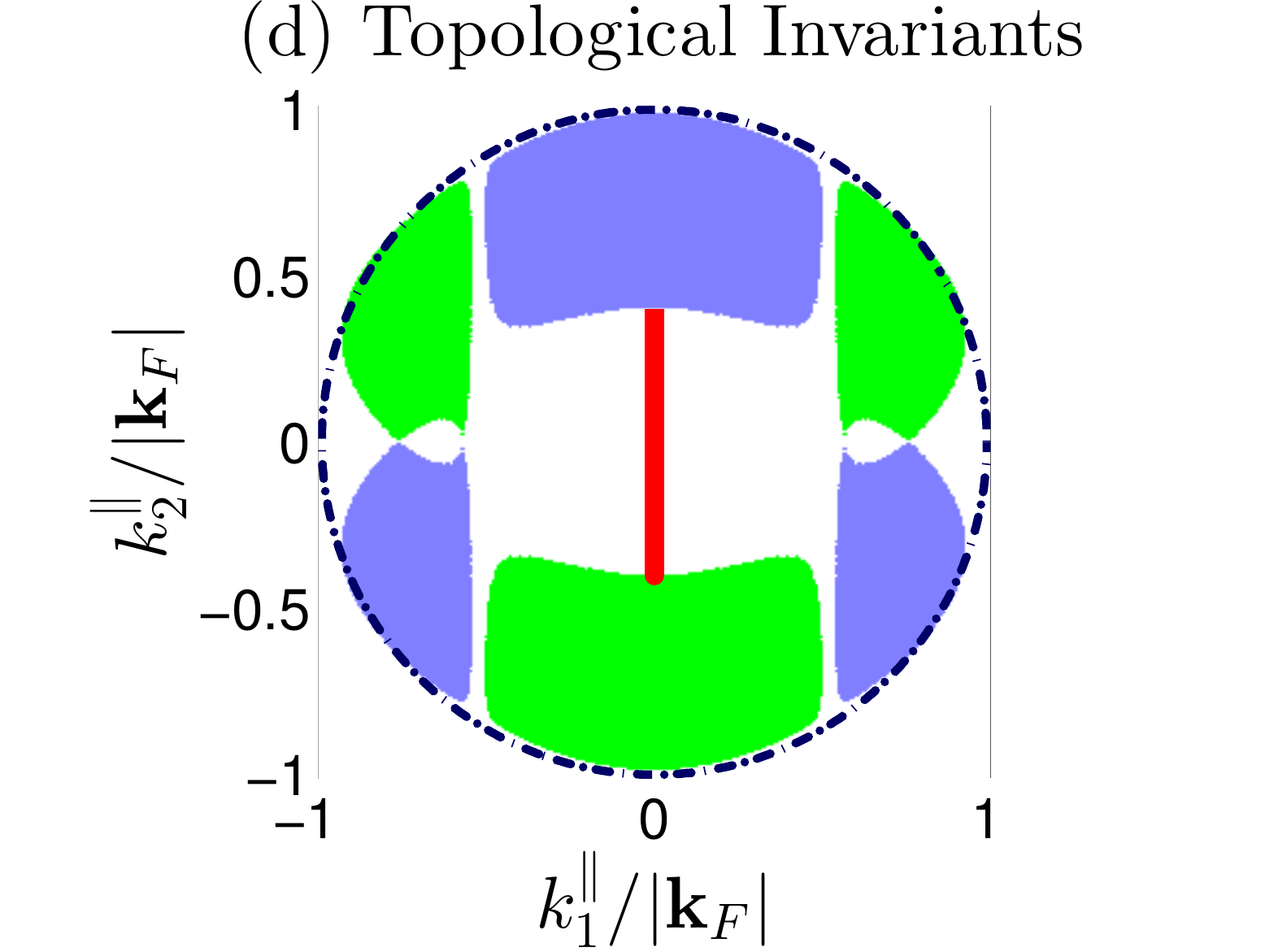} }}%
\subfloat{{\includegraphics[width=0.65\columnwidth]{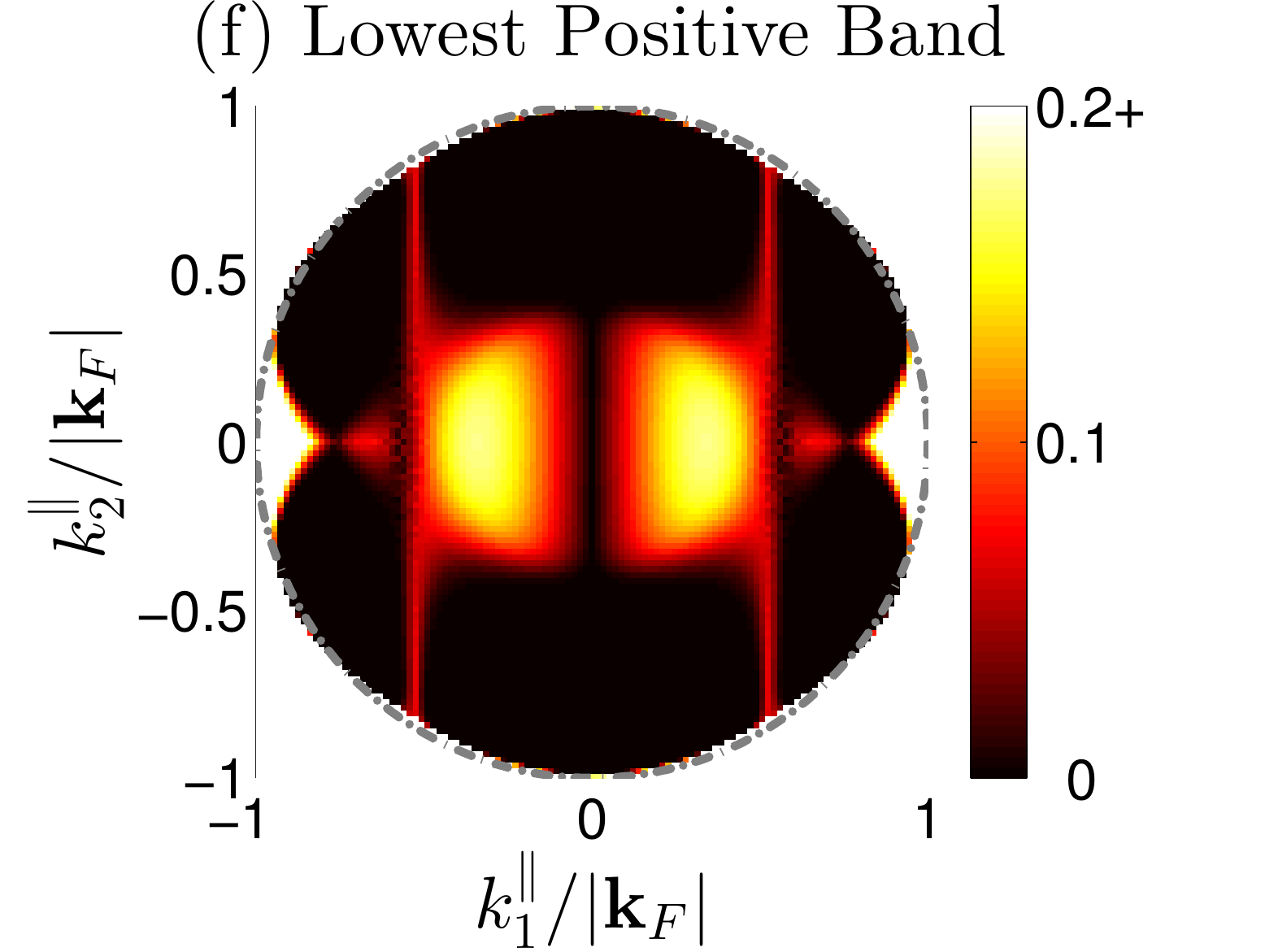} }}%
        \caption{\label{MoreTetragonalStuff}All plots are for the tetragonal point group $C_{4v}$ with $g_2 = 0$. See the caption of Fig. \ref{MoreCubicStuff}, here $r^\text{bulk}_\Delta = 0.69$ in (b) and (d) - (f).  The vertical magenta line in (b) denotes an ABS for which $\Upsilon_{\bf k} = \Upsilon_{\underline{\bf k}} = -1$.  The corresponding vertical red line in (d) indicates $W_{(111)} = -1$.}
\end{figure*}

The zero-bias conductances for the two $g_2$ values, are very dissimilar for surface normals in the $xy$-plane. With $g_2 = 0$, Fig. \ref{TetragonalSuppressionAndZBCP} (c), rather large conductances are seen for $0.69 \leq r^\text{bulk}_\Delta \leq 0.96$ in between the high symmetry axes ${\bf n} = (1,0,0)$ and ${\bf n} = (1,1,0)$, with the largest for $r^\text{bulk}_\Delta = 0.83$ and ${\bf n} \approx (1,0.44,0)$. The lines corresponding to $0 < r^\text{bulk}_\Delta \leq 0.55$ are (almost) degenerate due to all of them having smaller singlet to triplet ratios than the rather small difference between the maximum and minimum value of the SOC in the $xy$-plane of the Fermi surface. Only a few trajectories around the poles contribute to the ZBCPs. There are no ZBCPs for ${\bf n} = (1,1,0)$ but rather a large dome-like feature which is interestingly higher than the peaks for the (almost) degenerate lines. With $g_2 = 4$, Fig. \ref{TetragonalSuppressionAndZBCP} (d), the lines corresponding to $0.69 \leq r^\text{bulk}_\Delta \leq 0.96$  show a dip for ${\bf n} \approx (1,0.4,0)$ due to the higher order contributions in the SOC changing the shape of the nodal rings, causing their projection onto the surface to largely overlap for these singlet to triplet ratios.

The ABS are heavily affected by self-consistency and are, just like for $O$, spin polarized. In Fig. \ref{MoreTetragonalStuff} (a) the quantity $N^{(z)}({\bf k})$, Eq. \eqref{Nxyz}, is plotted in the $xy$-plane for a pure triplet order parameter and ${\bf n} = (1,0,0)$. The largest effect of self-consistency is seen for glancing trajectories and energies between approximately $|\epsilon|/\max(\Delta) \in [0.5, 1]$. The ABS in this range are not present in the non-self-consistent case. For a pure triplet order parameter $N^{(x)} = N^{(y)} = 0$.

The momentum-resolved zero-energy ABS, for $g_2 = 0$, $r^\text{bulk}_\Delta = 0.69$ and ${\bf n} = (1,1,1)$, is shown in Fig. \ref{MoreTetragonalStuff} (b). Here, not only the trajectories for which $\Upsilon_{\bf k} = -\Upsilon_{\underline{\bf k}} \not = 0$, colored green and yellow, give rise to ABS, but also trajectories for which $\Upsilon_{\bf k} = \Upsilon_{\underline{\bf k}} = -1$ and Eq. \eqref{Feq} holds, colored magenta. This magenta line is there due to the SOC vanishing along the high symmetry axis $\Gamma \rightarrow \text{Z}$, see Fig. \ref{TetragonalStuff} (a), combined with the SOC vector being perpendicular to this axis. This line is present for $g_2 = 4$ as well, but only for $0 \leq r^\text{bulk}_\Delta < 0.69$, whereas it is present for $0 \leq r^\text{bulk}_\Delta < 0.96$ with $g_2 = 0$, amongst the ratios investigated.

In the tunnel conductance spectra, plotted for ${\bf n} = (1,1,1)$ and $t_0 = 10^{-\frac{1}{2}}$ in Fig. \ref{MoreTetragonalStuff} (c), ZBCPs are seen for all scaled singlet to triplet ratios in the interval $r^\text{bulk}_\Delta \in (0,1)$, due to $\min|{\bf l}({\bf k})| = 0$. Unlike the ZBCPs in the tunnel conductance spectra for $O$ with $g_2 = 1.03$, Fig. \ref{MoreCubicStuff} (c), which emanate from valleys around $\epsilon = 0$, most of the ZBCPs here emanate from a large dome. The domes are a consequence of the magenta colored ABS together with the 'flatness' of the coherence functions in the denominator of the expression for the tunnel conductance when varying the momentum, such that trajectories with momenta in close vicinity to the ABS condition give rise to a large number of states that contribute considerably to the tunnel conductance.
With increasing $r^\text{bulk}_\Delta$ these states decrease in number. For $r^\text{bulk}_\Delta > 0.83$ they have completely disappeared and thus the dome is gone and the ZBCP emanates from a valley. 
%Even though the magenta line is present for $g_2 = 4$ as well, there are no domes in the corresponding spectra. 
%The reason is that the coherence functions are not as 'flat' as function of momentum in this case, hence the spectra show neither domes not valleys.

The topological invariants $N_{(111)}$ and $W_{(111)}$ are plotted in Fig. \ref{MoreTetragonalStuff} (d). Light blue/green corresponds to $N_{(111)} = \pm 1$ and white to trivial values of both invariants. The dashed circle is the projection of the spherical Fermi surface used in the quasiclassical calculations. The red line is given by $W_{(111)}=-1$. Thus states corresponding to solutions of Eq. \eqref{Feq} are directly related to the $\mathbb{Z}_2$ invariant being non-trivial, and are topologically protected as well.

In Fig. \ref{MoreTetragonalStuff} (e) the band structure is shown for $g_2 = 0$ and $r^\text{bulk}_\Delta = 0.69$ along the $k^\parallel_2$-axis with $k^\parallel_1 = 0$, and $L = 1.3\cdot 10^{4}$ layers. Note that the states corresponding to $W_{(111)}=-1$ are doubly degenerate on each surface. Just like for $O$ the zero-energy bands are given by the projection of the non-trivial values of the topological invariants, which can be seen in Fig. \ref{MoreTetragonalStuff} (f).

%%%%%%%%%%
\subsection{The Tetrahedral Point Group $T_d$}

\begin{figure}[t] \centering
%{\includegraphics[width=1.0\columnwidth]{Fig10_gray.pdf}}
{\includegraphics[width=1.0\columnwidth]{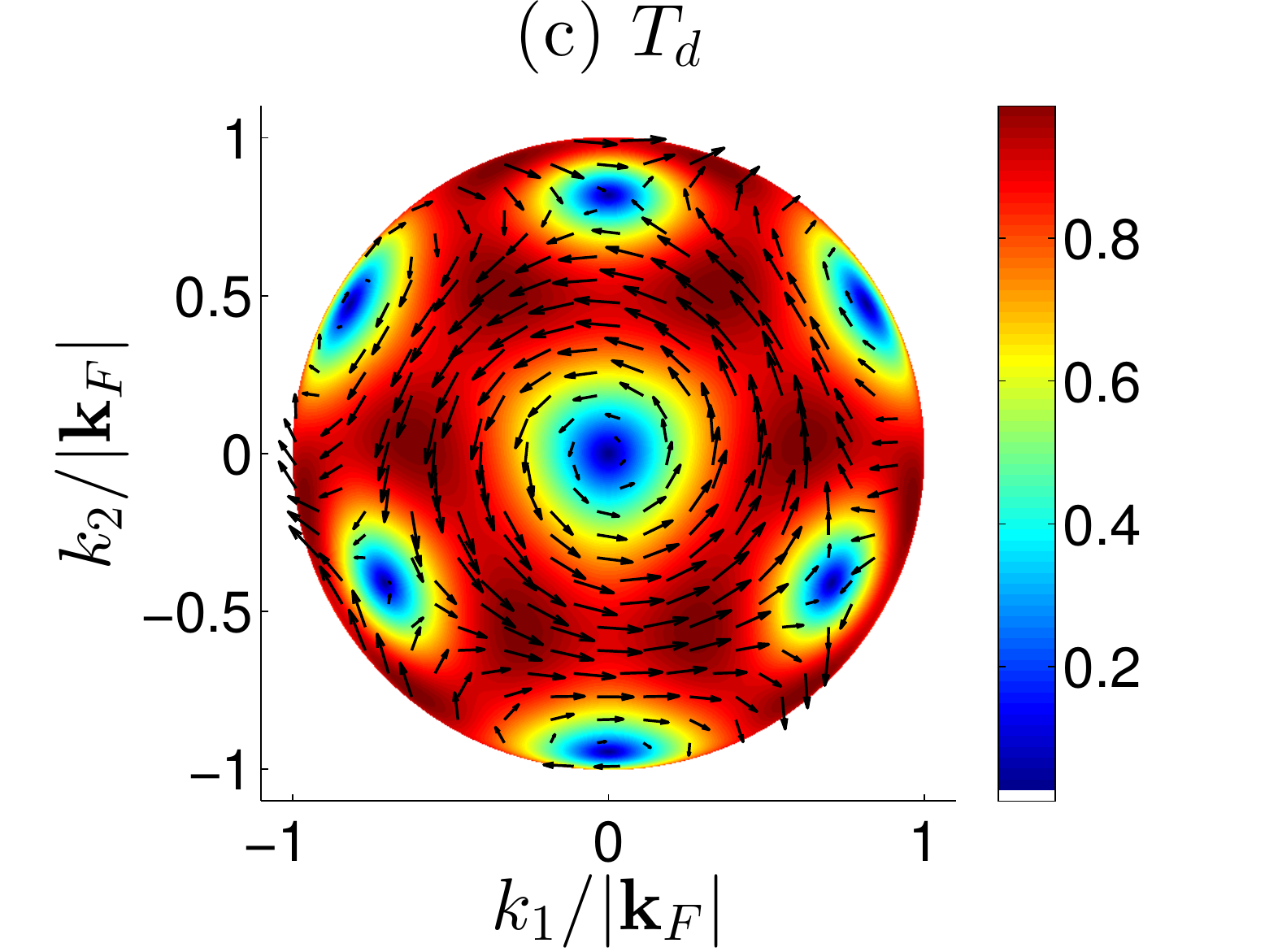}}
\caption{\label{SOCplotTd}The SOC vector, defined by Eq. \eqref{SOC_Td}. See the caption of Fig. \ref{SOCplotO}.}
\end{figure}

\begin{figure}[b] \centering
    \subfloat{{\includegraphics[width=0.5\columnwidth]{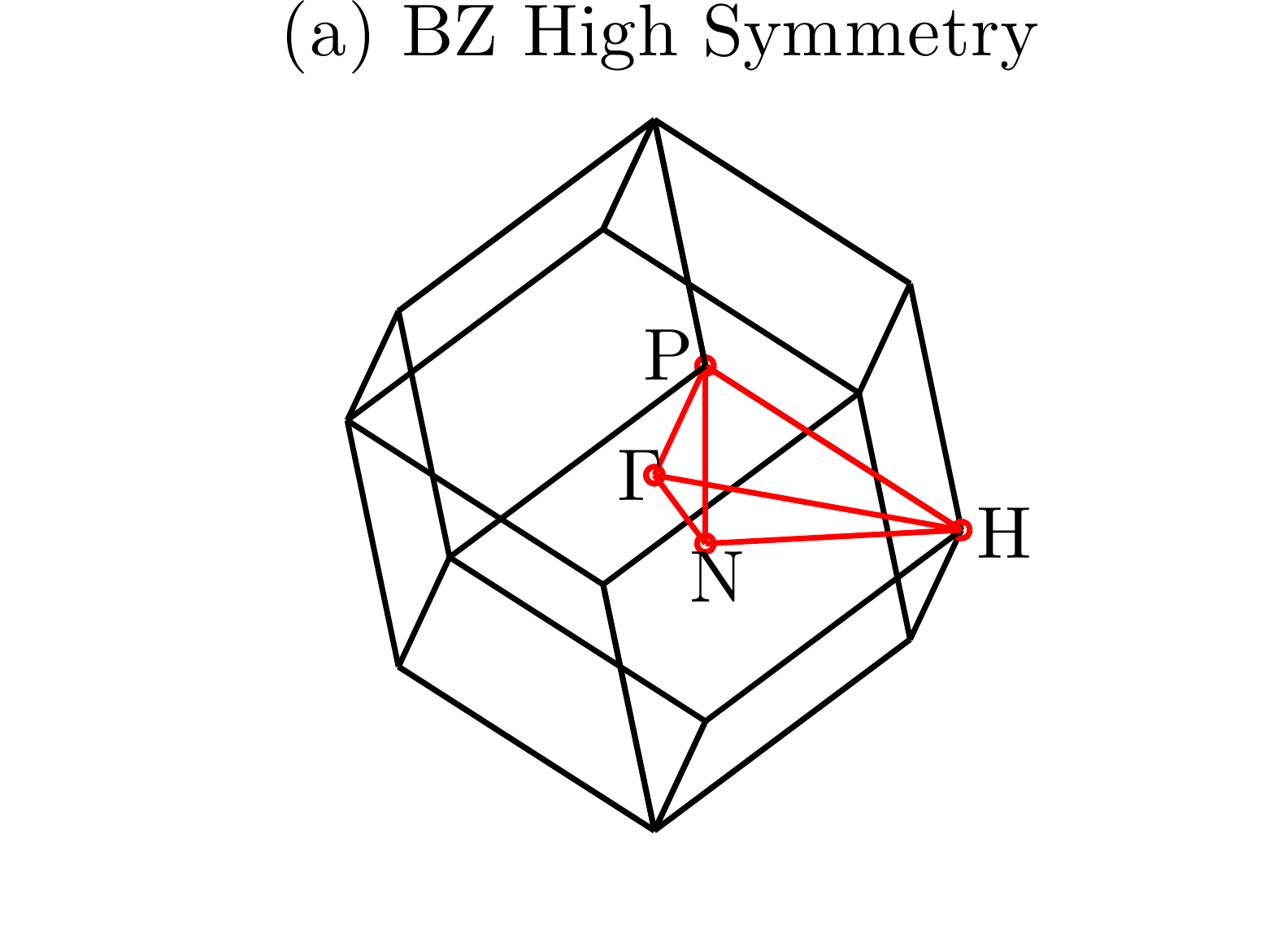} }}%
    \subfloat{{\includegraphics[width=0.5\columnwidth]{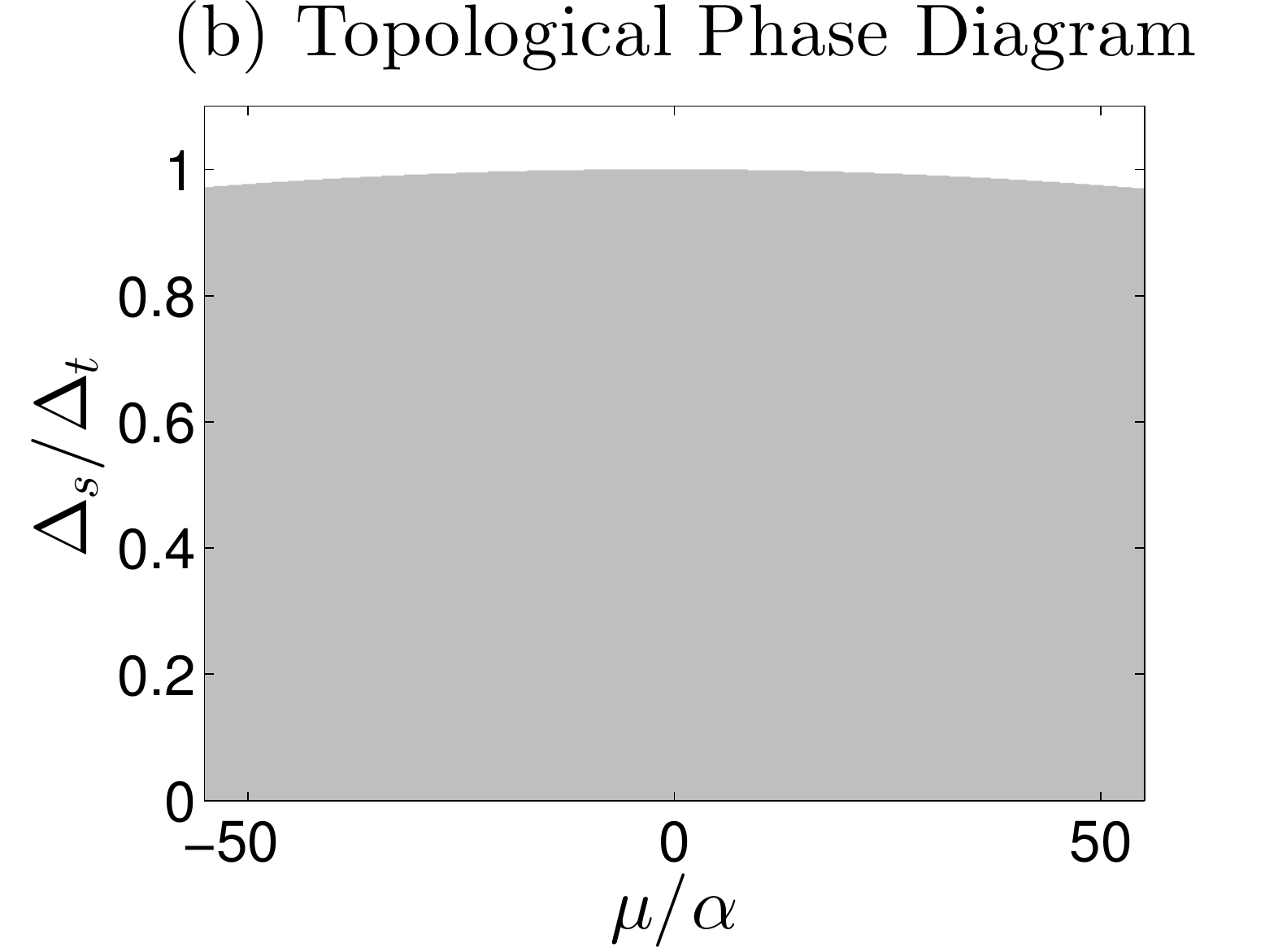} }}%
    \caption{\label{TetrahedralStuff}(a) The high symmetry points and axes in the BZ of a tetrahedral crystal. (b) The topological phase diagram for different values of the chemical potential. The Fermi surface is open (closed) for $\text{sign}[\mu] = \mp 1$.
White areas indicate a gapped phase with trivial topology, $(N_\mathcal{L}, \nu)= (0,0)$; grey a nodal phase with $N_\mathcal{L}=1$ [loop defined by Eq. \eqref{LoopPath}].
}%
\end{figure}

To next-nearest neighbors in the sum over Bravais lattice sites \cite{SamokhinAnnals} the SOC vector corresponding to the tetrahedral point group $T_d$ takes the form
\begin{equation}
\label{SOC_Td}
{\bf l}_{\bf k} = \begin{pmatrix} \sin(k_x)\left[\cos(k_z) - \cos(k_y) \right]  \\  \sin(k_y)\left[\cos(k_x) - \cos(k_z) \right]  \\  \sin(k_z)\left[\cos(k_y) - \cos(k_x)\right]  \end{pmatrix}
\end{equation}
with no free parameter $g_2$ in contrast with $O$ and $C_{4v}$. It is illustrated in Fig.~\ref{SOCplotTd}.
This SOC exhibits line nodes in the BZ along the paths $\Gamma \rightarrow \text{P} \rightarrow \text{H} \rightarrow \Gamma$ and $\text{P} \rightarrow \text{N}$ in Fig. \ref{TetrahedralStuff} (a). Just like for $C_{4v}$ the line nodes intersect the negative helical Fermi surface for all values of $\mu$, i.e. all Fermi surface geometries, given a BCC first order tight-binding dispersion. Thus $\min|{\bf l}({\bf k}^-_F)| = 0$ and there are no gapped phases with $\text{sign}[\Delta_-({\bf k}^-_F)] = -1$, which can be seen in the topological phase diagram in Fig. \ref{TetrahedralStuff} (b). Just like for $C_{4v}$ there are only two distinct topological phases; one gapped trivial, $(N_\mathcal{L}, \nu)= (0,0)$, and a nodal non-trivial, $N_\mathcal{L} = 1$, phase.

\begin{figure}[t] 
\begin{tabular}{ll}
    \subfloat{{\includegraphics[width=0.48\columnwidth,trim=30 -10 18 0, clip]{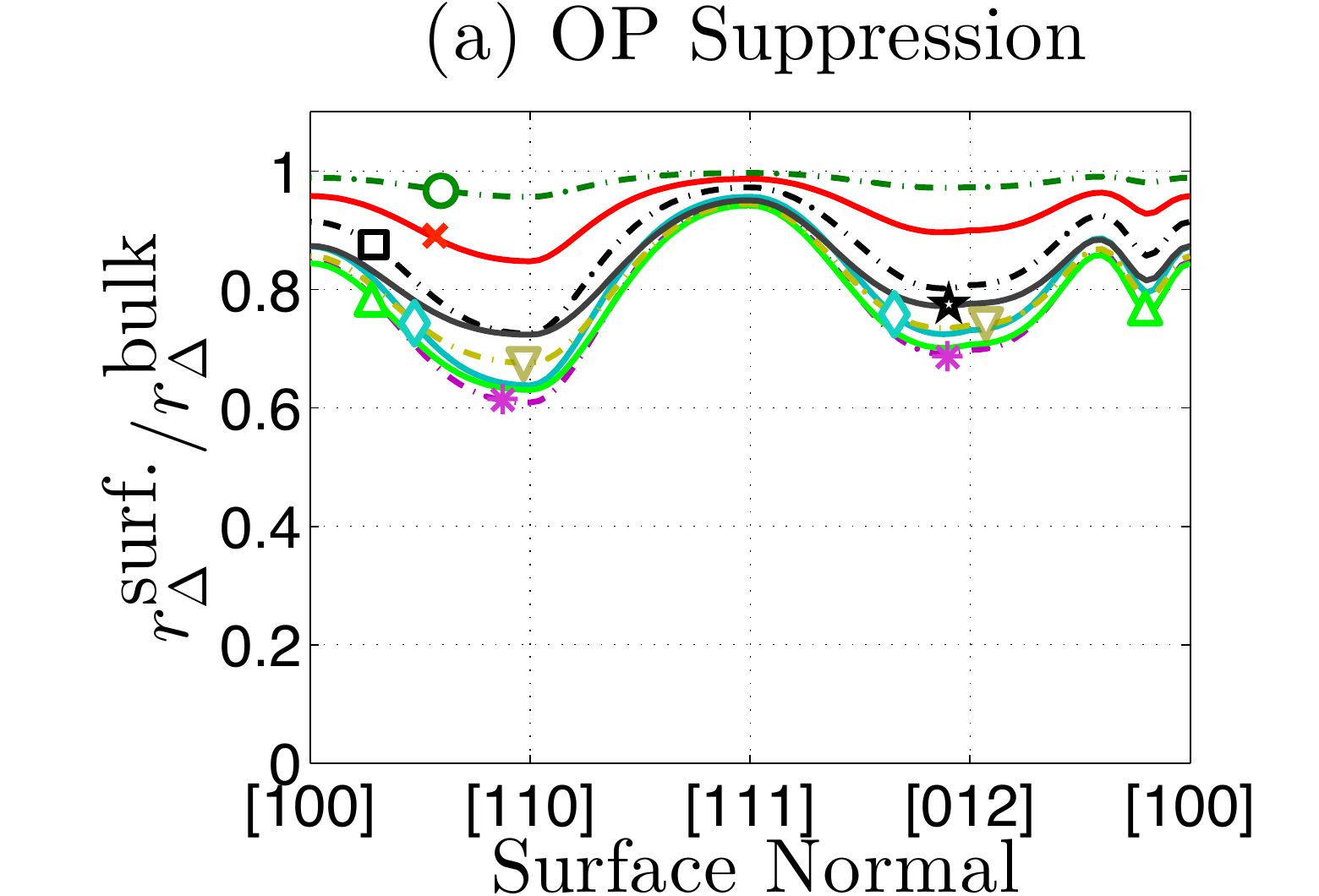} }} &
    \subfloat{{\includegraphics[width=0.52\columnwidth,trim=5 -10 0 0, clip]{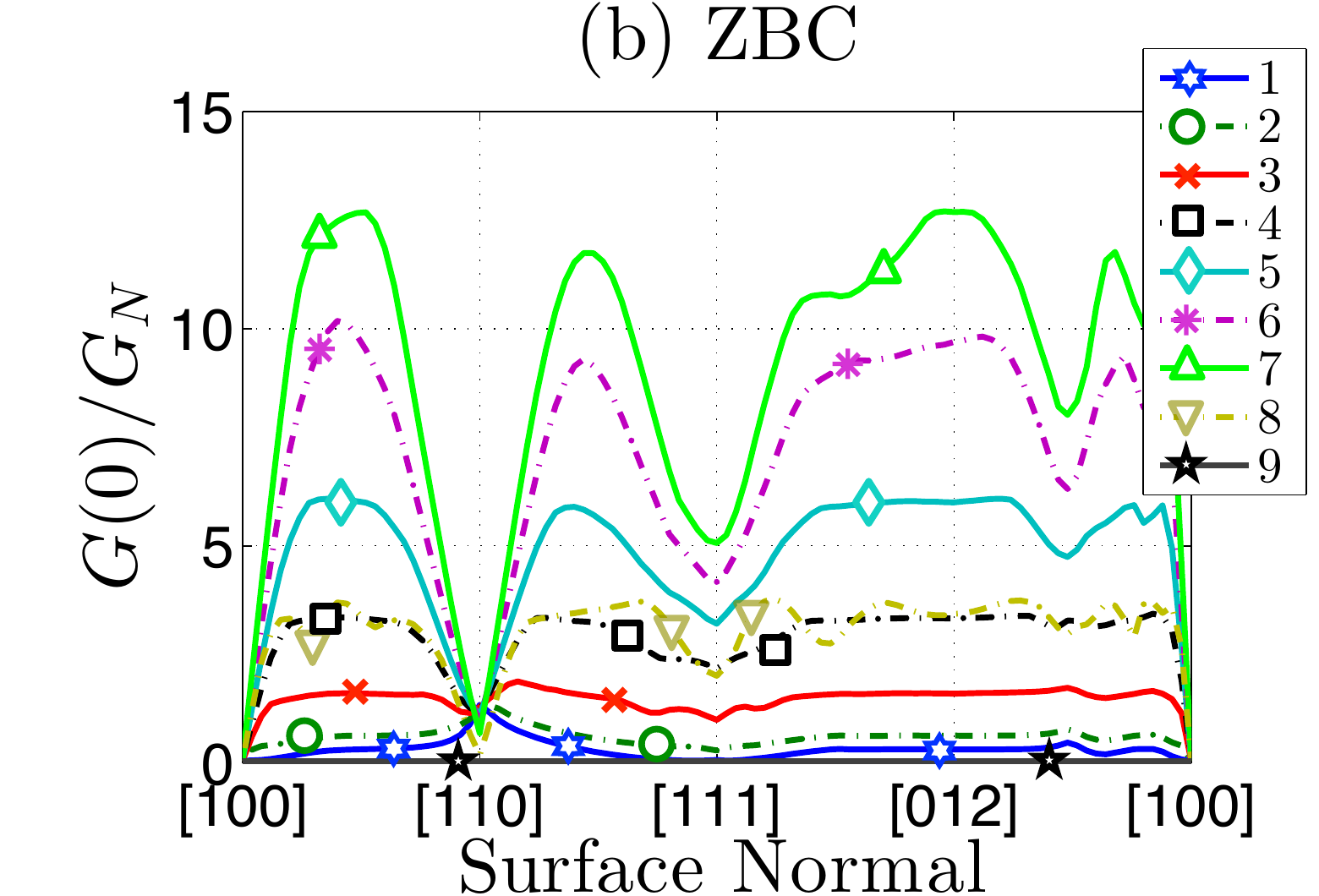} }}
\end{tabular}
\caption{\label{TetrahedralSuppressionAndZBCP}(a) The quantity $r^\text{surf.}_\Delta/r^\text{bulk}_\Delta = \left[ \Delta_s/\Delta_t \right]^\text{surf.} \cdot \left[\Delta_t/\Delta_s \right]^\text{bulk}$ as a measure of the order parameter surface suppression. (b) The zero-bias conductance for the same surface normals as in (a). The numbers in the legend holds for both plots and correspond to the columns in table \ref{RatiosPointGroups} showing the scaled singlet to triplet ratios.}
\end{figure}

Due to there being no free parameter to vary the self-consistent order parameter is only calculated with this single SOC vector for this point group. This is done for nine values of the scaled bulk singlet to triplet ratio, $r^\text{bulk}_\Delta \in [0, 1.1]$, with one active channel. The exact values are shown in table \ref{RatiosPointGroups}. The suppression of these order parameters is shown in Fig. \ref{TetrahedralSuppressionAndZBCP} (a) for the same range of surface normals as for the previous examined point groups. Here the largest suppression is for ${\bf n} = (1,1,0)$ and ${\bf n} = (0,1,2)$ and barely any suppression at all for ${\bf n} = (1,1,1)$.

In Fig. \ref{TetrahedralSuppressionAndZBCP} (b) the zero-bias conductance for these order parameters and surface normals are shown. Unsurprisingly the ZBC is very small for the high-symmetry axes ${\bf n} = (1,0,0)$ and ${\bf n} = (1,1,0)$, it is quite large, but still a local minima, for the high-symmetry axis ${\bf n} = (1,1,1)$, and larger still in between these surface normals.

\begin{figure*} \centering
%\subfloat{{\includegraphics[width=0.65\columnwidth]{Fig13a_gray.pdf} }}%
\subfloat{{\includegraphics[width=0.65\columnwidth]{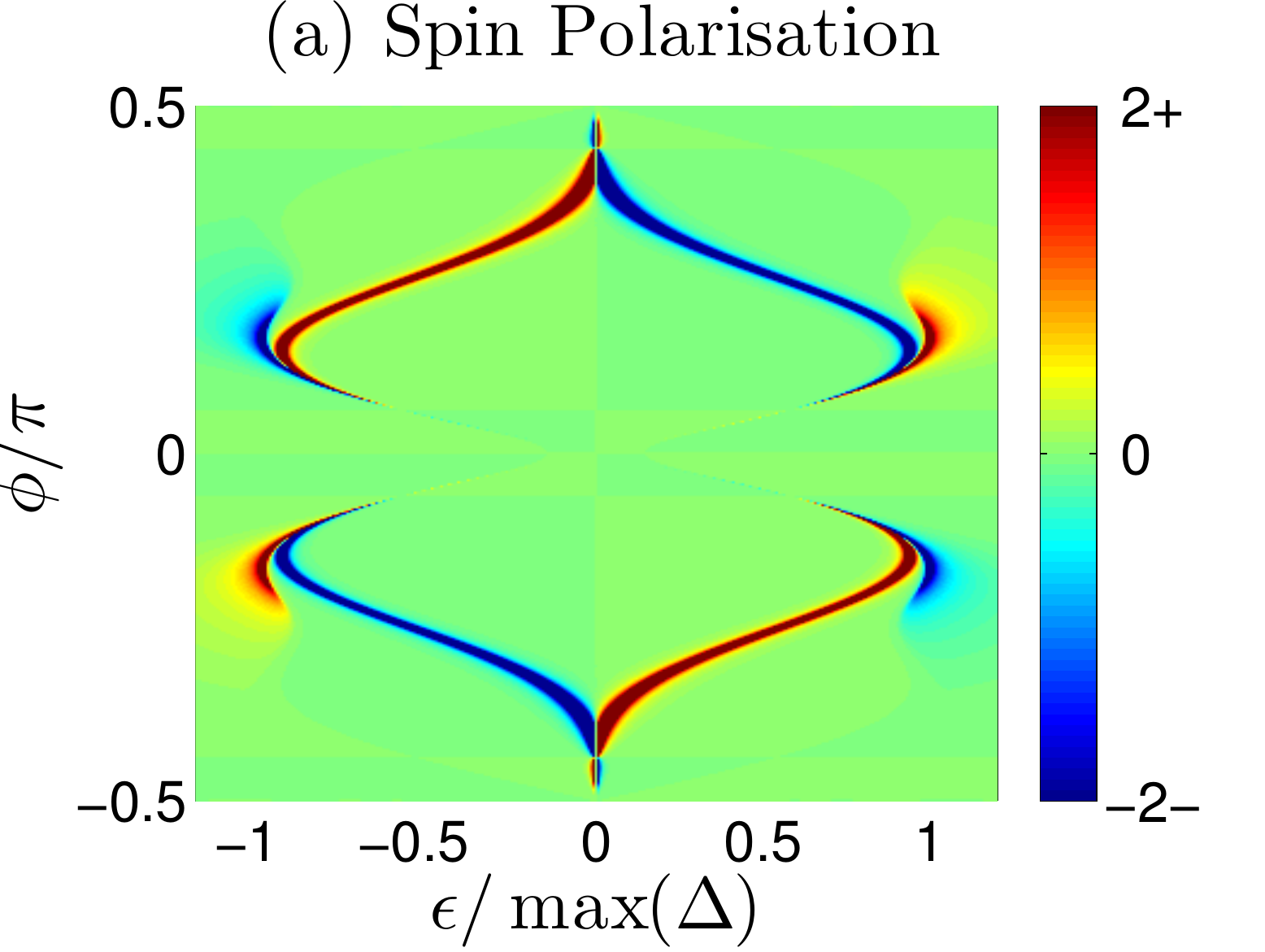} }}%
\subfloat{{\includegraphics[width=0.65\columnwidth]{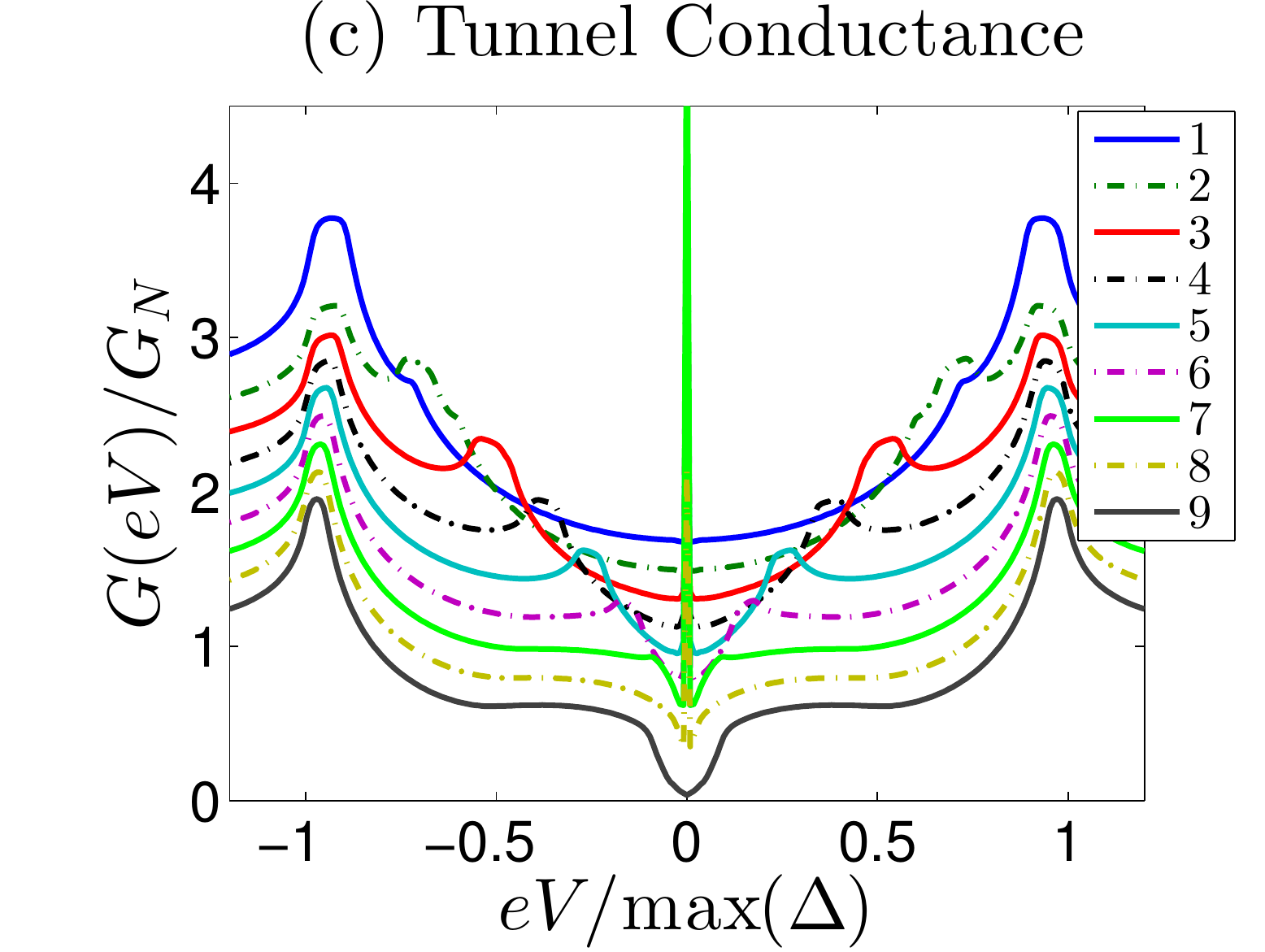} }}%
\subfloat{{\includegraphics[width=0.65\columnwidth]{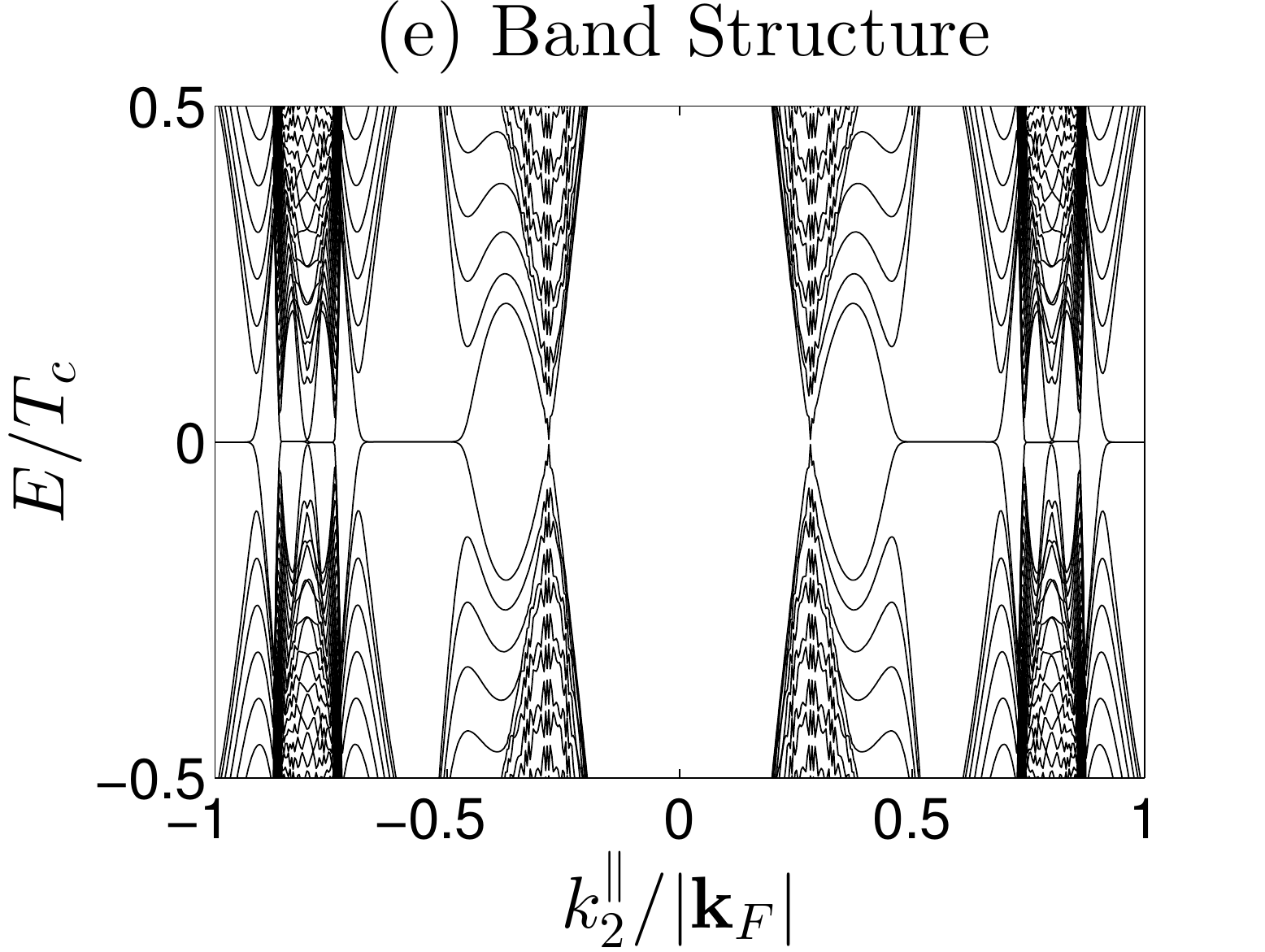} }}%
\\
\subfloat{{\includegraphics[width=0.65\columnwidth]{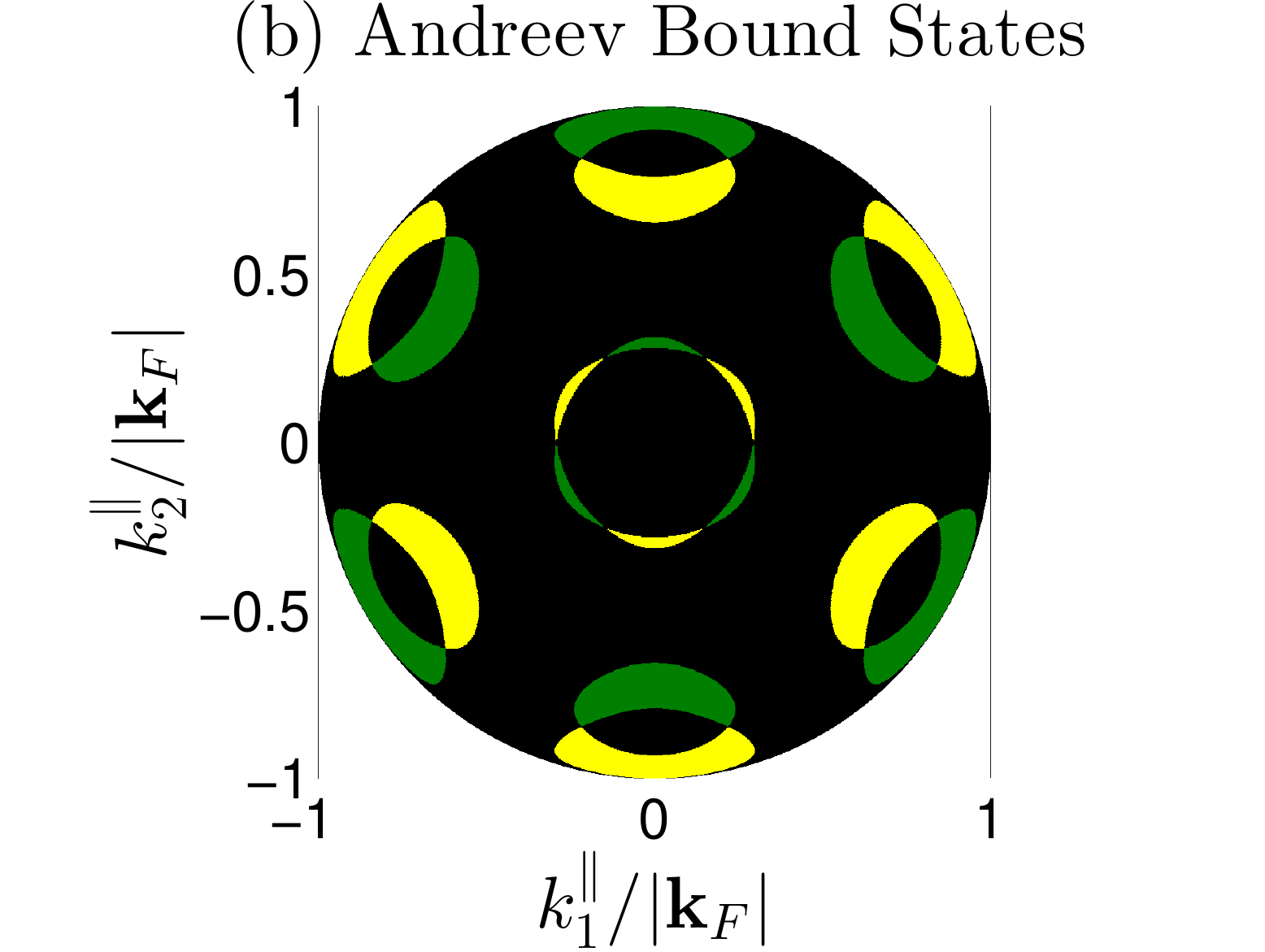} }}%
\subfloat{{\includegraphics[width=0.65\columnwidth]{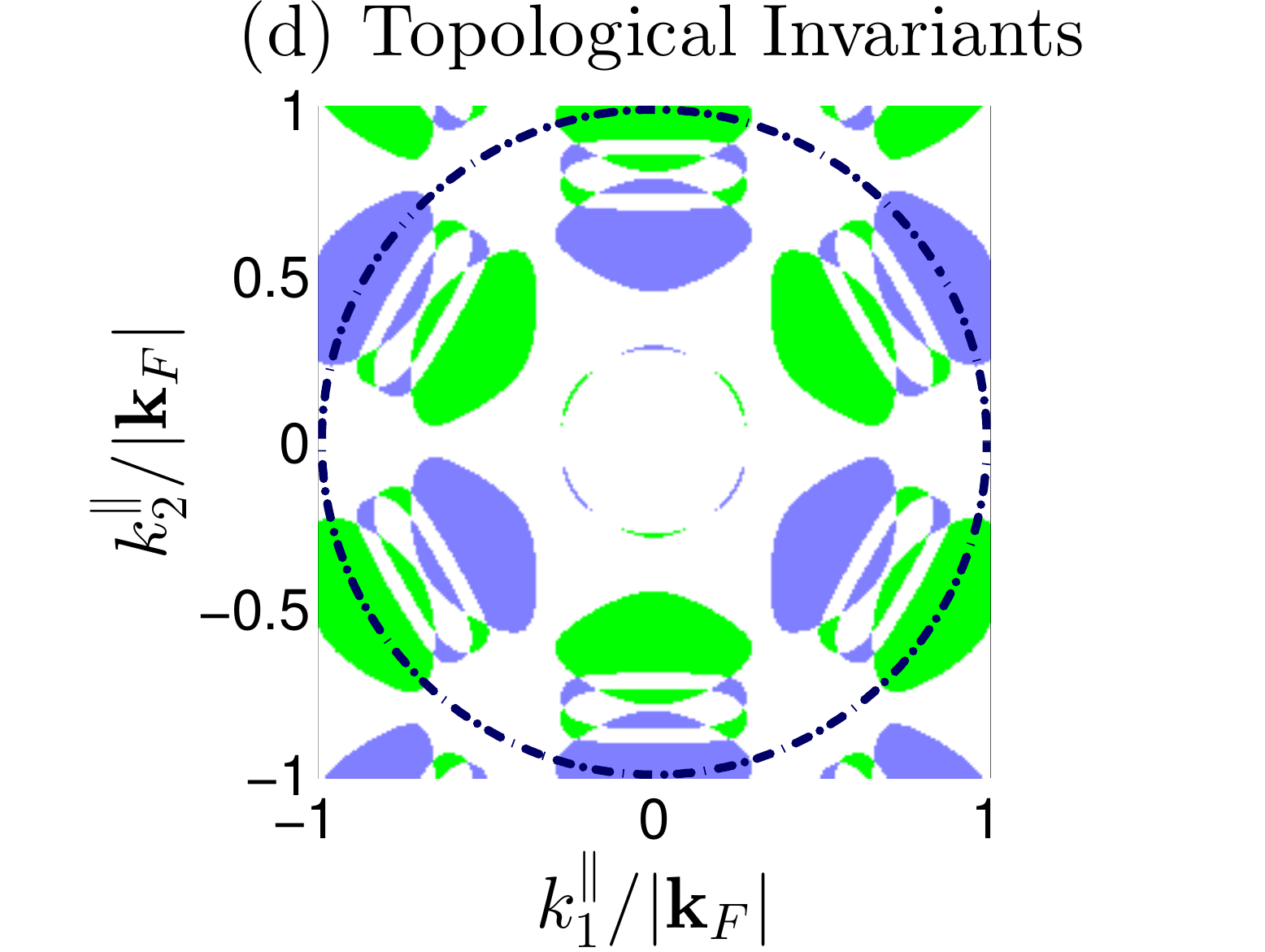} }}%
\subfloat{{\includegraphics[width=0.65\columnwidth]{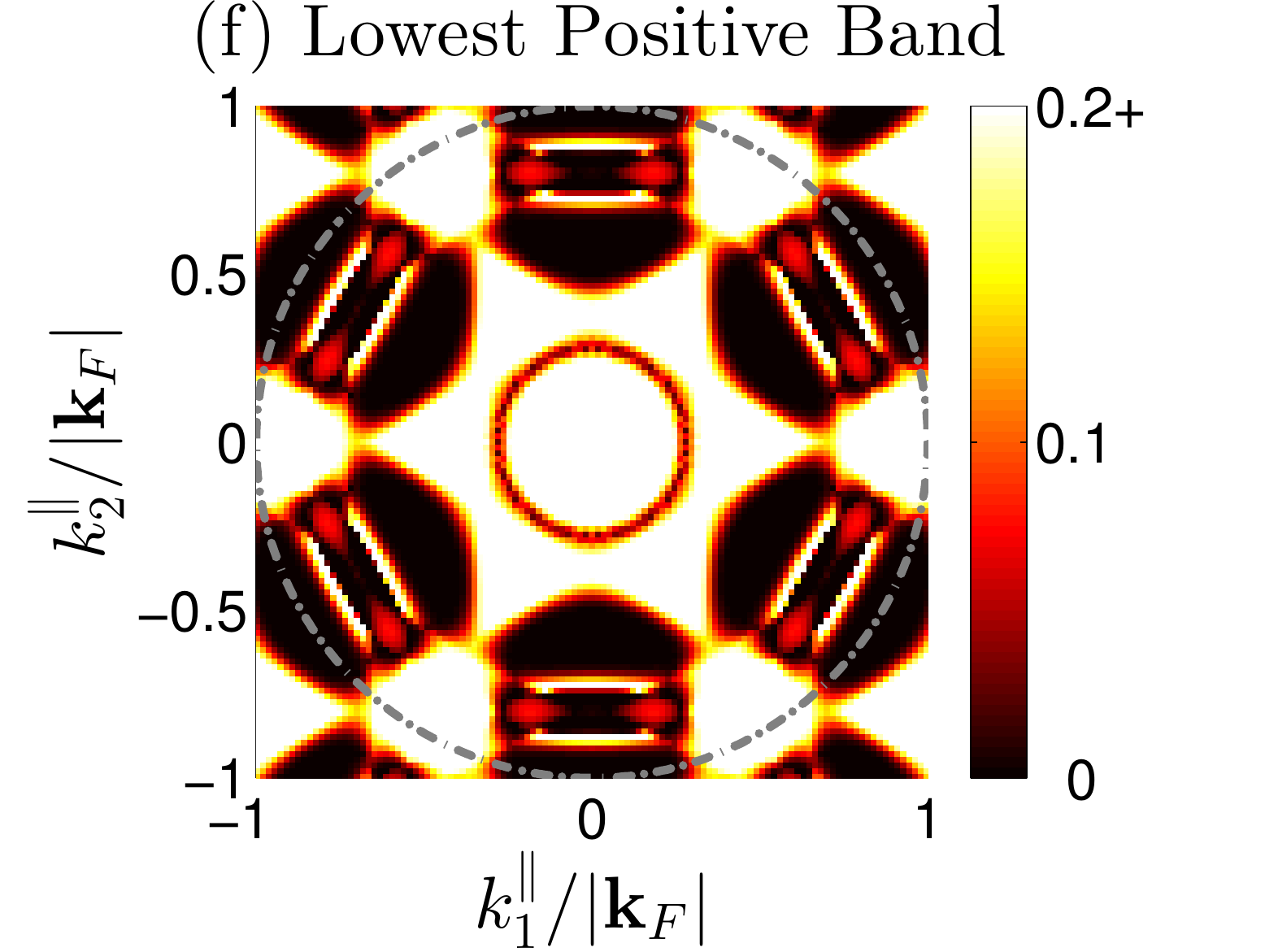} }}%
    \caption{\label{MoreTetrahedralStuff}All plots are for the tetrahedral point group $T_d$. See the caption of Fig. \ref{MoreCubicStuff}, here $r^\text{bulk}_\Delta = 0.69$ in (b) and (d) - (f). }
\end{figure*}

The ABS are spin polarized for this point group as well. In Fig. \ref{MoreTetrahedralStuff} (a) the quantity $N^{(z)}({\bf k})$, Eq. \eqref{Nxyz}, is plotted for a pure triplet order parameter and ${\bf n} = (1,0,0)$. For a pure triplet $N^{(x)} = N^{(y)} = 0$. Self-consistency does not drastically alter the ABS for this surface normal due to the states being predominantly located at small energies for glancing trajectories.

The momentum-resolved zero-energy ABS for $r^\text{bulk}_\Delta = 0.69$ are shown in Fig. \ref{MoreTetrahedralStuff} (b). The states in middle are from the non-overlapping parts of the projection of the nodal rings around the $\Gamma \rightarrow \text{P}$ high symmetry axis, and the ones around the edges of the disk from the projection of the nodal rings around $\Gamma \rightarrow \text{H}$ and $\Gamma \rightarrow \text{N}$.

Just like for the other point groups ZBCPs are seen in the tunnel conductance spectra, with $t_0 = 10^{-\frac{1}{2}}$, for 
${\bf n} = (1,1,1)$ and 
singlet to triplet ratios in the interval 
$\min|{\bf l}_{{\bf k}_F}| < \Delta_s/\Delta_t < \max|{\bf l}_{{\bf k}_F}|$, i.e. $0 < r^\text{bulk}_\Delta < 1$, see Fig. \ref{MoreTetrahedralStuff} (c). ABS given by Eq. \eqref{Feq} only appear for $0 \leq r^\text{bulk}_\Delta \leq 0.28 $, and then not for ${\bf k}_\parallel = 0$ which is the most important momentum when calculating the tunnel conductance \footnote{The angle-resolved tunnel conductance is proportional to the DOS weighted by $\cos(\alpha_{\bf k})$ as well as the transmission amplitude squared.}. Hence the ZBCPs emanating from valleys in the spectra.

The non-trivial values of the topological invariant $N_{(111)}$ ($W_{(111)}$ being trivial for this singlet to triplet ratio) is shown in Fig. \ref{MoreTetrahedralStuff} (d). The dashed circle is projection of the spherical Fermi surface used in the quasiclassical calculations. Compared to the other point groups considered the spherical Fermi surface approximation does not work as well due to the actual Fermi surface bulging out in the ${\bf k} = (1,1,1)$ direction. Furthermore, the BZ is not cubic and thus the line integral defining $N_{(lmn)}$ potentially goes through Fermi surfaces from adjacent BZs, which is precisely what happens for this surface normal. The slightly complicated structure near the circle thus stems from the partial overlap of the nodal rings of the adjacent Fermi surface combined with $N_{(lmn)}$ being additive.

Zero-energy flat bands are seen in the band structure for this point group and surface normal as well, shown for $r^\text{bulk}_\Delta = 0.69$ along the $k^\parallel_2$-axis, with $k^\parallel_1 = 0$ and $L = 1.3\cdot 10^4$ layers, in Fig. \ref{MoreTetrahedralStuff} (e).

As seen in the plot of the lowest positive band, Fig. \ref{MoreTetrahedralStuff} (f), the zero-energy states around the origin are somewhat patchy at this resolution, hence the slight gap at $k^\parallel_2 \approx \pm 0.28|{\bf k}_F|$ in Fig. \ref{MoreTetrahedralStuff} (e). Furthermore, there is a small gap at $k^\parallel_2 \approx \pm 0.8|{\bf k}_F|$ in Fig. \ref{MoreTetrahedralStuff} (e), but this is not an artifact of the lower resolution of the band structure compared to the topological invariant plot, Fig. \ref{MoreTetrahedralStuff} (d), as there is a small region separating $N_{(111)} = \pm 1$ at these momenta.

\section{Conclusions}

We have theoretically studied noncentrosymmetric superconductors self-consistently for the point groups $O$, $C_{4v}$, and $T_d$, with a closed Fermi surface. Four values of $g_2$, parameterizing the relative weight of first and second order contributions in the spin-orbit coupling (SOC), given by the Bravais lattice sum up to next-nearest neighbors, were chosen for $O$ in order to investigate all its gapped topological phases for a closed Fermi surface. The point groups $C_{4v}$ and $T_d$ were shown to have no gapped topological phases, yet two values of $g_2$ were chosen for $C_{4v}$ in order to study the effect of second order contributions in the SOC vector. For $T_d$ no higher order terms in the SOC were seen up next-nearest neighbors.

The reason for the existence of gapped topological phases for $O$ was shown to be due to the fact that the SOC only vanishes in the Brillouin zone at high symmetry \emph{points}, whereas the SOC vanishes at certain high symmetry \emph{axes} for $C_{4v}$ and $T_d$. It was shown for $O$ that the topology changes at the Lifshitz transition, i.e. at the transition point between an open and closed Fermi surface. This does not happen for $C_{4v}$ and $T_d$ and the Lifshitz transition is topologically seamless.

In the bulk it was shown that there are two distinct mixed states; with one or two nucleation channels. In both cases it was demonstrated that there is a possibility of a cross-over from dominating singlet to dominating triplet, or vice versa, with decreased temperature. Depending on the material this could be important if experiments are done at different temperatures. With two nucleation channels there is a possibility of a second phase transition at the subdominant critical temperature and it was shown by explicit construction that the subdominant channel for certain parameter values indeed has lower free energy. If this can be extended to more complicated Fermi surface geometries and parameter values remains to be seen.

The order parameter suppression's dependence on surface orientation and singlet to triplet order parameter ration was studied for a range of different surface normals. The suppression was seen to be highly dependent on surface orientation.

The Andreev bound states (ABS) are found to be spin polarized with different polarization axes for different singlet to triplet ratios. The order parameter suppression affects the ABS heavily for glancing trajectories and sub-gap energies close to the gap, and less for smaller energies. Zero-energy states are not affected by the calculated suppression. Thus the zero-bias conductance peaks are present in the non-self-consistent tunnel conductance as well. 

We showed that the zero-energy surface states are topological in nature. Thus it is clear that the calculated suppression should not affect the zero-energy states due to the gap not vanishing at any distance from the surface. If this can happen for other parameters and/or surface orientations is an open question.

\acknowledgements
We appreciate the highly stimulating atmosphere within the Hubbard Theory Consortium and during the annual ``Condensed Matter Physics in the City'' events in London. 
N.W. would like to thank Roland Grein for discussions in the early stages of the work, and Patric Holmvall for discussions in the later stages.
M.E. acknowledges support by EPSRC (Grant No. EP/J010618/1 and EP/N017242/1). N.W. acknowledges financial support by the Southeast Physics Network (SEPnet) for his Ph.D. study at Royal Holloway, University of London. 

\appendix
\section{Temperature Dependence of the Gap}
\label{SectionTemperatureDependence}

The self-consistency equation for the order parameter Eq. \eqref{op_explicit}, 
can be written symbolically in the form of a fixed point equation
\begin{eqnarray}
\label{FixedPointEq}
\boldsymbol{\Delta} &=& {\bf F}(\boldsymbol{\Delta})
\end{eqnarray}
where $\boldsymbol{\Delta} = (\Delta_s, \Delta_t)^T$, and the function ${\bf F}(\boldsymbol{\Delta})$ is simply a short-hand notation for the right hand side of Eq. \eqref{op_explicit}. 
Any $\boldsymbol{\Delta}$ that obeys eq. \eqref{FixedPointEq} is called a fixed point. 
Then a iteration scheme is employed to find a convergence to a fixed point.
This yields a series of points $\boldsymbol{\Delta}_{1}$, $\boldsymbol{\Delta}_{2}$, $\ldots $, which hopefully converges to a solution.
The procedure is said to have converged when the difference between iterations is sufficiently small
\begin{eqnarray}
\label{ConvergenceCriteria}
\frac{|\boldsymbol{\Delta}_{n+1} - \boldsymbol{\Delta}_{n} |}{|\boldsymbol{\Delta}_{n} |} &<& c
\end{eqnarray}
where the number $c$ is the convergence criterion.
In the bulk the fixed points can be obtained by computing ${\bf F}(\boldsymbol{\Delta})$ for a vast number of points. 

We illustrate the method for the case of two attractive channels.
Because the number of possible independent attractive fixed points is equal to the number of positive eigenvalues to the matrix $L$, one has for values of $(v_s,v_t,v_m)$ in the yellow oval in Fig. \ref{NucleationChannels} two nucleation channels. 
However, the subdominant channel does in general not nucleate at $T_\text{c}$, but at a lower temperature, $T^\text{sub.}_\text{c} < T_\text{c}$. Thus, if one follows the procedure in the previous paragraph for the initial guesses $\boldsymbol{\Delta}_0$ one will not see the possible transition to the subdominant channel. What is needed in this case is to calculate the order parameter with increasing temperature instead of decreasing. By computing a few iterations, $n\sim20$, at a sufficiently low temperature, say $T=0.1T_\text{c}$ (which must be smaller than $T^\text{sub.}_\text{c}$ obviously), for a number of random initial guesses, an attractive fixed point corresponding to the subdominant channel is obtained, and is denoted $\boldsymbol{\Delta}^\text{sub.}$. For the lowest temperature the initial guess will thus be $\boldsymbol{\Delta}^\text{sub.}$, and subsequent guesses $\boldsymbol{\Delta}_0(T+\delta T) = \boldsymbol{\Delta}_n(T)$. By calculating the order parameter this way, it will converge to the subdominant channel value until $T \leq T^\text{sub.}_\text{c}$. At $T=T^\text{sub.}_\text{c}$ the order parameter transitions to the dominant channel value due to it being the only attractive fixed point at these temperatures, unless the subdominant channel value at $T=T^\text{sub.}_\text{c}$ is zero, $\boldsymbol{\Delta}_n(T^\text{sub.}_\text{c}) = {\bf 0}$, in which case it will stay zero. In this manner, the subdominant critical temperatures are obtained. 

%%%
\begin{figure}[b] \centering
\includegraphics[width=1.0\columnwidth]{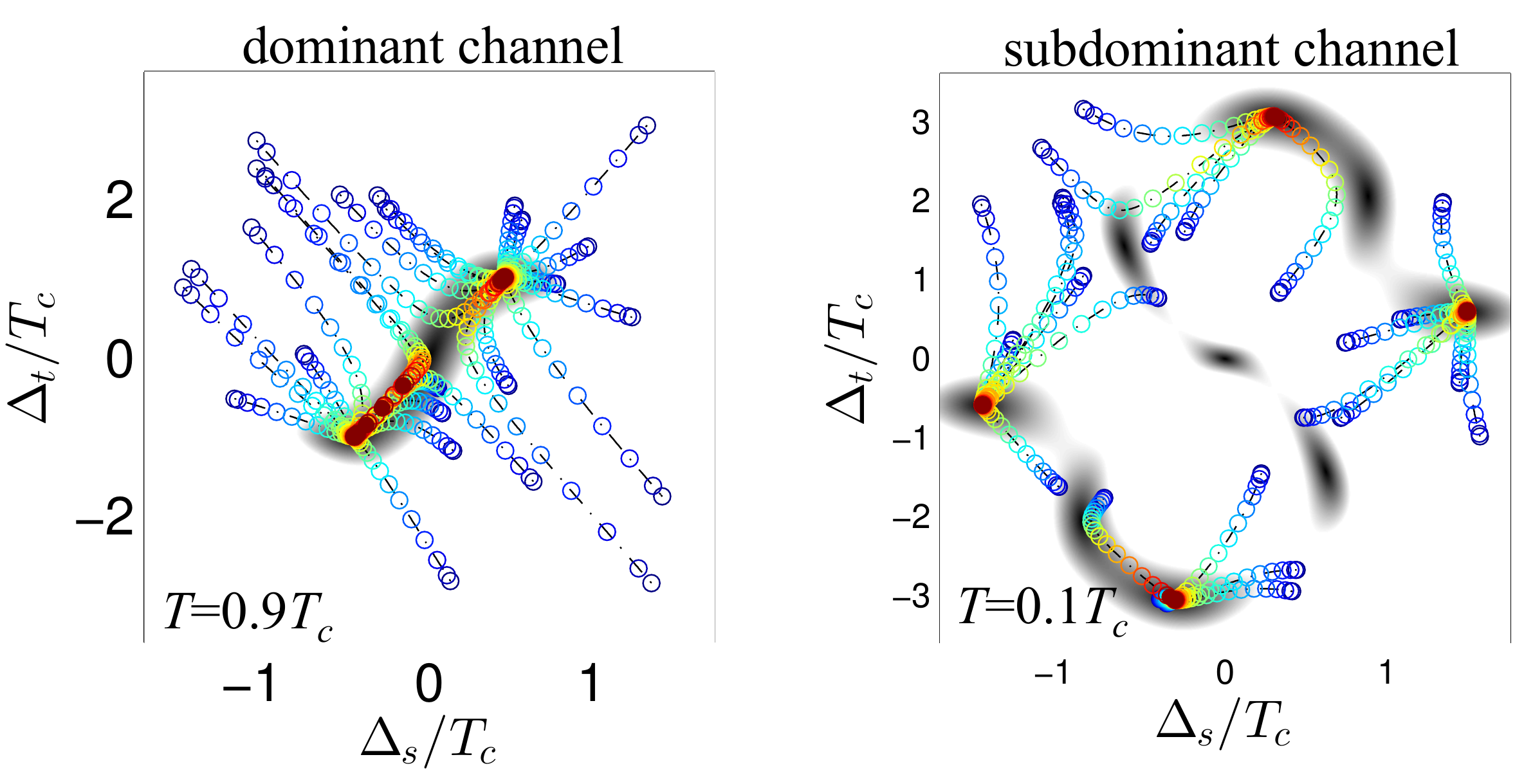}
\caption{
\label{FixedPointsDominant}
Two examples for convergence diagrams for a mixed order parameter with two active channels (dominant and subdominant), for point group $O$ with $g_2=0$, 
and $(v_s,v_t,v_m) = (\langle|{\bf l}_{\bf k}|^2\rangle, 1, -0.1\langle|{\bf l}_{\bf k}|^2\rangle)$ (ignoring normalization). 
See text for explanation.
}
\end{figure}
%%%

A choice of parameter values yielding an admixture of singlet and triplet with an attractive subdominant channel is e.g. $(v_s,v_t,v_m) = (1, 1/\langle|{\bf l}_{\bf k}|^2\rangle, -0.1)$ (ignoring normalization). 
In Fig.~\ref{FixedPointsDominant} examples for the fixed point iteration are shown for the point group $O$ (the plots look qualitatively similar for all other point groups).
The grey blobs correspond to the function $f(\boldsymbol{\Delta}) = |\boldsymbol{\Delta} - {\bf F}(\boldsymbol{\Delta})|$. Darker indicates smaller values of $f(\boldsymbol{\Delta})$, and pure black indicates the existence of a fixed point. The colored circles connected by lines show the convergence of 25 random initial guesses, $\boldsymbol{\Delta}_0$, progressing a number of iteration steps.
The colors of the circles indicate the iteration number $n$. Starting with dark blue for $n=0$, transitioning through cyan, green, yellow, and ending with red for $n=n_\text{max}$.
Any fixed point $\boldsymbol{\Delta}_{n_\text{max}}$ converges to is an attractive fixed point, however there are repulsive fixed points present in the subdominant channel.
Concentrating on the subdominant channel, Fig.~\ref{FixedPointsDominant}, one notices in addition to the attractive fixed points also two repulsive fixed points. From numerical investigations this seems to be a general feature, and the fixed points roughly fall on a parallelogram with the attractive fixed points at the vertices.

A criterion if a second phase transition exists can be obtained from the condensation energy, Eq.~\eqref{dW}.
Thus there is a second phase transition if it holds that
\begin{eqnarray}
\left[ |\Delta_s|^2 + 2 | \Delta_s \Delta_t|  \langle |{\bf l}_{\bf k}|\rangle + |\Delta_t|^2 \langle |{\bf l}_{\bf k}|^2\rangle \right]^\text{sub.}> \nonumber \\
\label{CondEnergyInequality}
\qquad \left[ |\Delta_s|^2 + 2 | \Delta_s \Delta_t|  \langle |{\bf l}_{\bf k}|\rangle + |\Delta_t|^2 \langle |{\bf l}_{\bf k}|^2\rangle  \right]^\text{dom.}\; .
\end{eqnarray}
at zero temperature. For certain parameters this is indeed the case. 
In general, how small $T^\text{sub.}_\text{c}$ can be without losing the second phase transition depends on the point group.
The key to get a second phase transition is to choose $(v_s,v_t,v_m)$ in such a way as to get a dominant channel with a large triplet component, as well as a rather large subdominant critical temperature. 

\section{Zero-bias Andreev bound states}
\label{SectionZBCPs}

%%%
\begin{figure*}[!pt] \centering
\includegraphics[width=2.0\columnwidth]{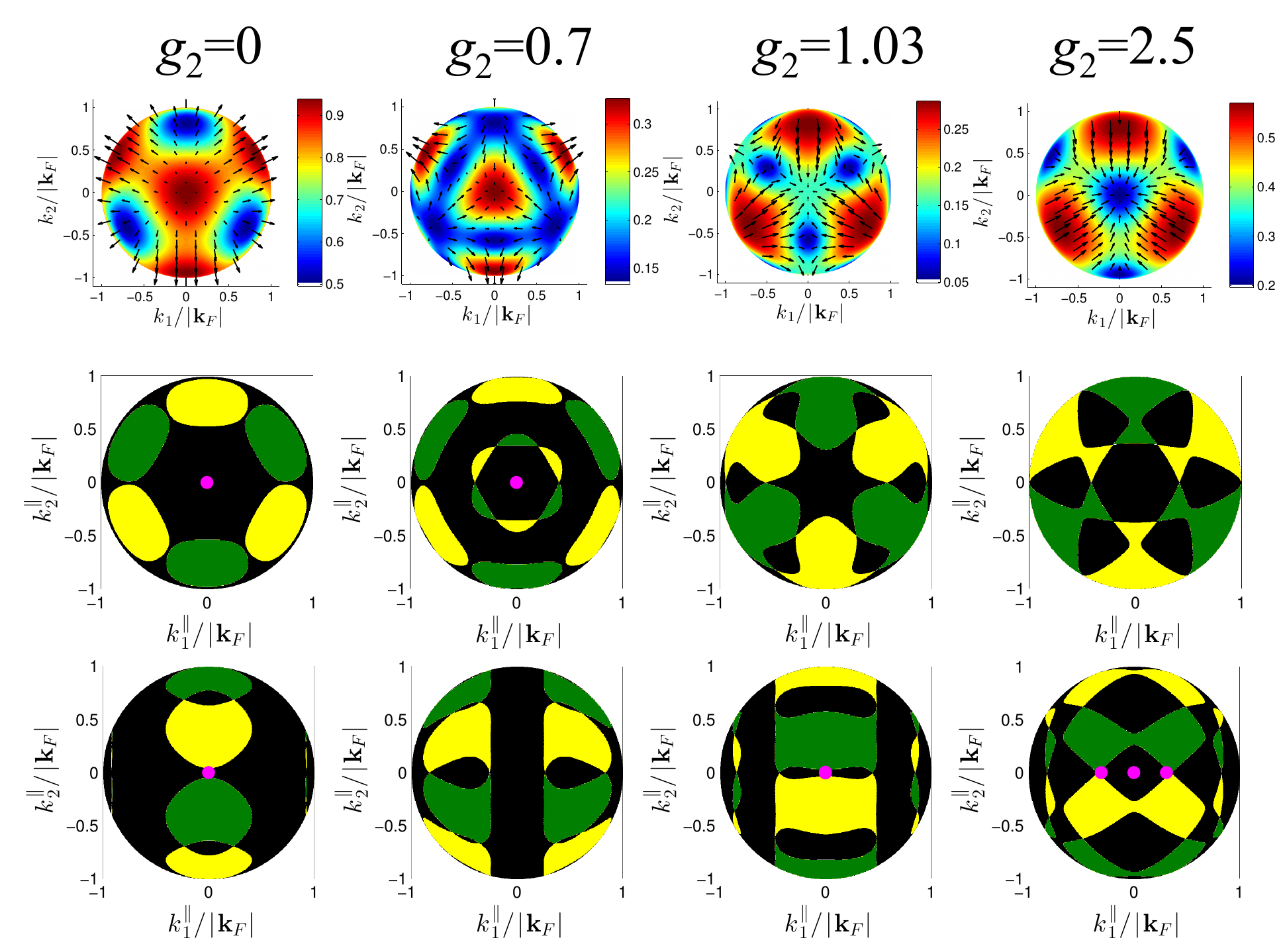}
\caption{\label{topolO}
The first row shows th SOC vector field projected on the ${\bf k}=(1,1,1)$ direction for the point group $O$ and indicated parameters $g_2$. The color coding corresponds to $|{\bf l}_{{\bf k}_F}|$, and the arrows show the direction of ${\bf l}_{{\bf k}_F}$ for selected points. 
Note the different color scales (blue and red correspond to nonzero local minima and local maxima).
The second and third rows show the angle resolved surface ABS at zero energy for surface normals ${\bf n} = (1,1,1)$ (second row) and ${\bf n}=(0,1,2)$ (third row). The green and yellow regions are given by solution to Eqs. \eqref{upsilon1-1} and \eqref{upsilon-11} respectively. The magenta colored dots correspond to solutions to Eqs. \eqref{upsilon0-1} - \eqref{upsilon-1-1}. The order parameters for the four columns from left to right correspond to the scaled singlet to triplet ratios 
$r^\text{bulk}_\Delta = \{0.74,0.71,0.66,0.70 \}$, respectively. Only one pairing channel is active, and $T=0.2T_\text{c}$.}
\end{figure*}
%%%

In the bulk, for zero energy and for real order parameter, the coherence functions take a particularly simple form,
\begin{eqnarray}
\gamma\left({\bf k}\right) &=& i \begin{pmatrix} \text{sgn} \left[\Delta_+\left({\bf k}\right)\right]t_+\left({\bf k}\right) & 0 \\ 0 & \text{sgn} \left[\Delta_-\left({\bf k}\right)\right]t_-\left({\bf k}\right)  \end{pmatrix} \; ,\\
\tilde \gamma\left(\underline{\bf k}\right) &=& -i \begin{pmatrix} \text{sgn} \left[\Delta_+\left(\underline{\bf k}\right)\right]t^*_+\left(-\underline{\bf k}\right) & 0 \\ 0 & \text{sgn} \left[\Delta_-\left(\underline{\bf k}\right)\right]t^*_-\left(-\underline{\bf k}\right)  \end{pmatrix}. \nonumber \\ 
\end{eqnarray}
The values for the parameters $(v_s, v_t, v_m)$ were chosen to yield positive singlet and triplet components, and we can therefore simplify the expressions further to
\begin{eqnarray}
\label{gammaUpsilon}
\gamma\left({\bf k}\right) &=& i \begin{pmatrix}  t_+\left({\bf k}\right) & 0 \\ 0 &\Upsilon_{\bf k}t_-\left({\bf k}\right)  \end{pmatrix} \; ,\\
\tilde \gamma\left(\underline{\bf k}\right) &=& -i\begin{pmatrix}  t^*_+\left(-\underline{\bf k}\right) & 0 \\ 0 &\Upsilon_{\underline{\bf k}}t^*_-\left(-\underline{\bf k}\right)  \end{pmatrix} \; ,
\end{eqnarray}
where $\Upsilon_{\bf k} \equiv \text{sgn} \left[\Delta_s/\Delta_t - |{\bf l}_{\bf k}|\right]$. 
We are interested in zero-bias states protected by topology, for which it suffices to discuss the 
non-self-consistent order parameter, i.e. bulk values all the way to the surface. 
Because $\Upsilon_{\bf k}$ and $\Upsilon_{\underline{\bf k}}$ can take three values each there are $3^2=9$ different cases to consider. They are listed below together with the equations for surface ABS, using the (real) reflection amplitude $0\le r\le 1$.
The solutions separate naturally into three groups. The first group
\begin{eqnarray}
\label{upsilon11} \Upsilon_{\bf k}=1\; , \Upsilon_{\underline{\bf k}} = 1 & : & r^2 + 1 = 0 \\
\label{upsilon10} \Upsilon_{\bf k}=1\; , \Upsilon_{\underline{\bf k}} = 0 & : & r^2 + 1 = 0 \\
\label{upsilon01}\Upsilon_{\bf k}=0\; , \Upsilon_{\underline{\bf k}} = 1 & : & r^2 + 1 = 0 
\end{eqnarray}
has no solutions.
Therefore, there can be \emph{no} ZBCPs for $r^\text{bulk}_\Delta > \max|{\bf l}_{{\bf k}_{\text{F}}} |$, because $\Upsilon_{\bf k} = +1 \; \forall {\bf k}$. 
The second group,
\begin{eqnarray}
\label{upsilon1-1}\Upsilon_{\bf k}=1\; , \Upsilon_{\underline{\bf k}} = -1 & : & r^4 -1= 0 \\
\label{upsilon-11}\Upsilon_{\bf k}=-1\; , \Upsilon_{\underline{\bf k}} = 1 & : & r^4 - 1 = 0 
\end{eqnarray}
requires $r=1$, as well as a sign change of $\Upsilon$ when reflected at the surface. 
The third group is
\begin{eqnarray}
\label{upsilon00}\Upsilon_{\bf k}&=&0\; , \Upsilon_{\underline{\bf k}} = 0  :  F\left(\phi_l,\theta_l,\phi_{\underline{l}},\theta_{\underline{l}}  \right) + \frac{2+r^2}{r^2} = 0\\
\label{upsilon0-1}\Upsilon_{\bf k}&=&0\; , \Upsilon_{\underline{\bf k}} = -1  :  F\left(\phi_l,\theta_l,\phi_{\underline{l}},\theta_{\underline{l}}  \right) +\frac{1}{r^2} = 0\\
\label{upsilon-10}\Upsilon_{\bf k}&=&-1\; , \Upsilon_{\underline{\bf k}} = 0  :  F\left(\phi_l,\theta_l,\phi_{\underline{l}},\theta_{\underline{l}}  \right) +\frac{1}{r^2} = 0\\
\label{upsilon-1-1}\Upsilon_{\bf k}&=&-1\; , \Upsilon_{\underline{\bf k}} = -1 :   F\left(\phi_l,\theta_l,\phi_{\underline{l}},\theta_{\underline{l}}  \right) +\frac{1+r^4}{2r^2} = 0,\nonumber \\
\end{eqnarray}
where $F\left(\phi_l,\theta_l,\phi_{\underline{l}},\theta_{\underline{l}}  \right) \in  [-1,1]$ is defined after Eq.~\eqref{Feq}.
Eq.~\eqref{upsilon00} has no solution for real $r$, and the remaining equations have solutions only for $r=1$.
Thus, there are only two classes of trajectories giving rise to ABS at zero energy, the first class given by Eqs.~\eqref{upsilon1-1}-\eqref{upsilon-11} and the second class given by Eqs.~\eqref{upsilon0-1}-\eqref{upsilon-1-1}.

Fig.~\ref{topolO} shows solutions to Eqs. \eqref{upsilon0-1} - \eqref{upsilon-1-1} and Eqs. \eqref{upsilon0-1} - \eqref{upsilon-1-1} for two surface normal directions.

\section{3D winding number}
\label{appendix3DWindingNumber}

The 3D winding number is given by \cite{PhysRevB.78.195125}
\begin{eqnarray}
\nu &=& \int_{\text{BZ}}\frac{\text{d}^3 k }{24\pi^2}\: \text{Tr}\left[ 
\epsilon^{abc} M_a M_b M_c \right] \; 
\end{eqnarray}
where $M_a =q^{-1}\partial_a q $.
Introducing the notation $q({\bf k}) = C_0({\bf k})\sigma_0 + {\bf C}({\bf k}) \cdot \boldsymbol{\sigma}$ with ${\bf C} = (C_1,C_2,C_3)$, and $q^{-1} = \left[C^2_0 - |{\bf C} |^2 \right]^{-1} \left( C_0\sigma_0 - {\bf C} \cdot \boldsymbol{\sigma} \right)$, and $R = \left[C^2_0 - |{\bf C} |^2 \right] $, 
and the $4\times 4$ matrix 
\begin{eqnarray}
Z&=& \begin{pmatrix} C_0 & C_1 & C_2 & C_3 \\ \partial_x C_0 & \partial_x C_1 & \partial_x C_2 & \partial_x C_3 \\ \partial_y C_0 & \partial_y C_1 & \partial_y C_2 & \partial_y C_3 \\ \partial_z C_0 & \partial_z C_1 & \partial_z C_2 & \partial_z C_3\end{pmatrix} \; .
\end{eqnarray}
%$Z_{ij}=\hat D_iC_j$, $i,j\in \left\{1,2,3,4\right\}$, where $\hat D_i$ are the components of the operator $\hat{\underline{D}}=(1,\partial_x,\partial_y,\partial_z)$,
we find the exact formula for the trace
\begin{eqnarray}
\label{innovation}
\text{Tr}\left[ \epsilon^{abc}M_a M_b M_c \right] &=& 12i R^{-2} \det \left (Z \right) .
\end{eqnarray}

\vspace{-0.5cm}

\bibliography{bibliography}{}

\end{document}